\documentclass[12pt,english]{article}
\usepackage[T1]{fontenc}
\usepackage{geometry}
\usepackage{color}
\usepackage{babel}
\usepackage{units}
\usepackage{amsmath}
\usepackage{amsthm}
\usepackage{amssymb}
\usepackage{setspace}
\usepackage{graphicx}
\usepackage{float}
\usepackage{rotfloat}

\usepackage{booktabs}
\usepackage[unicode=true,pdfusetitle,
bookmarks=true,bookmarksnumbered=false,bookmarksopen=false,
 breaklinks=false,pdfborder={0 0 1},backref=false,colorlinks=false]
 {hyperref}
\makeatletter
\theoremstyle{definition}
\newtheorem{defn}{\protect\definitionname}
\theoremstyle{plain}
\newtheorem{lem}{\protect\lemmaname}
\theoremstyle{plain}
\newtheorem{prop}{\protect\propositionname}
\theoremstyle{plain}
\newtheorem{thm}{\protect\theoremname}

\usepackage{babel}
\usepackage{babel}
\theoremstyle{plain}
\newtheorem{assumption}{\protect\assumptionname}\theoremstyle{plain}
\newtheorem{aalgorithm}{\protect\aalgorithmname}

\usepackage{caption}
\providecommand{\assumptionname}{Assumption}
\providecommand{\definitionname}{Definition}
\providecommand{\lemmaname}{Lemma}
\providecommand{\propositionname}{Proposition}
\providecommand{\theoremname}{Theorem}
\providecommand{\aalgorithmname}{Algorithm}

\makeatother

\providecommand{\definitionname}{Definition}
\providecommand{\lemmaname}{Lemma}
\providecommand{\propositionname}{Proposition}
\providecommand{\theoremname}{Theorem}

\begin{document}

\title{Sparse spanning portfolios and under-diversification \\
 with second-order stochastic dominance}

\author{Stelios Arvanitis, Olivier Scaillet, Nikolas Topaloglou}

\date{August 2024}

\maketitle
 
\begin{abstract}
We develop and implement methods for determining whether relaxing
sparsity constraints on portfolios improves the investment opportunity
set for risk-averse investors. We formulate a new estimation procedure
for sparse second-order stochastic spanning based on a greedy algorithm
and Linear Programming. We show the optimal recovery of the sparse
solution asymptotically whether spanning holds or not. From large
equity datasets, we estimate the expected utility loss due to possible
under-diversification, and find that there is no benefit from expanding
a sparse opportunity set beyond 45 assets. The optimal sparse portfolio
invests in 10 industry sectors  
and cuts tail risk when compared to a sparse mean-variance portfolio. 
On a rolling-window basis, the number of assets shrinks to 25 assets in crisis periods, while standard factor models cannot explain the performance of the sparse portfolios. 
\\[3mm] \textbf{Keywords and phrases}: Nonparametric estimation,
stochastic dominance, spanning, under-diversification, greedy algorithm,
Linear Programming. \\[3mm] \textbf{JEL Classification:} C13, C14,
C44, C58, C61, D81, G11.

\noindent {\scriptsize{}{}{}$^{a}$ Athens University of Economics
and Business, $^{b}$ University of Geneva and Swiss Finance Institute:
corresponding author; GFRI, Bd du Pont d'Arve 40, 1211 Geneva, Switzerland;
+41 22 379 88 16; olivier.scaillet@unige.ch, $^{c}$ IPAG Business
School and Athens University of Economics and Business. Acknowledgements: We thank D.\ Batista Da Silva,  M.\ Caner,  A.-P.\ Fortin,  Y.\ Li, P.-A.\ Phutphithak, X.\ Zheng, as well as participants at SFI research days, ESEM 2023, FinEML conference, ICEFI 2023, CFE 2023, and at the Department of Economics of AUEB Research Seminar Series for helpful copmments.}{\scriptsize\par}
\end{abstract}

\section{Introduction}

We know for decades that the diversification benefits measured by
the volatility of portfolio returns are limited when we invest beyond
10 to 20 assets; see e.g.\ Evans and Archer (1968), Klemkosky and
Martin (1975), Elton and Gruber (1977). Practitioners coin the term
over-diversification. At the opposite end of the spectrum, we often
observe under-diversification among households (Campbell (2006), Calvet,
Campbell and Sodini (2007)). It might be caused by information acquisition
costs (Van Nieuwerburgh and Veldkamp (2010)), overconfidence (Anderson
(2013)), solvency requirements (Liu (2014)), or overweighting low
probability events (Dimmock et al.\ (2021)). A characteristic-based
demand system might also explain why institutions and households hold
a small set of stocks (Koijen and Yogo (2019)).

Possible over-diversification contributes to motivating the recent
literature on sparse construction of mean-variance (MV) portfolios
within the Modern Portfolio Theory (Markowitz (1952)) through imposing
constraints on the portfolio weights; see e.g.\ Jagannathan and Ma
(2003), DeMiguel et al.\ (2009), Brodie et al.\ (2009), Fan, Zhang
and Yu (2012), Ao, Li and Zheng (2019), and Caner, Medeiros and Vasconscelos
(2023). Such a construction limits the impact of transaction costs,
and eases monitoring and risk management. It also achieves statistical
regularisation of the investment portfolio in the presence of ill-conditioned
large covariance matrices. Whether limitations of diversification
benefits beyond a given small number of assets still hold true when
we leave the MV paradigm is an open problem. This paper targets the
following questions: Is it possible to build a sparse portfolio of
dimension $q$ from a large set of assets of dimension $p$ so that
we cannot get further improvement from considering additional assets
in a second-order stochastic dominance (SSD) paradigm? If not, how
much do we lose by limiting ourselves to this sparse portfolio in
terms of expected utilities compatible with SSD? Can we design an
optimization algorithm to compute this sparse portfolio from available
data? Do we have the asymptotic statistical guarantee that we cannot
improve on the estimated expected utility loss due to under-diversification
by considering another sparse portfolio of the same fixed dimension?

The theory of stochastic dominance (SD) gives a systematic framework
for analyzing investor behavior under uncertainty (see Chapter 4 of
Danthine and Donaldson (2014) for an introduction oriented towards
finance). Stochastic dominance ranks portfolios based on general regularity
conditions for decision making under risk (Hadar and Russell (1969),
Hanoch and Levy (1969), and Rothschild and Stiglitz (1970)). SD uses
a distribution-free assumption framework which allows for nonparametric
statistical estimation and inference methods. We can see SD as a flexible
model-free alternative to MV dominance of Modern Portfolio Theory
(Markowitz (1952)). The MV criterion is consistent with Expected Utility
for elliptical distributions such as the normal distribution (Chamberlain
(1983), Owen and Rabinovitch (1983), Berk (1997)) but has limited
economic meaning when we cannot completely characterize the probability
distribution by its location and scale. Simaan (1993), Athayde and
Flores (2004), and Mencia and Sentana (2009) develop a mean-variance-skewness
framework based on generalizations of elliptical distributions that
are fully characterized by their first three moments. SD presents
a further generalization that accounts for all moments of the return
distributions without necessarily assuming a particular family of
distributions.

Second-order SD (SSD) spanning (Arvanitis et al.\ (2019)) is a model-free
alternative to MV spanning of Huberman and Kandel (1987) (see also
Jobson and Korkie (1989), De Roon, Nijman, and Werker (2001), Ardia, Laurent, and Sessinou (2024)). Spanning
occurs if introducing new securities or relaxing investment constraints
does not improve the investment possibility set for a given class
of investors. MV spanning checks if the MV frontier of a set of assets
is identical to the MV frontier of a larger set made of those assets
plus additional assets (Kan and Zhou (2012), Penaranda and Sentana
(2012)). Here we investigate such a problem for investors with risk-averse
preferences which are interested in the whole return distributions
generated by two sets of assets, a sparse subset of dimension $q$
(10 assets) with a limited number of assets coming from a much larger
set of dimension $p$ (500 assets).

The first contribution of the paper is to introduce the concept of
sparse SSD spanning. We propose a theoretical measure for sparse spanning
based on second-order stochastic dominance. For economic interpretation,
we provide a representation based on a class of concave utility
functions without assuming differentiability. When sparse SSD spanning
occurs, a risk-averse investor will not improve her expected utility
by shifting from the sparse subset to the larger investment opportunity
set. On the contrary, if it does not occur, the risk-averse investor
suffers an expected utility loss since we work with a subset instead
of the full set of assets. Hence we further provide a lower bound
that takes the interpretation of an optimal expected utility loss that cannot
be improved upon by any sparse subset made of $q$ assets. We know
that we suffer a loss because of the sparsity constraint but we cannot
do better though investing optimally in only $q$ assets under an
SSD criterion. To check sparse SSD spanning on data, we develop consistent
and feasible estimation procedures based on Linear Programming (LP)
and a greedy algorithm, namely the Forward Stepwise Selection (FSS) algorithm. We
use a finite set of increasing piecewise-linear functions, restricted
to the bounded empirical supports, that are constructed as convex
mixtures of appropriate ``ramp functions'' (in the spirit of Russel
and Seo (1989)) in our representation as in Arvanitis, Scaillet and
Topaloglou (2020a,b). For every such utility function, we solve two embedded
linear maximization problems. It is an improvement over the implementation
in Arvanitis and Topaloglou (2017) and Arvanitis, Scaillet and Topaloglou
(2020b) where they formulate the empirical counterpart in terms of
Mixed-Integer Programming (MIP) problems. MIP problems are NP-complete,
and far more difficult to solve. Our numerical approximations are
simple and fast since they are based on standard LP. They suit better
computationally intensive optimisation methods, which otherwise become
quickly computationally demanding in empirical work on large data
sets. Those formulations are reminiscent of the LP programs developed
in the early papers of testing for SSD efficiency of a given portfolio
by Post (2003) and Kuosmanen (2004) (see also Scaillet and Topaloglou
(2010)).

Since we aim at a sparse solution computed from a large dimensional
problem, we rely on a greedy optimisation algorithm. We use a discrete
combinatorial algorithm for maximizing a function subject to a cardinality
constraint. It starts with the empty set, and then adds elements to
it in $r$ iterations. In each iteration, the algorithm adds to its
current solution the single element increasing the value of this solution
by the most, i.e., the element with the largest marginal value with
respect to the current solution. In the context of submodular maximization
(see Buchbinder and Feldman (2018) for a survey), this simple FSS algorithm checking for incremental gain at each step using
nested models is usually referred to simply as ``the greedy algorithm''.
In the case of submodular functions, it returns a solution that is
provably within a constant factor of the optimum (Nemhauser, Wolsey
and Fisher (1978)), and it turns out to be the best approximation
ratio possible for the problem (Nemhauser and Wolsey (1978)). Submodular
functions have a natural diminishing return property: adding an element
to a larger set results in smaller marginal increase in the value
of the function compared to adding the element to a smaller set. Submodular functions share also a natural sub-additivity property: for two disjoint sets, submodularity implies sub-additivity (but the converse is not true). In the context of risk measures, sub-additivity is related to the notion of diversification, a desirable property to be classified as a coherent risk measure (Artzner at al.\ (1998)). The lower partial moments that we use below to construct our test statistics have a type of put option pay-off and are measure of downside risk (Bawa (1975), Fishburn (1977)). They are known to be coherent risk measures and, as such, benefit from the sub-additivity property. Bian
et al.\ (2017) extend guarantee results of the greedy algorithm for
cardinality constrained maximization of non-submodular nondecreasing
set functions, in particular nondecreasing standard LP problems with
non-degenerate basic feasible solution (Bertsimas and Tsitsiklis (1997),
Ch.\ 3) that we implement in our empirics.

We choose that approach over penalization methods currently used for
building sparse MV portfolios for two reasons. First, we wish to bound
the relative error without any assumptions on the underlying sparsity
for the true parameter. It is useful to show the consistency of our
empirical strategy irrespective of sparse spanning being present or
absent. Our proof relies on the recent work of Elenberg et al.\ (2018)
(see Das and Kempe (2011) for the linear regression case). Contrary
to prior work in the MV setting, we require neither assumptions on
the sparsity of the underlying problem nor i.i.d.\ returns. We establish multiplicative approximation
guarantees from the best-case sparse solution. Our results improve
over previous work by providing bounds on a solution that is guaranteed
to match the desired sparsity and cannot be further decreased. Convex
methods for linear regressions such as the standard LASSO objective
(Tibshirani (1996)) require strong assumptions on the model and the
data, such as the unrepresentable condition on the parameter vector
and i.i.d.\ data (Zhao and Yiu (2006), Meinshausen and Buhlmann (2006)),
in order to provide exact sparsity guarantees on the recovered solution
(see Zhang (2009) for use of these assumptions in greedy least squares
regression). More specifically, when the number $r$ of iterations
is equal to $r=q\ln T$, $T$ being the time-series sample size, we
show that the algorithm provides a consistent estimate of the bound
of the expected utility loss computed from financial returns satisfying
a mixing condition. Mixing holds true for many time series models
such as ARMA models as well as several GARCH and stochastic volatility
processes (see Francq and Zakoian (2011) for several examples). It
allows us to build a path of the estimated bound as a function of
the sparsity constraint $q$, and verify when we have a sufficiently
large $q$ to get sparse SSD spanning, namely when the bound vanishes.
Second, the only input we need is the sparsity number $q$ of assets.
Hence, we avoid the selection problem of a tuning parameter, namely
the regularization parameter in penalization methods. As discussed in Brodie et al.\ (2009), a portfolio selection with a LASSO approach regulates the amount of shorting. 
In our setting, we use short-sales constraints which corresponds to using an implicit  large regularisation  parameter for the LASSO penalty. Our numerical approach based on a greedy algorithm however does not require the true portfolio to be sparse, and a large regularisation  parameter is not required for developing valid statistical inference. As a by-product,
our approach also provides a selection algorithm for sparse MV spanning
under multivariate normality using the equivalence with sparse SSD
spanning for elliptical distributions. It allows to bypass the regularization
of ill-conditioned estimates of large covariance matrices (see e.g.\ Fan,
Liao, and Shi (2015), Ledoit and Wolf (2017)). 

The second contribution of the paper aims at checking on large datasets of equity returns whether sparse SSD holds or not. We find that there is no benefit from expanding a sparse opportunity set beyond 45 assets. The optimal sparse portfolio invests in 10 industry sectors  and cuts tail risk when compared to a sparse MV portfolio. On a rolling-window basis, the number of assets shrinks to 25 assets in crisis periods, while standard factor models cannot explain the performance of the sparse portfolios. .

The paper is organized as follows. In Section
2, we establish our probabilistic framework, and review the definition
of SSD. In Section 3, we define the relevant concept of sparse SSD
spanning and provide convenient functional representations. In Section 4, we construct an estimate of the bound for sparse SSD spanning by using empirical
analogues.
We exploit the limiting distribution of the empirical process underlying
the estimator which has the form of a Gaussian process. Our estimation
strategy builds on LP and an FSS algorithm. We show the
asymptotic optimal recovery of the sparse solution, namely statistical
approximation guarantee of the greedy algorithm output for a given
$q$ when $T$ becomes large. In Section 5, we describe the numerical
implementation aspects of our empirical procedures.  In Section 6, we analyze large
datasets of equity returns to study whether sparse SSD holds or not 
and compare with results given by the construction of sparse MV portfolios
with the MAXSER approach of Ao, Li and Zheng (2019).
We provide concluding remarks  in Section
7. We provide our proofs and the list of factors used in the empirical application
in the Appendix. The Online Appendix discusses the concept of approximate sparse spanning. Given
a fixed support dimension, it specifies the low dimensional portfolio
set that comes closer (in an appropriate sense defined later on in
the paper) spanning the high dimensional one. It also gathers Monte Carlo experiments  to assess the finite sample properties of our procedure for sparse SSD
spanning.

\section{Background-Second Order Stochastic Dominance}

\noindent We describe our limiting economy for a large number
of financial assets. We denote the financial returns by a process
$X^{\infty}$ living in $\ell^{\infty}\left(\mathbb{N},\mathbb{R}\right)$,
which is the space of bounded real valued sequences equipped with
the uniform metric. $X^{(i)}$ denotes the $i^{\text{th}},\:i\in\mathbb{N}$
coordinate, $X$ denotes the projection of $X^{\infty}$ in the first
$p$ coordinates, and $\mathbb{P}$ denotes the distribution of $X^{\infty}$.
We suppress dependence on $p$ for brevity.
%$F$ denotes the cdf of
%the distribution of the random vector $X$. 

\noindent We introduce the associated portfolio weights with
short-sales constraints. Short-sales constraints on the asset allocation
promote sparsity (Brodie et al.\ (2009)); our approach can be used to trace further patterns of (desired) sparsity. The set $\Lambda_{\infty}$
is a non-empty subset of the $\mathbb{N}$-simplex $\left\{ \lambda\in\mathbb{R}^{\mathbb{N}}:\lambda_{i}\geq0,i\in\mathbb{N},\sum_{i=0}^{\infty}\lambda_{i}=1\right\}$, and for $p\in\mathbb{N}$,
$\Lambda=\left\{ \lambda\in\Lambda_{\infty},\sum_{i=0}^{p-1}\lambda_{i}=1\right\} $
denotes the $p-1$ dimensional unit sub-simplex of $\Lambda_{\infty}$
and $K$ is a non-empty closed subset of $\Lambda$.\footnote{Positive portfolio weights summing to one induce compactness of the parameter space $\Lambda$ which facilitates proofs. If short sales are allowed, we can alternatively assume that portfolio weights lie inside compact sets because of lending restrictions. The unit ball $\ell^{1}$-type restrictions imposed by the simplex consideration along with the Lipschitz continuity of $(x)_{+}$ imply distributional robustness: the expectations are equivalent to expectations w.r.t.\ the worst case distribution in a Wasserstein neighbourhood of the underlying distribution; see Theorem 1 of Gao, Chen, and Kleywert (2017).} In the present context,
$X$ is a random vector of financial returns for $p$ base assets,
while $\Lambda$ represents a set of portfolios formed on $X$. The
process $X^{\infty}$ idealizes the high dimensional situation in
the limiting case where $p\rightarrow +\infty$. Our first assumption
specifies probabilistic properties for $X^{\infty}$. It requires
mild moment existence conditions (bounded sequence of first order moments), and
a lower bound on the associated supports consistent with non-logarithmic
returns. There, $\mathrm{supp}$ denotes support of the distribution of the random variable involved, $\bar{\text{co}}$ denotes the closure of the convex
hull, and $(x)_{+}:=\max(0,x)$. Given the restrictions that its elements satisfy, $\Lambda_{\infty}$ is considered topologized by the $l_{1}$
norm and $\lambda,\:\kappa$ denote generic elements of $\Lambda_{\infty}$.

\noindent \begin{assumption}

\noindent \textbf{\label{assu:UMom}}$\max_{0<i\leq+\infty}\mathbb{E}\left[\left|X^{(i)}\right|\right]<+\infty$.
$Z:=\bar{\mathrm{co}}\left[\cup_{i}\mathrm{supp}\left(X^{(i)}\right)\right]$
and $\inf Z>-\infty$. 

\end{assumption}

The assumption implies that for any $\lambda\in\Lambda_{\infty}$, $\sum_{i=0}^{\infty}\lambda_{i}X^{(i)}$
is a well defined random variable and that the 
Lower Partial Moment Differential (LPMD) defined by \begin{eqnarray} D\left(z,\kappa,\lambda,\mathbb{P}\right):=\mathbb{E}\left[\left(z-\sum_{i=0}^{\infty}\kappa_{i}X^{(i)}\right)_{+}\right]-\mathbb{E}\left[\left(z-\sum_{i=0}^{\infty}\lambda_{i}X^{(i)}\right)_{+}\right] \label{LPMD}, \end{eqnarray}
is also bounded (and hence exists) and continuous in $z,\lambda,\kappa$.\footnote{This is due to the monotonicity of the integral, $\mathbb{E}\left[\left|\sum_{i=0}^{\infty}\lambda_{i}X^{(i)}\right|\right]\leq\max_{i}\mathbb{E}\left[\left|X^{(i)}\right|\right]<+\infty$. The assumption implies that the partial moment $\mathbb{E}\left[\left(z-\sum_{i=0}^{\infty}\lambda_{i}X^{(i)}\right)_{+}\right]$
is continuous in $z,\:\lambda$ via dominated convergence, and that
it is also bounded in $\lambda$ for any $z$, even though $\Lambda_{\infty}$
is not ($l_{1}$-) totally bounded. Along with the Lipschitz continuity
property of $\left(\cdot\right)_{+}$, it also implies that for any
$\lambda\neq\kappa$, 
\begin{equation*}\label{eq:hhh}
\sup_{z\in\mathbb{R}}\left|\mathbb{E}\left[\left(z-\sum_{i=0}^{\infty}\kappa_{i}X^{(i)}\right)_{+}\right]-\mathbb{E}\left[\left(z-\sum_{i=0}^{\infty}\lambda_{i}X^{(i)}\right)_{+}\right]\right|\leq\max_{i}\mathbb{E}\left[\left|X^{(i)}\right|\right]\sum_{i=0}^{\infty}\left(\kappa_{i}+\lambda_{i}\right).
\end{equation*}} The LPMD (\ref{LPMD}) takes the interpretation of the difference between two coherent risk measures evaluated on the portfolios at hand. Assumption \ref{assu:UMom} facilitates the definition
of SSD for the constructed portfolios: 
\begin{defn}
\label{def:SD}$\kappa$ SSD dominates $\lambda$,
written $\kappa\underset{\text{SSD}}{\succeq}\lambda$, iff $D\left(z,\kappa,\lambda,\mathbb{P}\right)\leq0$,
for all $z\in Z$. 
\end{defn}
The definition is simply an adaptation of the usual SSD relation in
our high dimensional framework. Using the classical Russell and Seo
(1989) utility representations, we obtain the well known result that
$\kappa\underset{\text{SSD}}{\succeq}\lambda$ iff the former is preferred
to the latter by every increasing and concave utility. Thus, SSD exemplifies
universal choices w.r.t.\ every insatiable and risk averse investor.

\section{Sparse SSD Spanning\label{sec:Sparse-SSD-Spanning}}

Arvanitis et al.\ (2018) define the notion of SSD Spanning as an extension
of the MV analogue. It involves comparison of portfolio sets
that are not necessarily singletons. 
\begin{defn}[SSD Spanning]
\label{def:Span} $K\underset{\text{SSD}}{\succeq}\Lambda$ iff $\forall\lambda\in\Lambda,\:\exists\kappa\in K:\kappa\underset{\text{SSD}}{\succeq}\lambda$. 
\end{defn}
If the sets are not related by inclusion and $K\underset{\text{SSD}}{\succeq}\Lambda$,
then we have necessarily $K\underset{\text{SSD}}{\succeq}K\cup\Lambda$.
Furthermore, spanning would be trivial if $K\supseteq\Lambda$ were
allowed. Hence, we can always consider that $K$ lies inside $\Lambda$.
Spanning admits an economic interpretation  when
$K\subseteq\Lambda$; it means that extension of the investment opportunity
set from $K$ to $\Lambda$ does not improve investment possibilities
for any risk averter. Hence, no spanning means that the extension
contains a non dominated element. It  is formalized as follows: $K\underset{\text{SSD}}{\nsucceq}\Lambda$
iff $\exists\lambda\in\Lambda:\forall\kappa\in K,\:\kappa\underset{\text{SSD}}{\nsucceq}\lambda$,
i.e., $\lambda$ is maximal (efficient) w.r.t.\ $K$.

Under some further structure on $K$, SSD spanning admits an empirically
useful characterization involving a saddle-type point of the LPMDs. 
\begin{lem}
\label{lem:Span_Char}Under Assumption \ref{assu:UMom}, $K\underset{\text{SSD}}{\succeq}\Lambda$ iff 
$\sup_{\Lambda}\inf_{K}\sup_{z\in Z}D\left(z,\kappa,\lambda,\mathbb{P}\right)\leq0$. 
\end{lem}
We extend the notion in the high dimensional setting, by also allowing\textcolor{black}{{}
a potentially unknown} low dimensional investment opportunity set
to SSD span a high dimensional superset. In order to formally define
this and extend it to the limiting case where $p\rightarrow +\infty$,
we introduce the following notation for the support of a portfolio
set: $\mathrm{csupp}\left(K\right):=\#\left\{ i:\kappa_{i}\neq0,\kappa\in K\right\} $.
By construction $\mathrm{csupp}\left(\Lambda\right)=p$. We suppose
that as $p\rightarrow+\infty$ and $\lim_{p\rightarrow +\infty}\Lambda=\Lambda_{\infty}$,
where the limit is interpreted in the Painleve-Kuratowski convergence
mode. The sequence $\left(\Lambda\right)_{p}$ is by construction
monotone increasing. 
\begin{defn}[Sparse Spanning SSD]
\label{def:S_Span_SemiStrong}For some fixed $q$, there exists a
$K\subset\Lambda$ with $\mathrm{csupp}\left(K\right)\leq q$ and such
that $K\underset{\text{SSD}}{\succeq}\Lambda$. 
\end{defn}
\textcolor{black}{Definition \ref{def:S_Span_SemiStrong} generalizes
Definition \ref{def:Span} in a twofold manner. First, it allows for
a limiting high dimensional setting thus providing the proper framework
for addressing the empirical  questions listed
in the introduction. Second, it only prescribes the existence of a
``low-dimensional'' spanning subset of $\Lambda$, whereas for the
original definition the spanning subset is exogenously given. It implies
that any procedure designed to test whether SS-SSD holds would have
to search for a spanning set inside the collection of ``low-dimensional''
subsets of $\Lambda$. It is useful even in the case where SS-SSD
does not hold. As the following paragraph suggests, such a procedure,
if consistent, would end up with a sparse portfolio set that ``comes
as close as possible'' to SSD span its high dimensional universe
of portfolios. }

As in Lemma \ref{lem:Span_Char}, we obtain a useful characterization
of SS-SSD by considering the collection
 $\mathcal{L}_{p,q}=\left\{ K\subset\Lambda:K\text{ closed},0<\mathrm{csupp}\left(K\right)\leq q\right\} $:\footnote{
When $\Lambda$ is itself a simplicial complex, then $\mathcal{L}_{p,q}$
is also a simplicial complex of dimension $q-1$. Then and if $p\geq2q$,
$\mathcal{L}_{p,q}$ has a geometric realization as a sub-simplex
of the standard $p-1$ simplex (see the Geometric Realization Theorem
in Edelsbrunner (2014)).}
\begin{lem}
\label{lem:Characterizations}Under Assumption \ref{assu:UMom}, 
%suppose
%moreover that $\Lambda$ is closed in the Euclidean topology. Then,
 for large enough $p$ and fixed $q$,
SS-SSD is equivalent to $\inf_{\mathcal{L}_{p,q}}\sup_{\Lambda}\inf_{K}\sup_{z\in Z}D\left(z,\kappa,\lambda,\mathbb{P}\right)\leq0$, and
the latter is equivalent to \break
 \textup{$\inf_{\mathcal{L}_{\infty,q}}\sup_{\Lambda_{\infty}}\inf_{K}\sup_{z\in Z}D\left(z,\kappa,\lambda,\mathbb{P}\right)\leq0$.} 
\end{lem}
The possibility of interchanging the order of appearance of the optimization
operators in the characterization $\inf_{\mathcal{L}_{p,q}}\sup_{\Lambda}\inf_{K}\sup_{z\in Z}D\left(z,\kappa,\lambda,\mathbb{P}\right)\leq0$
to $\sup_{z\in Z}\sup_{\Lambda}\inf_{\mathcal{L}_{p,q}}\inf_{K}$
will greatly facilitate numerical aspects as well as the derivations
of limiting properties for the empirical procedures. It actually holds
via the use of appropriate minimax theorems and the extension of our
assumption framework. 
\begin{lem}
\label{lem:Helpful_Compactness}Suppose that Assumption \ref{assu:UMom}
 holds.
 %and that $\Lambda$ is closed in the Euclidean
%topology.
 Then, \textup{for all $p$,} 
\[
\inf_{\mathcal{L}_{p,q}}\sup_{\Lambda}\inf_{K}\sup_{z\in Z}D\left(z,\kappa,\lambda,\mathbb{P}\right)=\sup_{z\in Z}\sup_{\Lambda}\inf_{\mathcal{L}_{p,q}}\inf_{K}D\left(z,\kappa,\lambda,\mathbb{P}\right).
\]
\end{lem}
We have $\sup_{\Lambda}\inf_{\mathcal{L}_{p,q}}\inf_{K}D\left(z,\kappa,\lambda,\mathbb{P}\right) =\inf_{\mathcal{L}_{p,q}}\inf_{K}\mathbb{E}\left[\left(z-\sum_{i=0}^{\infty}\kappa_{i}X_{t}^{(i)}\right)_{+}\right] $ \break $-\inf_{\Lambda}\mathbb{E}\left[\left(z-\sum_{i=0}^{\infty}\lambda_{i}X_{t}^{(i)}\right)_{+}\right]$, given an arbitrary threshold $z$, so that we can separate the optimizations w.r.t.\ the ``parameter
sets'' $\Lambda$ and $\mathcal{L}_{p,q}\times K$. It is useful
especially in the case where we approximate the outer optimization
over $Z$ by some discretization, as  in our empirical numerical implementations.

\section{Sparse Optimization: Greedy Algorithm and Statistical Guarantees\label{sec:Sparse-Optimization-Sparse-Ident}}

In this section, and given the latency of $\mathbb{P}$,
we are interested in the empirical approximation of the element of
$\mathcal{L}_{p,q}$ that approximately spans $\Lambda$ for a fixed
$q$, and the subsequent estimation of the associated diversification loss $M\left(\Lambda,\mathcal{L}_{p,q},\mathbb{P}\right)$. We employ the empirical analogues of the functionals that
characterize spanning, and design the sparse optimization involved
via a greedy algorithm. We establish consistency using
the results on statistical guarantees by Elenberg et al.\ (2018). We
derive the usual parametric $\sqrt{T}$ rate and the limiting distribution,
and construct a conservative inferential procedure based on fast subsampling.

Consider the sequence $\left(X_{t}^{\infty}\right)_{t\in\mathbb{Z}}$
where for all $t$, $X_{t}^{\infty}\underset{\text{d}}{=}X^{\infty}$
and $\underset{\text{d}}{=}$ denotes equality in distribution. Suppose,
that for some $p$, a sample of $\left(X_{t}\right)_{t=1,\dots,T}$
is available from the sequence $\left(X_{t}\right)_{t\in\mathbb{Z}}$.
Denote with $\mathbb{P}_{T}$ its empirical distribution function
in $\mathbb{R}^{p}$ (in what follows $\mathbb{P}$ also identifies
the distribution of $X_{0}$ in $\mathbb{R}^{p}$ without inconsistency
due to the Daniel-Kolmogorov Theorem). We approximate $D\left(z,\kappa,\lambda,\mathbb{P}\right)$
by $D\left(z,\kappa,\lambda,\mathbb{P}_{T}\right)$ and design a procedure
that evaluates $\inf_{\mathcal{L}_{p,q}}\sup_{\Lambda}\inf_{K}\sup_{z\in Z}D\left(z,\kappa,\lambda,\mathbb{P}_{T}\right)$.

Given Lemma \ref{lem:Helpful_Compactness}, we design our empirical
procedure as follows: for fixed $q$, formulate the empirical
optimization problem 
$\sup_{z\in Z}\sup_{\Lambda}\inf_{\mathcal{L}_{p,q}}\inf_{K}D\left(z,\kappa,\lambda,\mathbb{P}_{T}\right)$, as 
\begin{equation}
\begin{array}{c}
M\left(\Lambda,\mathcal{L}_{p,q},\mathbb{P}_{T}\right):=\sup_{z\in Z}\left[\mathcal{K}\left(\Lambda,\mathcal{L}_{p,q},z,\mathbb{P}_{T}\right)-\mathcal{J}\left(\Lambda,z,\mathbb{P}_{T}\right)\right],\\
\mathcal{K}\left(\Lambda,\mathcal{L}_{p,q},z,\mathbb{P}_{T}\right):=\inf_{\mathcal{L}_{p,q}}\inf_{K}\frac{1}{T}\sum_{t=0}^{T}\left(z-\sum_{i=0}^{\infty}\kappa_{i}X_{t}^{(i)}\right)_{+}\\
\mathcal{J}\left(\Lambda,z,\mathbb{P}_{T}\right):=\inf_{\Lambda}\frac{1}{T}\sum_{t=0}^{T}\left(z-\sum_{i=0}^{\infty}\lambda_{i}X_{t}^{(i)}\right)_{+}.
\end{array},\label{eq:opt_help}
\end{equation}

Given $\mathcal{K}\left(\Lambda,\mathcal{L}_{p,q},z,\mathbb{P}_{T}\right)$
the numerical technology evaluating $\mathcal{J}\left(\Lambda,z,\mathbb{P}_{T}\right)$
and $M\left(\Lambda,\mathcal{L}_{p,q},\mathbb{P}_{T}\right)$ is the
same as the one employed in the SD literature in the low dimensional
settings. The main issue here is to design a procedure that evaluates
$\mathcal{K}\left(\Lambda,\mathcal{L}_{p,q},z,\mathbb{P}_{T}\right)$.
The outer optimization there involves searching over low dimensional
subsets of $\Lambda$. As explained in the introduction, we favor a
procedure based on a greedy algorithm that approximates $\mathcal{K}\left(\Lambda,\mathcal{L}_{p,q},z,\mathbb{P}_{T}\right)$
over procedures based on penalization. We use the FSS Algorithm (Algorithm 2 in Elenberg
et al.\ (2018)). Let us denote $r_{T}\left(q\right)$ the number
of iterations performed.

\begin{aalgorithm} \label{alg:FS}Forward Stepwise Selection (see
p.\ 3542 of Elenberg et al.\ (2018)).\linebreak{}
Inputs: the sparsity Parameter $q<p$, the \# of iterations $r_{T}\left(q\right)$,
for a given set $S$ the set function $2^{p}\rightarrow\mathbb{R}$
defined as 
\[
f\left(S\right):=\inf_{\mathrm{csupp}(S)\leq q}\frac{1}{T}\sum_{t=0}^{T}\left(z-\sum_{i=0}^{\infty}\kappa_{i}X_{i,t}\right)_{+}.
\]

\begin{enumerate}
\item[a] Choose the initial set $S_{0}$, 
\item[b] for $i=1,\dots,r_{T}\left(q\right)$ do, 
\item[c] $s:=\arg\max_{j\in\left[p\right]/S_{i-1}}f\left(S_{i-1}\cup\left\{ j\right\} \right)-f\left(S_{i-1}\right)$, 
\item[d] $S_{i}:=S_{i-1}\cup\left\{ s\right\} $. 
\end{enumerate}
\end{aalgorithm}

The last step (d), for $i=r_{T}\left(q\right)$, returns $\mathcal{K}^{\mathrm{FS}}\left(\Lambda,\mathcal{L}_{p,q},z,\mathbb{P}_{T},r_{T}\left(q\right)\right)$,
namely the numerical approximation of $\mathcal{K}\left(\Lambda,\mathcal{L}_{p,q},z,\mathbb{P}_{T}\right)$
in (\ref{eq:opt_help}) by the greedy algorithm. The next section
provides details on the numerical aspects of the three optimizations
appearing in (\ref{eq:opt_help}), including the implementation of
FSS.

We examine the issues of consistency, rates
of convergence and limiting distribution of $M\left(\Lambda,\mathcal{L}_{p,q},\mathbb{P}_{T}\right)$
given $\mathcal{K}^{\mathrm{FS}}\left(\Lambda,\mathcal{L}_{p,q},z,\mathbb{P}_{T},r_{T}\left(q\right)\right)$.
Our analysis depends on the asymptotic behavior of the empirical
process
 $\sqrt{T}D\left(z,\kappa,\lambda,\mathbb{P}_{T}-\mathbb{P}\right)$, and of the process
$G_{T}\left(z,\kappa,\lambda\right):=\sqrt{T}\left[g\left(z,\lambda,\mathbb{P}_{T}\right)-g\left(z,\lambda,\mathbb{P}\right)\right]^{T}\left(\kappa-\lambda\right)$,
as well as of the empirical moment process\linebreak $\frac{1}{T}\sum_{t=0}^{T}\left(z_{T}-\sum_{i=0}^{\infty}\lambda_{i}X_{t}^{(i)}\right)_{+}$,
where the subdifferential $g\left(z,\lambda,\mathbb{Q}\right):=\mathbb{E}_{\mathbb{Q}}\left[X\mathbb{I}\left\{ z\geq\sum_{i=0}^{\infty}\lambda_{i}X^{(i)}\right\} \right]$
and $\mathbb{E}_{\mathbb{Q}}$ denotes integration w.r.t.\ the measure $\mathbb{Q}$.
Specifically, consistency is facilitated if the first and the second
processes are asymptotically tight over appropriate subsets of parameters,
and the third process (locally) uniformly converges to its population counterpart.
This behavior depends on stationarity and mixing rates for the returns
process involved as well as a stricter moment existence condition
compared to Assumption \ref{assu:UMom}.

\begin{assumption} \label{assu:Mix} $\left(X_{t}^{\infty}\right)_{t\in\mathbb{Z}}$
is strictly stationary and absolutely regular with mixing coefficients
$\left(\beta_{m}\right)_{m\in\mathbb{N}}$ that satisfy $\beta_{m}\sim b^{m}$
for some $b\in\left(0,1\right)$, as $m\rightarrow\infty$, and $\max_{0<i\leq+\infty}\mathbb{E}\left[\left|X^{(i)}\right|^{2+\varepsilon}\right]<+\infty$,
for some $\varepsilon>0$. \end{assumption}

The stationarity, ergodicity and mixing rates conditions as well as
the moment existence condition hold for several geometrically ergodic
(finite dimensional), linear as well as GARCH type models with values
in Euclidean spaces. Those are frequently employed in empirical finance
with data consistent parameter restrictions; see Francq and Zakoian
(2011). Using the Daniell-Kolmogorov Theorem we have that stationarity
and mixing rates hold for the $\left(X_{t}^{\infty}\right)_{t\in\mathbb{Z}}$
process whenever they hold uniformly over the collection of finite
dimensional parts of the process. Thereby, they hold whenever the
finite dimensional parts of the process are consistent with the aforementioned
models with uniform parameter restrictions.

We obtain the following limit theory; let $\ell^{\infty}\left(Z\times\Lambda^{\infty}\times\Lambda^{\infty}\right)$
denote the space of real valued bounded functions on $Z\times\Lambda^{\infty}\times\Lambda^{\infty}$
equipped with the $\sup$ norm. We use $\rightsquigarrow$
to denote weak convergence. 
\begin{thm}
\label{thm:ELT}Suppose that Assumptions \ref{assu:UMom}, \ref{assu:Mix}
hold. Then, (a)  $\frac{1}{T}\sum_{t=0}^{T}\left(z_{T}-\sum_{i=0}^{\infty}\lambda_{i}X_{t}^{(i)}\right)_{+}\rightsquigarrow\mathbb{E}\left[\left(z-\sum_{i=0}^{\infty}\lambda_{i}X_{t}^{(i)}\right)_{+}\right]$,
for any $z,z_{T}\in Z$ with $z_{T}\rightarrow z$, and uniformly in $\Lambda^{\infty}$. Furthermore, suppose also that $\frac{\ln p}{\sqrt{T}}\rightarrow 0$. Then  as $T\rightarrow\infty$, with $\kappa\in\tilde{\Lambda}_{\left(\left\lfloor q\left(\ln T+1\right)\right\rfloor \right)}$
(b) $\sqrt{T}D\left(z,\kappa,\lambda,\mathbb{P}_{T}-\mathbb{P}\right)\rightsquigarrow\mathcal{G}\left(z,\kappa,\lambda\right)$,
in $\ell^{\infty}\left(Z\times\Lambda^{\infty}\times\Lambda^{\infty}\right)$, where $\mathcal{G}\left(z,\kappa,\lambda\right)$
is a zero mean Gaussian process with covariance kernel defined by
\begin{equation*}
\begin{array}{c}
\begin{array}{c}
\mathcal{V}\left[\left(z_{1},\kappa_{1},\lambda_{1}\right),\left(z_{2},\kappa_{2},\lambda_{2}\right)\right]:=\sum_{t\in\mathbb{Z}}\mathrm{Cov}\left[\mathcal{I}\left(z_{1},\kappa_{1},\lambda_{1},X_{0}\right),\mathcal{I}\left(z_{2},\kappa_{2},\lambda_{2},X_{t}\right)\right]\end{array}\end{array},\label{eq:kernel}
\end{equation*}
where $\mathcal{I}\left(z,\kappa,\lambda,X_{t}\right):=\left(z-\sum_{i=0}^{\infty}\kappa_{i}X_{t}^{(i)}\right)_{+}-\left(z-\sum_{i=0}^{\infty}\lambda_{i}X_{t}^{(i)}\right)_{+}$.
%\textcolor{red}{NOT NEEDED?$G_{\lambda}$ has a.s.\ continuous sample paths w.r.t.\ the product
%topology of $Z$ for the first argument and $l_{1}$ for the second.}
Finally, (c) \[\lim\sup_{T\rightarrow\infty}\mathbb{E}[\sup_{z}\sup_{\Lambda_{\left(\left\lfloor q\left(\ln T+1\right)\right\rfloor \right)}}G_{T}\left(z,\kappa,\lambda\right)]<\infty.\] 
\end{thm}
The condition $\frac{\ln p}{\sqrt{T}}\rightarrow 0$ that appears in the final pair of results of the theorem is somewhat stricter than the usual $\frac{\ln p}{T}\rightarrow 0$ that appears in the literature, it however facilitates standard rates and limiting Gaussianity for the empirical processes involved and thus the results that go beyond consistency. It ensures that the bracketing entropy of $\Lambda$ grows at an appropriate rate in order for tightness to hold in the limit. For the notion of the bracketing
entropy numbers of a metric space, see Section 5 of Andrews (1994)
and Ch.\ 2 of van der Vaart and Wellner (1996). In our context, it
corresponds to the mapping that keeps track of the logarithm of the
minimal number of $\delta$-brackets (w.r.t.\ the $l_{1}$ norm) of
real sequences with absolutely convergent series needed to cover the
particular neighborhood, for each $\delta>0$.

We furthermore use an assumption concerning a property
of restricted strong convexity (see Ch.\  9 of Wainwright (2019) for reviewing restricted strong convexity in high-dimensional statistics) for the LPMs as $p\rightarrow +\infty$.  In this section,
we also denote $\Lambda$ with $\Lambda_{p}$ whenever it is important
to keep track of the dimension of the portfolio space.
For $p\gg m\in\mathbb{N}$, we denote the set
 $\left\{ \left(\lambda,\lambda^{\star}\right)\in\Lambda_{p}\times\Lambda_{p}:\mathrm{csupp}\left(\lambda\right)\leq m,\mathrm{csupp}\left(\lambda^{\star}\right)\leq m,\mathrm{csupp}\left(\lambda-\lambda^{\star}\right)\leq m\right\} $
with $\Lambda_{\left(m\right)}$. $\tilde{\Lambda}_{\left(m\right)}$
denotes the set obtained by keeping the first component $\lambda$
of the pairs $\left(\lambda,\lambda^{\star}\right)$ that define the
elements of $\Lambda_{\left(m\right)}$.

\begin{assumption}{{[}Restricted Strong Convexity
(RSC){]}} \label{assu:SCS}$X$ has a continuous density $f$. 
%Furthermore, 
%$\mathbb{E}\left(z-\sum_{i=0}^{\infty}\kappa_{i}X_{0,i}\right)_{+}$
%is twice differentiable w.r.t.\ any $\kappa$ appearing in some pair
%of $\Lambda_{\left(\left\lfloor q\left(\ln\left(T+1\right)\right)\right\rfloor \right)}$
%for all $z\in Z$.
For $m_{\left\lfloor q\left(\ln\left(T+1\right)\right)\right\rfloor }$
denoting the supremum 
%and $M_{\left\lfloor q\left(\ln\left(T+1\right)\right)\right\rfloor }$ the infimum
over $\Lambda_{\left(\left\lfloor q\left(\ln\left(T+1\right)\right)\right\rfloor \right)}$,
of the smallest 
%and the largest 
eigenvalues of the Hessian matrix
of $\mathbb{E}\left(z-\sum_{i=0}^{\infty}\kappa_{i}X_{0,i}\right)_{+}$ as a function of $\kappa$,
we have that as $T\rightarrow\infty$, 
%$\frac{m_{\left\lfloor q\left(\ln\left(T+1\right)\right)\right\rfloor }}{M_{\left\lfloor q\left(\ln\left(T+1\right)\right)\right\rfloor }}\ln T\rightarrow+\infty$
$m_{\left\lfloor q\left(\ln\left(T+1\right)\right)\right\rfloor }\ln T\rightarrow+\infty$,
locally uniformly in $Z-\left\{\inf{Z}\right\}$.
\end{assumption}

%Let us characterize that assumption on an example, which shows that
%this assumption is mild. For the $\left\lfloor q\left(\ln\left(T+1\right)\right)\right\rfloor $-dimensional,
Due to Assumption \ref{assu:UMom}, the existence of the continuous density $f$, Theorem 1 of Savare (1996) and
given the distributional derivative of $\left(x\right)_{+}$ (see
p.\ 1 in Savare (1996)), we obtain that $\mathbb{E}\left(z-\sum_{i=0}^{\infty}\kappa_{i}X_{0,i}\right)_{+}$
is twice differentiable and the Hessian assumes the form $\int_{\mathbb{R}^{q}}XX^{T}\delta\left(z-\kappa^{T}X\right)f\left(X\right)dX$,
where $\delta$ denotes the Dirac Delta function. Using Example 27
in Estrada and Kanwal (2012), the latter equals $\int_{z=\kappa^{T}X}XX^{T}f\left(X\right)dX$. The eigenvalue restriction part of Assumption \ref{assu:SCS} thus follows whenever the minimum eigenvalue of $V$, the second moment matrix of $X$,
%Suppose
%now that $f$ is a normal density. Then, using the results of Cong
%et al.\ (2017) (see their Algorithm 2), we have that the Hessian takes
%the form $C_{z,\kappa}\mathbb{E}\left(\left[\left(\text{Id}_{q}-\frac{1}{\Delta}V\kappa\kappa^{T}\right)X+\frac{1}{\Delta}V\kappa z\right]\left[\left(\text{Id}_{q}-\frac{1}{\Delta}V\kappa\kappa^{T}\right)X+\frac{1}{\Delta}V\kappa z\right]^{T}\right)$,
%where $V$ is the second moment matrix of $X$, $C_{z,\kappa}>0$
%is an integration constant depending on both $z$ and $\kappa$ and
%$\Delta=\kappa^{T}V\kappa$. A simple calculation along with the constraint
%$z=\kappa^{T}X$, yields that the Hessian equals $C_{z,\kappa}V$.
%Given that $\frac{\sup_{z,\kappa}m}{\inf_{z,\kappa}M}\geq\sup_{z,\kappa}\frac{m}{M}$,
is dominated by $\ln T$ as
$T\rightarrow\infty$, due to the Cauchy's eigenvalue interlacing theorem (see for example Hwang (2004)). It means that we can also accommodate second moment matrices that become asymptotically singular at a slowly
varying rate. Controlling the asymptotic singularity of the Hessian is related to the geometric interpretation provided in Remark 4 of Elenberg et al.\ (1998). By doing so, we  control the way the set function values diminish via adding individual features through a stepwise procedure compared to adding multiple features at once as in a LASSO penalization, approximating the diminishing return property of a submodular function. The analysis in Par.\ 5  of Kim and Pollard (1990) implies that the same control suffices when $f$ is continuously differentiable, which is more in line with the bounded support framework of our applications. When $V$ has a Kac-Murdock-Szego type Toeplitzian structure (Trench (1999)), where $V_{i,j}=v^{|i-j|},\:i,j=1,\dots,p$ for $v\in [0,1)$, Assumption \ref{assu:SCS} holds trivially since the minimum eigenvalue of the matrix is then uniformly bounded in $p$  (Trench (1999), p.\ 182). Such matrices appear in zero mean normalised autoregressive progresses. In the case of the zero mean spiked identity model (Example 7.18 in Wainwright (2019)), where $V=\mathbf{Id}+\mu \mathbf{1}\mathbf{1}'$ for some $\mu\in [0,1)$, the results in the aforementioned example imply that Assumption \ref{assu:SCS} holds for every fixed value of $\mu$ as well as when $\mu$ converges to one with $T$ at a slower than logarithmic rate. It fails to hold whenever $\mu$ becomes asymptotically null with faster rates. When $X$ is not necessarily zero mean, then the variational representation of the minimum eigenvalue $\lambda_{\min}(A)$, of a p.d.\ matrix $A$, implies that Assumption \ref{assu:SCS} holds whenever $\min\left\{\lambda_{\min}(\mathbb{E}(X)\mathbb{E}(X)'),\lambda_{\min}(\mathrm{Var}(X))\right\}\ln T\rightarrow\infty$. Due to $\Lambda$ being simplicial, no restricted smoothness conditions for the maximum eigenvalue of the Hessian are required. Hence, the assumption can accommodate without such restrictions covariance matrices for models arising in strong factor settings where the largest eigenvalue is of the same order as the number of assets. 

The RSC assumption along with Assumption \ref{assu:Mix} imply
analogous RSC properties for the empirical LPMs $\frac{1}{T}\sum_{t=0}^{T}\left(z-\sum_{i=0}^{\infty}\kappa_{i}X_{t}^{(i)}\right)_{+}$
with probability converging to one (w.h.p.). It enables the use of
the results of Elenberg et al.\ (2018) on statistical guarantees
for the FSS Algorithm. Using the above, we first obtain the following consistency result. 
\begin{thm}
\label{thm:Guarantees}Suppose that Assumptions \ref{assu:UMom},
\ref{assu:Mix}, \ref{assu:SCS}, hold, and that $\frac{\ln p}{\sqrt{T}}\rightarrow 0$. Then, as $T\rightarrow\infty$,
and uniformly in $Z$, 
\begin{equation}
\mathcal{K}^{\mathrm{FS}}\left(\Lambda,\mathcal{L}_{p,q},z,\mathbb{P}_{T},q\ln T\right)\rightsquigarrow\mathcal{K}\left(\Lambda^{\infty},\mathcal{\mathcal{L}}_{\infty,q},z,\mathbb{P}\right).\label{eq:cons}
\end{equation}
Consequently, $M^{\mathrm{FS}}\left(\Lambda,\mathcal{\mathcal{L}}_{p,q},\mathbb{P}_{T},q\ln T\right)\rightsquigarrow M\left(\Lambda_{\infty},\mathcal{\mathcal{L}}_{\infty,q},\mathbb{P}\right)$,
where $M^{\mathrm{FS}}\left(\Lambda,\mathcal{\mathcal{L}}_{p,q},\mathbb{P}_{T},r_{T}\left(q\right)\right):=\sup_{z\in Z}\left[\mathcal{K}^{\mathrm{FS}}\left(\Lambda,\mathcal{\mathcal{L}}_{p,q},z,\mathbb{P}_{T},r_{T}\left(q\right)\right)-\mathcal{J}\left(\Lambda,z,\mathbb{P}_{T}\right)\right]$. 
\end{thm}
Whenever Assumption \ref{assu:SCS} holds for some $q^{\star}\in\mathbb{N}$,
Theorem \ref{thm:Guarantees} implies then that the mapping $q\rightarrow M^{\mathrm{FS}}\left(\Lambda,\mathcal{K}_{p,q},\mathbb{P}_{T},q\ln T\right)$
converges in probability to $q\rightarrow M\left(\Lambda_{\infty},\mathcal{K}_{\infty,q},\mathbb{P}\right)$
uniformly in $q\leq q^{\star}$. 

Theorem 2 holds whether we have sparse spanning or not at the limit.
We do not need to assume sparsity in the population. The statistical
guarantee result of Theorem 2 is a strong advantage of the greedy
algorithm over penalization methods.

We are further occupied with the determination of the rates of convergence
and the distributional limit for the deviation $M^{\mathrm{FS}}\left(\Lambda,\mathcal{\mathcal{L}}_{p,q},\mathbb{P}_{T},q\left(\ln T\right)^{\epsilon}\right)-M\left(\Lambda^{\infty},\mathcal{\mathcal{L}}_{\infty,q},\mathbb{P}\right)$,
that gauges the gap between $M^{\mathrm{FS}}\left(\Lambda,\mathcal{\mathcal{L}}_{p,q},\mathbb{P}_{T},q\left(\ln T\right)^{\epsilon}\right)$,
which is returned by the greedy algorithm on the data, and the limit
$M\left(\Lambda^{\infty},\mathcal{\mathcal{L}}_{\infty,q},\mathbb{P}\right)$.
To this end, we augment $r_{T}$ to $q\left(\ln T\right)^{\epsilon}$ for some arbitrary $\epsilon>1$, in
order to facilitate arguments that estimate the rate of the approximation of the infimum of $\frac{1}{T}\sum_{t=0}^{T}\left(z-\sum_{i=0}^{\infty}\kappa_{i}X_{t}^{(i)}\right)_{+}$
over the empirical solution in $\mathcal{\mathcal{L}}_{p,q}$, by
the infimum of $\mathbb{E}\left[\left(z-\sum_{i=0}^{\infty}\kappa_{i}X_{t}^{(i)}\right)_{+}\right]$
over the population solution.  Given the second result of Theorem \ref{thm:ELT}, we obtain standard
rates and a distributional limit defined as a saddle type point of
a zero mean Gaussian process, using among others the generalized Delta
method applicable due to the Hadamard directional differentiability of the optimization functionals that appear in the definition of spanning (see C\'arcamo et al.\ (2020)). 
\begin{thm}
\label{thm:delta}Suppose that Assumptions \ref{assu:UMom}, \ref{assu:Mix},
\ref{assu:SCS} hold, and that $\frac{\ln p}{\sqrt{T}}\rightarrow 0$. 
Then as $T\rightarrow\infty$, 
\begin{equation}
\sqrt{T}\left(M^{\mathrm{FS}}\left(\Lambda,\mathcal{\mathcal{L}}_{p,q},\mathbb{P}_{T},q\left(\ln T\right)^{\epsilon}\right)-M\left(\Lambda_{\infty},\mathcal{\mathcal{L}}_{\infty,q},\mathbb{P}\right)\right)\rightsquigarrow\sup\inf_{\left(z,\lambda,\kappa\right)\in\Gamma}\mathcal{G}\left(z,\lambda,\kappa\right),\label{eq:limit}
\end{equation}
where $\mathcal{G}\left(z,\lambda,\kappa\right)$ is a zero mean
Gaussian process with covariance kernel defined by 
\begin{equation*}
\begin{array}{c}
\begin{array}{c}
\mathcal{V}\left[\left(z_{1},\lambda_{1},\kappa_{1}\right),\left(z_{2},\lambda_{2},\kappa_{2}\right)\right]:=\sum_{t\in\mathbb{Z}}\mathrm{Cov}\left[\mathcal{I}\left(z_{1},\lambda_{1},\kappa_{1},X_{0}\right),\mathcal{I}\left(z_{2},\lambda_{2},\kappa_{2},X_{t}\right)\right]\end{array}\end{array},\label{eq:kernel-1}
\end{equation*}
$\mathcal{I}\left(z,\lambda,\kappa,X_{t}\right)$ as in Theorem \ref{thm:ELT},
and $\Gamma:=\arg\max_{z\in Z,\lambda\in\Lambda_{\infty}}\min_{\mathrm{csupp}\left(\kappa\right)\leq q}D\left(z,\lambda,\kappa,\mathbb{P}\right)$. 
\end{thm}
%Since $\Lambda_{\infty}$ is separable, there are no measurability
%problems with the definition of \break $\sup\inf_{\left(z,\lambda,\kappa\right)\in\Gamma}\mathcal{G}\left(z,\lambda,\kappa\right)$.
%Furthermore, for any $\lambda^{\star}\in\Lambda_{\infty}$, $\mathcal{G}_{\lambda^{\star}}\left(z,\kappa\right)=\mathcal{G}\left(z,\lambda^{\star},\kappa\right)$.
%Regarding CO, $\mathbb{E}\left(z-\sum_{i=0}^{\infty}\kappa_{i}X_{0,i}\right)_{+}$
%is twice differentiable w.r.t.\ $z$ and the second derivative assumes
%the form $\int_{\mathbb{R}^{q}}\delta\left(z-\kappa^{T}X\right)f\left(X\right)dX$.
%It is equal to the probability attributed by $f$ to the hyperplane $z=\sum_{i=0}^{\infty}\kappa_{i}X^{(i)}$
%which is positive if $X$ has a non degenerate covariance matrix.
%Under normality, the condition is thus guaranteed by the aforementioned
%limiting behavior on $V$ that also guarantees Assumption \ref{assu:SCS}.
%For CM, Theorem 4.5 of Beer and Lucchetti (1991) says that compactness
%of the set of minimizers is a topologically generic property in the sense of Baire
%category. Hence, it is expected to hold at least for a dense subset
%of $Z$, due to Assumption \ref{assu:UMom} and dominated convergence.
%The exclusion of the trivial threshold from the considerations is
%innocuous since $\mathcal{G}\left(\inf Z,\lambda,\kappa\right)$ is
%identically zero. CO and CM imply that $\Gamma-\left\{ \inf Z\right\} \times\Lambda_{\infty}\times\tilde{\Lambda}_{\left(q\right)}$
%is compact and thereby the generalized Delta method is applicable.

In practice, for fixed $p$, and as long as $p>q\ln T$, $\epsilon$ can be chosen conveniently close to 1, so that $p>q(\ln T)^{\epsilon}$ and the method is applicable.  Theorem \ref{thm:delta} allows for the construction of a feasible inferential
procedure based on subsampling in the spirit of  Linton et al.\ (2014) (see also Linton et al.\ (2005)) that approximates the asymptotic quantiles
of the limit in (\ref{eq:limit}). To get a viable numerical strategy, we design the subsampling technique to  avoid the costly
numerical search of the FSS  algorithm inside each subsample. To
this end, let $\kappa_{z,T}$ denote the solution of $\inf_{\mathrm{csupp}\left(\kappa\right)\leq q}\frac{1}{T}\sum_{t=0}^{T}\left(z_{t}-\sum_{i=0}^{\infty}\kappa_{i}X_{t}^{(i)}\right)_{+}$
over $\mathcal{L}_{p,q}$. Denote with $\Gamma^{\star}$
the subset of $\Gamma$ that contains the triplets at which some accumulation
point of $\kappa_{z,T}$ appears. Let $0<b_{T}\leq T$, and consider
the subsamples from the original observations $(X_{j})_{j=t,\ldots t+b_{T}-1}$
for all $t=1,2,\ldots,T-b_{T}+1$. For $\alpha\in\left(0,1\right)$,
denote with $q_{T,B_{T}}\left(1-\alpha\right)$ the $1-\alpha$ quantile
of the subsample empirical distribution of $\left(\sqrt{b_{T}}\left(\sup_{Z\times\Lambda_{p}}D\left(z,\kappa_{z,T},\lambda,\mathbb{P}_{t,b_{T}}\right)-M^{\mathrm{FS}}\left(\Lambda,\mathcal{\mathcal{L}}_{p,q},\mathbb{P}_{T},q\left(\ln T\right)^{\epsilon}\right)\right)\right)_{t=1,\dots,T-b_{T}+1}$,
where $\mathbb{P}_{t,b_{T}}$ denotes the empirical distribution of
$(X_{j})_{j=t,\ldots t+b_{T}-1}$ and we use the same $\kappa_{z,T}$ across subsamples. Hence, we get a fast subsampling method (Hong and Scaillet (2006)) that we use in our empirics to build confidence intervals for the estimated diversification loss.

Our final result depends on a condition on the elements
of $\Gamma^{\star}$ that avoids limiting degeneracies (Condition
ND below). They would imply poor higher order properties for
the conservative inference that we consider in Proposition \ref{prop:subs_1}. We say that
a triplet in $\Gamma^{\star}$ is trivial if the variance of $\mathcal{G}$
there is zero. We have triviality when the first element
of the triplet is $\inf Z$. It is also the case when $\lambda$ coincides
with the $\text{\ensuremath{\kappa}}$ appearing in the triplet. Then,
$\lambda$ is by construction an efficient element of $\Lambda_{\infty}$
that is also $q$-sparse. Whenever the elements of $X_{p}$ are linearly
independent for $p$ larger than the maximum desired value of $q$
for the analysis at hand, trivialities can occur only if SS-SSD holds.
This linear independence holds for Gaussian returns for example. 
\begin{prop}
\label{prop:subs_1} Suppose that (Condition ND) for
the given $q$, $\Gamma^{\star}$ contains at least
one non trivial triplet. Under the premises of Theorem \ref{thm:delta},
if $b_{T}\rightarrow\infty$, $\frac{b_{T}}{T}\rightarrow0$ and $\alpha<\frac{1}{2}$,
then we get the conservative result: 
\begin{equation}
\liminf_{T\rightarrow\infty}\mathbb{P}\left[M\left(\Lambda_{\infty},\mathcal{\mathcal{L}}_{\infty,q},\mathbb{P}\right)\in\left(Z_{M^{\mathrm{FS}}}(q),M^{\mathrm{FS}}\left(\Lambda,\mathcal{\mathcal{L}}_{p,q},\mathbb{P}_{T},q\left(\ln T\right)^{\epsilon}\right)+ \frac{q_{T,B_{T}}\left(1-\alpha\right)}{\sqrt{T}}\right)\right]\geq1-\alpha,\label{eq:cons-1}
\end{equation}
where $Z_{M^{\mathrm{FS}}}(q):=\max(0,M^{\mathrm{FS}}\left(\Lambda,\mathcal{\mathcal{L}}_{p,q},\mathbb{P}_{T},q\left(\ln T\right)^{\epsilon}\right)-\frac{q_{T,B_{T}}\left(1-\alpha\right)}{\sqrt{T}})$. If moreover there exists a unique $q$-sparse element
of $\Lambda$ that appears in every triplet in $\Gamma^{\star}$, then
we get the exact result:
\begin{equation}
\lim_{T\rightarrow\infty}\mathbb{P}\left[M\left(\Lambda_{\infty},\mathcal{\mathcal{L}}_{\infty,q},\mathbb{P}\right)\in\left[Z_{M^{\mathrm{FS}}}(q),M^{\mathrm{FS}}\left(\Lambda,\mathcal{\mathcal{L}}_{p,q},\mathbb{P}_{T},q\left(\ln T\right)^{\epsilon}\right)+ \frac{q_{T,B_{T}}\left(1-\alpha\right)}{\sqrt{T}}\right]\right]=1-\alpha.\label{eq:exact}
\end{equation}
\end{prop}
When spanning does not hold, then a form of non-degeneracy of $\mathbb{P}$ suffices for ND; for any optimal $K\in\mathcal{L}_{\infty,q}$, it suffices that no random variables in $X$ exist, corresponding to coordinates outside the support of $K$, that are obtainable as linear combinations of elements of $X$ that correspond to coordinates in the support of $K$. When spanning holds, the same non-degeneracy condition suffices, as long as there are triples in $\Gamma^{\star}$ that correspond to $z\neq\inf Z$. The latter can be achieved by trimming $z$ to be greater than or equal to an arbitrarily close number above the infimum of the minima of the empirical supports of the elements of $X$. It comes at the cost of weakening the SD relation. The non-degeneracy condition can be empirically tested via augmenting the vector of the elements of $X$ that correspond to coordinates in the support of the FS solution for the optimal $K$, with a single at a time element of $X$ outside $K$ and performing rank tests for the relevant $(q+1)\times (q+1)$ empirical covariance matrices-see for example Robin and Smith (2000). 
%non-degeneracy of $\mathbb{P}$ suffices for ND.}Under linear independence ND would hold whenever every
%$q$-sparse efficient element is matched by an efficient element of
%appropriately large support compared to the maximum desired level
%of $q$ for the underlying analysis. \textcolor{blue}{ND is weaker than the covariance matrix non-degeneracy conditions associated with every element in $D$ corresponding to a triplet in $\Gamma^{\star}$; see for example the conditions in Definition 3.1.(iii)-(iv) of Canay (2010) in the context of Empirical Likelihood inference for models of moment inequalities. Using the triplets $(z,\lambda,\kappa)$ corresponding to the empirical optimizers of the spanning criterion, ND can be empirically tested } 

The evaluation of the subsample quantile has small computational burden
since we avoid the costly sparse optimization w.r.t.\ $\kappa$
inside each subsample. Usually, $Z$ is approximated by some finite
discretization and optimization w.r.t.\ $\lambda$ is performed via
linearization of the SD conditions and the use of LP methods. Then,
the computational cost of sparse optimization is avoided and the asymptotic
results in (\ref{eq:cons-1})-(\ref{eq:exact}) hold as long as the
discretized set converges to a dense subset of $Z$.

In the special case where the problem $M\left(\Lambda_{\infty},\mathcal{\mathcal{L}}_{\infty,q},\mathbb{P}\right)$
has a unique optimizer - possible only if SS-SSD does not hold, (\ref{eq:limit}) implies asymptotic normality. It occurs
whenever the maximal expected utility difference between an efficient
element of $\Lambda_{\infty}$ and its approximate counterpart of
dimension $q$ occurs at a unique Russell-Seo utility for a unique
pair of efficient-approximate efficient portfolios. In such a case,
we can exploit normality to obtain a result like (\ref{eq:exact}).
A feasible normality result requires a consistent estimator for the
limiting variance. It can be obtained via a subsampling methodology
that does not involve subsample optimizations, as long as stricter
moment conditions hold for $X_{0}$, and a non-degeneracy condition
for the covariance kernel of $\mathcal{G}$ holds in some neighborhood
of the optimizer.

Proposition \ref{prop:subs_1} also implies an obvious conservative Kolmogorov-Smirnov testing procedure for the null hypothesis of $q$ sparse spanning based on subsampling. The null hypothesis is rejected iff zero does not lie inside the confidence interval. Given a finite set $Q\subset \mathbb{N}^{\star}$, it is also easy to use the result above in order to test more complicated hypotheses; e.g.\ the hypothesis that sparse spanning holds for at least some $q\in Q$ would be rejected iff zero lies outside the associated confidence interval for $\max\left\{q\in Q\right\}$. An interesting extension would be the construction of a test for the null of SS-SSD spanning when $q$ is allowed to diverge with rates dominated by the logarithm of the sample size.

\section{Numerical Implementation}

For $q<p$, we consider the following empirical
optimization problem 
\begin{equation}
\sup_{z\in Z}\sup_{\Lambda}\inf_{\mathcal{L}_{p,q}}\inf_{K}D\left(z,\kappa,\lambda,\mathbb{P}_{T}\right),\label{eq:empi_reg}
\end{equation}

The utility class interpretation of Arvanitis, Scaillet and Topaloglou (2020a,b) implies that we can represent (\ref{eq:empi_reg}) in terms of expected utility as:
\begin{eqnarray*}
\sup_{\boldsymbol{\lambda}\in\Lambda;u\in\mathcal{\mathcal{U}}}\inf_{\mathcal{L}_{p,q}}\inf_{\kappa\in\mathrm{K}}\mathbb{E}_{\mathbb{P}_{T}}\left[u\left(X^{\mathrm{T}}\boldsymbol{\lambda}\right)-u\left(X^{\mathrm{T}}\boldsymbol{\kappa}\right)\right],\label{eq:EU form of metric}
\end{eqnarray*}
with $\mathcal{U}:=\left\{ u\in\mathcal{C}^{0}:u(y)=\text{\ensuremath{\int}}_{\underline{x}}^{\overline{x}}v(x)r(y;x)dx\;v\in\mathcal{V}\right\}$,
$\mathcal{V}:=\left\{ v:\mathcal{X}\rightarrow\mathbb{R}_{+}:\int_{\mathcal{X}}v\left(x\right)=1\right\}$, and $
r(y;x):=(y-x)1(y\leq x),\;(x,y)\in\mathcal{X}^{2}$.

The set $\mathcal{U}$ is comprised of normalized, increasing, and concave utility functions that are constructed as convex mixtures of elementary Russell et Seo (1989) ramp functions $r(y;x),\:x\in\mathcal{X}$. This representation is used in the numerical implementation via
\begin{equation*}\sup_{u\in\mathcal{\mathcal{U}}}\left(\mathbb{\sup_{\boldsymbol{\lambda}\in\mathrm{\Lambda}}}\mathbb{E}_{\mathbb{P}_{T}}\left[u\left(X^{\mathrm{T}}\boldsymbol{\lambda}\right)\right]-\sup_{\mathcal{L}_{p,q}}\mathbb{\sup_{\boldsymbol{\kappa}\in\mathrm{K}}}\mathbb{E}_{\mathbb{P}_{T}}\left[u\left(X^{\mathrm{T}}\boldsymbol{\kappa}\right)\right]\right). \label{eq:empi_reg2}
\end{equation*}

We approximate every element of $\mathcal{U}$ with arbitrary prescribed accuracy using a finite set of increasing and concave piecewise-linear functions in the following way:

For $N_{1},N_{2}$ integers greater than
or equal to $2$, first, $\mathcal{X}$ is partitioned into $N_{1}$
equally spaced values as $\underline{x}=z_{1}<\cdots<z_{N_{1}}=\overline{x}$,
where $z_{n}:=\underline{x}+\frac{n-1}{N_{1}-1}(\overline{x}-\underline{x})$,
$n=1,\cdots,N_{1}$. Second, $[0,1]$ is partitioned as $0<\frac{1}{N_{2}-1}<\cdots<\frac{N_{2}-2}{N_{2}-1}<1$.
Using these partitions, an approximate optimization problem is considered:
\begin{eqnarray}
\sup_{u\in\underline{\mathcal{\mathcal{U}}}}\left(\mathbb{\sup_{\boldsymbol{\lambda}\in\mathrm{\Lambda}}}\mathbb{E}_{\mathbb{P}_{T}}\left[u\left(X^{\mathrm{T}}\boldsymbol{\lambda}\right)\right]-\mathbb{\sup_{\boldsymbol{\kappa}\in\mathrm{K}}}\mathbb{E}_{\mathbb{P}_{T}}\left[u\left(X^{\mathrm{T}}\boldsymbol{\kappa}\right)\right]\right),\label{eq:eta underbar}\end{eqnarray}
where $\qquad$  $\underline{\mathcal{U}}:=\left\{ u\in\mathcal{C}^{0}:u(y)=\sum_{n=1}^{N_{1}}v_{n}r(y;z_{n})\:v_{n}\text{\ensuremath{\in}}V\right\},$ and the set of allowable weights $V:=\left\{ v\in\left\{ 0,\frac{1}{N_{2}-1},\cdots,\frac{N_{2}-2}{N_{2}-1},1\right\} ^{N_{1}}:\sum_{n=1}^{N_{1}}v_{n}=1\right\}$.

By construction, every $u\in\underline{\mathcal{\mathcal{U}}}$ consists of at most $N_{2}$ linear line segments with endpoints at $N_{1}$ possible outcome levels. Furthermore, $\underline{\mathcal{\mathcal{U}}}\subset\mathcal{U}$, which is finite as it has $N_{3}:=\frac{1}{(N_{1}-1)!}\prod_{i=1}^{N_{1}-1}(N_{2}+i-1)$ elements and subsequently \eqref{eq:eta underbar} approximates \eqref{eq:empi_reg2} from below as the partitioning scheme is refined; $(N_{1},N_{2}\rightarrow\infty)$. Then, for every $u\in\underline{\mathcal{\mathcal{U}}}$, the two embedded utility maximization problems in (\ref{eq:eta underbar}) can be solved using LP. Consider 
$
c_{0,n}:=\sum_{m=n}^{N_{1}}\left(c_{1,m+1}-c_{1,m}\right)z_{m}$, 
$c_{1,n}:=\sum_{m=n}^{N_{1}}w_{m}$, and $
\mathcal{N}:=\left\{ n=1,\cdots,N_{1}:v_{n}>0\right\} \bigcup\left\{ N_{1}\right\}$.
Then, for any given $u\in\underline{\mathcal{\mathcal{U}}}$,
$\mathbb{\sup_{\boldsymbol{\lambda}\in\mathrm{\Lambda}}}\mathbb{E}_{\mathbb{P}_{T}}\left[u\left(X^{\mathrm{T}}\boldsymbol{\lambda}\right)\right]$
is the optimal value of the objective function of the following LP
problem in canonical form:
$
\max T^{-1}\sum_{t=1}^{T}y_{t}$ 
s.t.\ $y_{t}-c_{1,n}X_{t}^{\mathrm{T}}\boldsymbol{\lambda}\leq c_{0,n}$, $t=1,\cdots,T$, $n\in\mathcal{N}$,
$\sum_{i=1}^{M}\lambda_{i}=1$, 
$\lambda_{i}\geq0,\:i=1,\cdots,M$, and 
$y_{t}$ free, $t=1,\cdots,T$.
The LP problem always has a feasible solution and has $\mathcal{O}(T+N)$ variables and constraints. In the empirical application, we take $N_{1}=10$ and $N_{2}=5$. Thus, we end up with $N_{3}=\frac{1}{9!}\prod_{i=1}^{9}(4+i)=715$
distinct utility functions and $2N_{3}=1\,430$ small LP problems, which is time manageable with modern-day computer hardware and
solver software. We use a desktop PC with a 3.6 GHz, 24-core Intel i7 processor,
with 128 GB of RAM, using MATLAB and GAMS with the Gurobi optimization
solver. We start with an empty set, and then we gradually increase the number of assets adding 1 asset at a time until we find a set $K\subset\Lambda$ with $\mathrm{csupp}\left(K\right)\leq q$ and such
that $K\underset{\text{SSD}}{\succeq}\Lambda$. In each iteration, we search for the asset that increases (\ref{eq:empi_reg}) the most. 

\label{Sub_alg} The overall procedure consists of the following steps: \\
For $w=1$ to $q$:
\begin{enumerate}
\item If $w=1$, we search for the single asset that maximizes the value of (\ref{eq:empi_reg}). 
\item For $1 < w <q$, we solve (\ref{eq:empi_reg}) for each additional asset, and we keep the subset $K$ with dimension $w$, that maximizes (\ref{eq:empi_reg}).
\item If we find  a spanning set  $K$ inside the collection of all possible  subsets of $\mathrm{\Lambda}$ with dimension $w$, then the procedure stops. 
\item Else, if $w=q $ or the maximum amount of iterations $q\ln\lfloor T+1\rfloor$ is reached, we end up with a sparse portfolio set $K$ that "comes as close as possible" to SSD spanning its high dimensional universe of portfolios, and we evaluate the utility loss.
\end{enumerate}

Given the output of the last step of the procedure above, and since in the empirical applications $p$ is fixed, the optimal $q$, i.e., the one that provides the portfolio that comes closest in eliminating the empirical utility loss, can be readily estimated. To do so, and if the output of step 4 does not already imply zero optimal empirical utility loss, we may continue for $w>q $ up to $p$.

\section{Empirical Application}

We analyze large datasets of equity returns. We investigate the performance of our strategy based on the S\&P 500 index constituents, and we compare the results with the sparse mean-variance efficient portfolios (MAXSER) of 
Ao, Li, and Zheng (2019). We consider the period from January 1981 to December 2020, namely a total of 480 monthly return observations.

\subsection{In-Sample Analysis}

\subsubsection{Diversification Loss}

Starting with the empty set, we implement our sparse dominance methodology as described above, adding one element at a time in $r$ iterations. In each iteration, the algorithm 
adds to its current solution the single element decreasing the value of this solution by the most, i.e., 
the element with the largest marginal value with respect to the current solution. The target is to get the optimal portfolio with size $q$ that yields the minimal empirical diversification loss. 
Whenever the latter is zero, a sparse portfolio of support $q$ is built from a large set of assets of support $p$ 
which cannot be improved in terms of expected utility from the consideration of additional assets (full diversification).
\begin{singlespace}
\begin{figure}[H]
\caption{\footnotesize{}{}{}{}The upper panel plots the diversification loss w.r.t.\ the number of assets for the SS-SSD optimal portfolios. The lower panel plots the diversification loss of the optimal MAXSER portfolio with respect to the SS-SSD portfolio with zero loss, and the upper bound of a 95\%  and 90\% one-sided confidence intervals (CI).}
\centering{}\includegraphics[scale=1,width=1\textwidth]{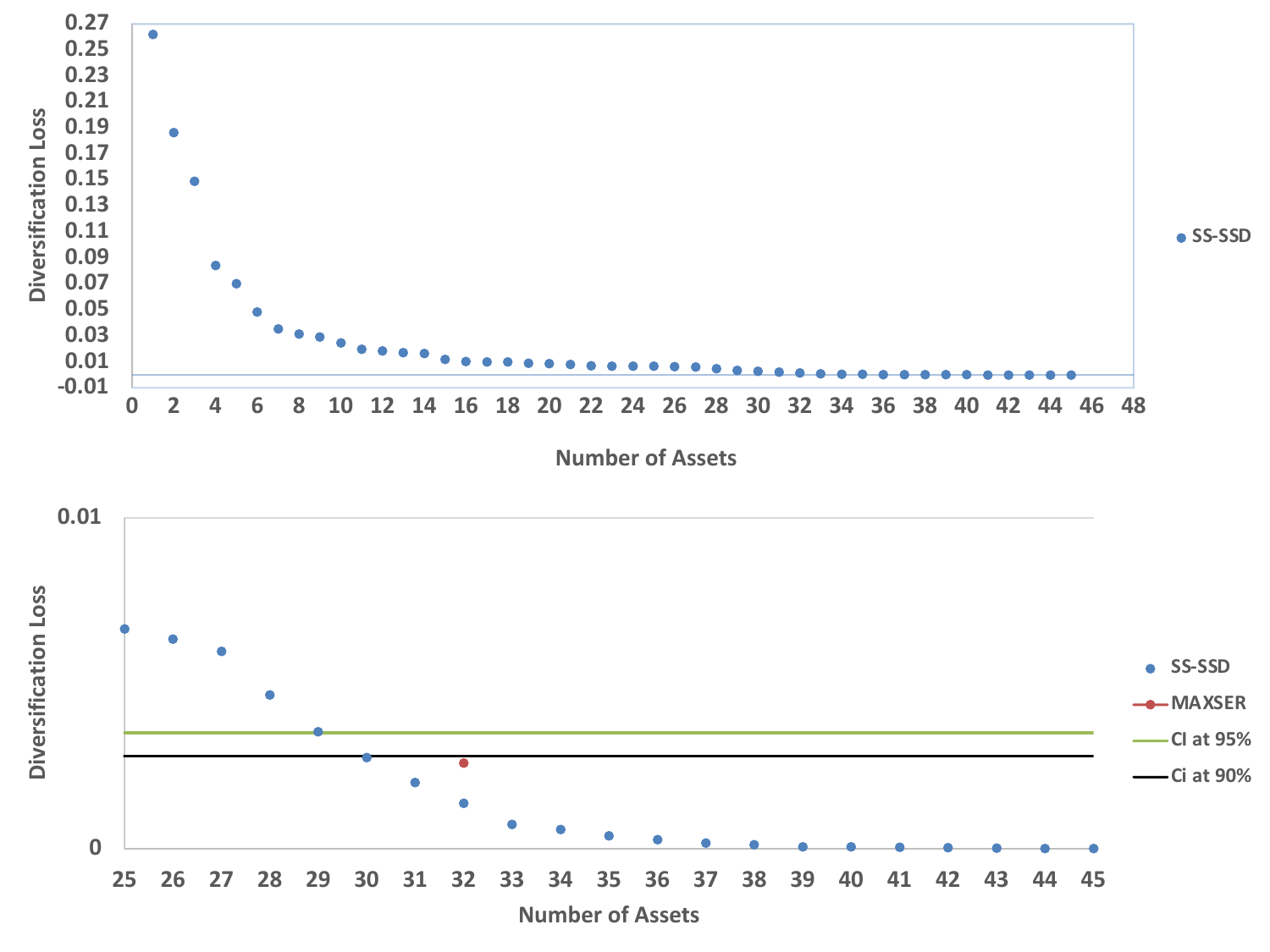}

\label{Fig1} 
\end{figure}
\end{singlespace}

In Figure \ref{Fig1}, we observe that the number of assets that yield zero diversification loss is 45 (upper panel)\footnote{Analogous analysis has been done for the FTSE100 constituents as well as the 49 Industry portfolios of Kenneth French. For the FTSE100, we get a subset $K$ with size $q=25$ that yields zero diversification loss, while, for the 49 Industry portfolios, the size is 13 assets.}. In the same figure (lower panel), we also observe that the MAXSER portfolio of Ao, Li, and Zheng (2019) consists of 32 assets.\footnote{The tuning parameter $\lambda$ in MAXSER is the regularisation parameter in the LASSO penalization. To determine it, we use the 10-fold cross-validation procedure, which is described in Section 1.5.1.\ of their paper.}  We evaluate the diversification loss, namely the estimated expected utility loss, of the optimal MAXSER portfolio with respect to the  SSD portfolio  with the smallest number of stocks reaching the zero bound. In the same graph, the upper bound of a 95\% as well as 90\% one-sided confidence intervals (CI) corresponding to the portfolio reaching the zero bound (45 assets) are additionally reported. We build them with the fast subsampling method  of Section \ref{sec:Sparse-Optimization-Sparse-Ident} whose validity is established in Proposition \ref{prop:subs_1}.  We observe that the diversification loss of the MAXSER portfolio is between the loss of the  SS-SSD portfolio for $q=32$ and the 90\% confidence interval. 

\subsubsection{Performance Summary of the Optimal Portfolios}

We compare the in-sample performance of the MAXSER and the SS-SSD optimal portfolios as well as the $1/N$ (equally-weighted) portfolio with $N=p=500$.
We compute the first four moments of portfolio returns (Average, Standard Deviation, Skewness and Kurtosis), 
as well as a number of commonly used parametric performance measures for portfolios:  the Sharpe Ratio, the Downside Sharpe Ratio of Ziemba
(2005), the 95\% Value-at-Risk (with a positive sign for a loss), the 95\% Expected Shortfall (with a positive sign for a loss), the Upside Potential and Downside Risk (UP) ratio of Sortino
and van den Meer (1991), the Opportunity Cost, and the Certainty Equivalent return (CEQ).

The definition of Downside Sharpe Ratio uses the downside variance
(or more precisely the downside risk) defined as 
$
\sigma_{P_{-}}^{2}=\frac{\sum_{t=1}^{T}(R_{t})_{-}^{2}}{T-1},
$
where $(R_{t})_{-}$ is the return
of portfolio $P$ at day $t$ which is below zero (i.e., those
with losses).
Given that the total variance equals twice the downside variance $2\sigma_{P_{-}}^{2}$,
the Downside Sharpe Ratio is given by 
$
\mathrm{S}_{P}=\frac{\bar{R}_{P}-\bar{R}_{f}}{\sqrt{2}\sigma_{P-}},
$
where $\bar{R}_{P}$ is the average period return of portfolio $P$
and $\bar{R_{f}}$ is the average risk free rate. 

The UP ratio compares the upside potential to the shortfall
risk over a specific target (benchmark): 
$
\textrm{UP ratio}=\frac{\frac{1}{T}\sum_{t=1}^{T}(R_{P,t}-R_{f,t})_+}{\sqrt{{\frac{1}{T}\sum_{t=1}^{T}((R_{f,t}-R_{P,t})_+)^{2}}}},
$
where $R_{P,t}$ is the realized monthly return of the portfolio $P$ for the
out-of-sample period, $T$ is the number of experiments performed,
and $R_{f,t}$ is the monthly return of the benchmark (the riskless asset).
The numerator equals the average excess return over the benchmark
reflecting the upside potential while the denominator measures the
downside risk (i.e., shortfall risk over the benchmark).

Both the Downside Sharpe and UP Ratios are viewed to
be more appropriate measures of performance than the typical Sharpe
Ratio given the asymmetric return distribution of assets.

Moreover, we compute a certainty equivalent return (CEQ) for the three portfolios based on the exponential and power utility functions: 
$
\mathbb{E}_{\mathbb{P}_{T}}[u(1+R_{\mathrm{P}})]=u(1+ CEQ).
$ For its calculation, exponential and power utility functions are
used, consistent with second degree stochastic dominance.
For the coefficient of risk aversion alternative values are employed.

Finally, the Opportunity Cost $\theta$ of Simaan (2013) is used, which is a useful measure
for the economic significance of the performance difference of two
portfolios. It is defined as the return that needs to be added
to (or subtracted from) the MAXSER portfolio return $R_{\mathrm{MAXSER}}$,
so that the investor is indifferent (in utility terms) between the
the two different portfolios: $
\mathbb{E}_{\mathbb{P}_{T}}[u(1+R_{\mathrm{MAXSER}}+\theta)]=\mathbb{E}_{\mathbb{P}_{T}}[u(1+R_{\mathrm{SS-SSD}})].
$
A positive (negative) Opportunity Cost implies that the investor is
better (worse) off if  he invests in the SS-SSD over the MAXSER portfolio.
We use the same type of definition for the Opportunity Cost for the $1/N$ portfolio.
Given that this measure takes into account the entire probability
distribution of asset returns, it is suitable to evaluate strategies
even when the asset return distribution is not normal. Again we use exponential and power utility functions under alternative values
for the coefficient of risk aversion.

Table \ref{t-insample} reports the performance and risk measures of the in-sample performance of
the three portfolios.  They allow to finer distinguish the differences between the portfolios.
We observe that the Average as well as the Standard Deviation for the SS-SSD portfolio are higher that those of the MAXSER portfolio,  while the Sharpe Ratio is slightly lower. It is expected, since the Sharpe Ratio is the maximization target in the construction of the MAXSER portfolio. The Skewness is less negative and the Kurtosis is higher. Although the Sharpe Ratio of the SS-SSD portfolio is slightly lower, the Downside Sharpe Ratio as well as the UP Ratio are higher.
The VaR and the Expected Shortfall (with a positive sign for a loss) are lower as expected when investors want to mitigate the impact of large losses. 
The SS-SSD portfolio targets and achieves a transfer of probability mass from the left to the right tail of the return distribution when compared to the MAXSER portfolio.
We also observe that both the SS-SSD and the MAXSER portfolios outperform the $1/N$ portfolio in all performance and risk measures.
The CEQ is higher for the SS-SSD portfolio, and the Opportunity Cost is always positive, 
indicating that a positive return should be added in the MAXSER or the $1/N$ portfolio to achieve the same expected return with the SS-SSD portfolio. 
The CEQ for the optimal SS-SSD with $q = 45$ range from 0.958\% to 1.183\% for the exponential utility and from 1.184\% to 6.214\% for the power utility.  For the optimal MAXSER portfolio with $q = 32$, we get a CEQ between 0.855\% and 1.060\% for the exponential utility  and an Opportunity Cost between 1.060\% and 5.536\% for the power utility. For the $1/N$ portfolio, we get a CEQ between 1.016\% and 0.670\% for the exponential utility  and between  1.016\% and  5.243\% for the power utility.  

In order to economically quantify the diversification loss in Figure \ref{Fig1},  we also report in Table \ref{t-insample}  the performance measures for $P(5)$ and $P(10)$ i.e., optimal portfolios $P(q)$ with cardinality constraints $q=5$ and $q=10$ assets.
We observe that both portfolios exhibit significantly worse performance than the SS-SSD and MAXSER portfolios. Moreover, we get a very negative Skewness and positive Kurtosis, while the VaR and Expected Shortfall are huge.
We also calculate the CEQ as well as the Opportunity Cost $\theta$:  $
\mathbb{E}_{\mathbb{P}_{T}}[u(1+R_{\mathrm{P(q)}}+\theta)]=\mathbb{E}_{\mathbb{P}_{T}}[u(1+R_{\mathrm{SS-SSD}})].
$
For $q=5$,  the CEQ ranges from 0.496\% to 0.815\% for the exponential utility and from 0.916\% to 4.257\% for the power utility, while the Opportunity Cost ranges from 0.042\% to 0.062\% for the exponential utility and from 0.051\% to 0.122\% for the power utility. For $q=10$, the CEQ ranges from 0.629\% to 0.928\% for the exponential utility and from 1.019\% to 4.910\% for the power utility, while the Opportunity Cost for the diversification loss ranges from 0.047\% to 0.068\% for the exponential utility and from 0.069\% to 0.145\% for the power utility.  We observe differences in the CEQ between the various portfolios around 0.1\%, 1\%, or 1.5\% depending on the chosen utility function. They correspond to around 1.2\%, 12\%, or 18\% on an annual basis.

\begingroup
\setlength{\tabcolsep}{6pt}
\renewcommand{\arraystretch}{0.5}
\begin{table}[H]
\caption{In-sample performance: risk and performance measures}
\label{t-insample}%
\begin{center}
\scalebox{0.8}{
\begin{tabular}{lccccc}
\toprule[1.5pt]
 & {\small{}{}{}{}{}{}{}{}{}{}{}{}{}SS-SSD}  & {\small{}{}{}{}{}{}{}{}{}{}{}{}{}MAXSER}   & {\small{}{}{}{}{}{}{}$1/N$}  & {\small{}{}{}{}{}{}{}$P(5)$} & {\small{}{}{}{}{}{}{}$P(10)$}\tabularnewline
\midrule
{\small{}{}{}{}{}{}{}{}{}{}{}{}{}{}{}{}{}
Measures}  &  &   & \tabularnewline
\midrule 
{\small{}{}{}{}{}{}{}{}{}{}{}{}{}{}{}{}{}Average}  & {\small{}{}{}{}{}{}{}{}{}{}0.0129 } & {\small{}{}{}{}{}{}{}{}{}{}0.0126}  &  {\small{}{}{}{}{}{}0.0133 } & {\small{}{}{}{}{}{}{}{}{}{}0.0114}  &  {\small{}{}{}{}{}{}0.0118 } \tabularnewline
{\small{}{}{}{}{}{}{}{}{}{}{}{}{}{}{}{}{}Standard Deviation}  & {\small{}{}{}{}{}{}{}{}{}0.0331} &  {\small{}{}{}{}{}{}{}{}0.0314}    & {\small{}{}{}{}{}{}0.0458 } & {\small{}{}{}{}{}{}{}{}{}{}0.0480}  &  {\small{}{}{}{}{}{}0.0481 } \tabularnewline
{\small{}{}{}{}{}{}{}{}{}{}{}{}{}{}{}{}{}Skewness}  &{\small{}{}{}{}{}{}{}{}-0.1986} &  {\small{}{}{}{}{}{}{}{}{}-0.2122}  &  {\small{}{}{}{}{}{}{} -0.2689 }& {\small{}{}{}{}{}{}{}{}{}{}-0.5747}  &  {\small{}{}{}{}{}{}-0.5025 } \tabularnewline
{\small{}{}{}{}{}{}{}{}{}{}{}{}{}{}{}{}{}Kurtosis}  &{\small{}{}{}{}{}{}{}{}{}1.7595}  & {\small{}{}{}{}{}{}{}{}{}1.2521}  &  {\small{}{}{}{}{}{}{}{}{}{} 2.9690 } & {\small{}{}{}{}{}{}{}{}{}{}5.4234}  &  {\small{}{}{}{}{}{}3.2736 } \tabularnewline
{\small{}{}{}{}{}{}{}{}{}{}{}{}{}{}{}{}{}Sharpe Ratio}  & {\small{}{}{}{}{}{}{}0.3904}  &  {\small{}{}{}{}{}{}{}0.4013}  &  {\small{}{}{}{}{}{}{}0.2899 } & {\small{}{}{}{}{}{}{}{}{}{}0.2199}  &  {\small{}{}{}{}{}{}0.2353 } \tabularnewline
{\small{}{}{}{}{}{}{}{}{}{}{}{}{}{}{}{}{}Downside Sharpe Ratio} & {\small{}{}{}{}{}{}{}{}{}0.6613}  & {\small{}{}{}{}{}{}{}0.6036}   & {\small{}{}{}{}{}0.4453 }  & {\small{}{}{}{}{}{}{}{}{}{}0.3825}  &  {\small{}{}{}{}{}{}0.4195 } \tabularnewline
{\small{}{}{}{}{}{}{}{}{}{}{}{}{}{}{}{}{}Value-at-Risk}  & {\small{}{}{}{}{}{}{}{}{}0.0396}  & {\small{}{}{}{}{}{}{}{}{}0.0430}  &  {\small{}{}{}{}{}{}{} 0.0615 } & {\small{}{}{}{}{}{}{}{}{}{}0.0683}  &  {\small{}{}{}{}{}{}0.0672 } \tabularnewline
{\small{}{}{}{}{}{}{}{}{}{}{}{}{}{}{}{}{}Expected Shortfall}  & {\small{}{}{}{}{}{}{}0.0617} & {\small{}{}{}{}{}{}{}0.0651}  &  {\small{}{}{}{}{}{} 0.0959 } & {\small{}{}{}{}{}{}{}{}{}{}0.1196}  &  {\small{}{}{}{}{}{}0.0983 } \tabularnewline
{\small{}{}{}{}{}{}{}{}{}{}{}{}{}{}{}{}{}UP ratio}  & {\small{}{}{}{}{}{}{}{}0.9019} & {\small{}{}{}{}{}{}{}{}0.8739}    & {\small{}{}{}{}{}{}{}{}0.7780 } & {\small{}{}{}{}{}{}{}{}{}{}0.0655}  &  {\small{}{}{}{}{}{}0.7102 } \tabularnewline
{\small{}{}{}{}{}{}{}{}{}{}{}{}{}{}{}{}{}Certainty Equivalent}  &  &  & \tabularnewline
\emph{\small{}{}{}{}{}{}{}{}{}{}{}{}{}{}{}{}{}Exponential Utility}{\small{}{}{}{}{}{}{}{}{}{}{}{}{}{}{}{}{}}  &  &    & && \tabularnewline
{\small{}{}{}{}{}{}{}{}{}{}{}{}{}{}{}{}{}ARA=2}  & {\small{}{}{}{}{}{}{}{}{}{}1.183\%} & {\small{}{}{}{}{}{}{}{}{}{}{}1.060\%}  &    {\small{}{}{}{}{}{}{}{}1.016\%}  & {\small{}{}{}{}{}{}{}{}{}{}0.815\%}  &  {\small{}{}{}{}{}{}0.978\% }  \tabularnewline
{\small{}{}{}{}{}{}{}{}{}{}{}{}{}{}{}{}{}ARA=4}  & {\small{}{}{}{}{}{}{}{}{}{}1.071\%} & {\small{}{}{}{}{}{}{}{}{}{}{}0.958\%}  &   {\small{}{}{}{}{}{}{}{}0.898\%}   & {\small{}{}{}{}{}{}{}{}{}{}0.706\%}  &  {\small{}{}{}{}{}{}0.831\% } \tabularnewline
{\small{}{}{}{}{}{}{}{}{}{}{}{}{}{}{}{}{}ARA=6}  & {\small{}{}{}{}{}{}{}{}{}{}0.958\%}  & {\small{}{}{}{}{}{}{}{}{}{}{}0.855\%}  &   {\small{}{}{}{}{}{}{}{}0.670\%}   & {\small{}{}{}{}{}{}{}{}{}{}0.496\%}  &  {\small{}{}{}{}{}{}0.629\% } \tabularnewline
\emph{\small{}{}{}{}{}{}{}{}{}{}{}{}{}{}{}{}{}Power Utility}{\small{}{}{}{}{}{}{}{}{}{}{}{}{}{}{}{}{}}  &  &   &  && \tabularnewline
{\small{}{}{}{}{}{}{}{}{}{}{}{}{}{}{}{}{}RRA=2}  & {\small{}{}{}{}{}{}{}{}{}{}1.184\%}  &  {\small{}{}{}{}{}{}{}{}{}{}1.060\%}  &    {\small{}{}{}{}{}{}{}{}{}1.016\%}   & {\small{}{}{}{}{}{}{}{}{}{}0.916\%}  &  {\small{}{}{}{}{}{}1.019\% } \tabularnewline
{\small{}{}{}{}{}{}{}{}{}{}{}{}{}{}{}{}{}RRA=4}  &  {\small{}{}{}{}{}{}{}{}{}{}3.638\%} & {\small{}{}{}{}{}{}{}{}{}{}{}3.250\%}  &     {\small{}{}{}{}{}{}{}{}{}3.125\%}  & {\small{}{}{}{}{}{}{}{}{}{}1.801\%}  &  {\small{}{}{}{}{}{}2.464\% } \tabularnewline
{\small{}{}{}{}{}{}{}{}{}{}{}{}{}{}{}{}{}RRA=6}  & {\small{}{}{}{}{}{}{}{}{}{}6.214\%} & {\small{}{}{}{}{}{}{}{}{}{}{}5.536\%} &      {\small{}{}{}{}{}{}{}{}{}5.242\%}  & {\small{}{}{}{}{}{}{}{}{}{} 4.257\%}  &  {\small{}{}{}{}{}{}4.910\% }  \tabularnewline
{\small{}{}{}{}{}{}{}{}{}{}{}{}{}{}{}{}{}Opportunity Cost}  &  &  & \tabularnewline
\emph{\small{}{}{}{}{}{}{}{}{}{}{}{}{}{}{}{}{}Exponential Utility}{\small{}{}{}{}{}{}{}{}{}{}{}{}{}{}{}{}{}}  &  &    & && \tabularnewline
{\small{}{}{}{}{}{}{}{}{}{}{}{}{}{}{}{}{}ARA=2}  & & {\small{}{}{}{}{}{}{}{}0.090\%}  &    {\small{}{}{}{}{}{}{}{}{}0.081\%} & {\small{}{}{}{}{}{}{}{}{}{}0.062\%}  &  {\small{}{}{}{}{}{}0.068\% } \tabularnewline
{\small{}{}{}{}{}{}{}{}{}{}{}{}{}{}{}{}{}ARA=4}  & & {\small{}{}{}{}{}{}{}{}0.087\%}  &   {\small{}{}{}{}{}{}{}{}{}0.075\%} & {\small{}{}{}{}{}{}{}{}{}{}0.051\%}  &  {\small{}{}{}{}{}{}0.056\% }\tabularnewline
{\small{}{}{}{}{}{}{}{}{}{}{}{}{}{}{}{}{}ARA=6}  & & {\small{}{}{}{}{}{}{}{}0.077\%}  &   {\small{}{}{}{}{}{}{}{}{}0.061\%}   & {\small{}{}{}{}{}{}{}{}{}{}0.042\%}  &  {\small{}{}{}{}{}{}0.047\% } \tabularnewline
\emph{\small{}{}{}{}{}{}{}{}{}{}{}{}{}{}{}{}{}Power Utility}{\small{}{}{}{}{}{}{}{}{}{}{}{}{}{}{}{}{}}  &  &   &  && \tabularnewline
{\small{}{}{}{}{}{}{}{}{}{}{}{}{}{}{}{}{}RRA=2}  & &  {\small{}{}{}{}{}{}{}{}0.093\%}  &    {\small{}{}{}{}{}{}{}{}{}0.072\%}  & {\small{}{}{}{}{}{}{}{}{}{}0.051\%}  &  {\small{}{}{}{}{}{}0.069\% }  \tabularnewline
{\small{}{}{}{}{}{}{}{}{}{}{}{}{}{}{}{}{}RRA=4}  & & {\small{}{}{}{}{}{}{}{}{}0.156\%}  &    {\small{}{}{}{}{}{}{}{}{}{}0.121\%} & {\small{}{}{}{}{}{}{}{}{}{}0.092\%}  &  {\small{}{}{}{}{}{}0.112\% }  \tabularnewline
{\small{}{}{}{}{}{}{}{}{}{}{}{}{}{}{}{}{}RRA=6}  & & {\small{}{}{}{}{}{}{}{}{}0.189\%} &     {\small{}{}{}{}{}{}{}{}{}{}0.162\%} & {\small{}{}{}{}{}{}{}{}{}{}0.122\%}  &  {\small{}{}{}{}{}{}0.145\% }\tabularnewline
\bottomrule[1.5pt]
\end{tabular}}
\end{center}
\noindent {\small{}{}{}{}{}{}{}{}{}{}{}{}{}{}{}{}{}}%
\noindent\parbox[c][0.7\totalheight][s]{1\textwidth}{%
\noindent {\footnotesize\begin{singlespace}Entries report the risk and performance
measures (Sharpe Ratio, Downside Sharpe Ratio, VaR, ES, UP Ratio, Opportunity Cost and Certainty Equivalent) for the SS-SSD, the MAXSER, the P(5) and P(10) optimal portfolios (with cardinality constraints $q=5$ and $q=10$) as well as for the $1/N$ portfolio.
The data cover the period from January, 1980
to December, 2020.\end{singlespace}}}
\end{table}
\endgroup

Finally, Table \ref{Inweights} reports the average and standard deviation of asset weights of the major Industries selected by each one of the two portfolios. We observe that both portfolios are well diversified and invest in almost the same Industries,  with different overall weights. The table also exhibits the \% of overlap, between assets selected by SS-SSD and MAXSER. The overlap is high, since both strategies select assets from the same industries. Moreover, Table \ref{SkewKurt} shows the average skewness and kurtosis of the assets selected by both strategies. We observe that the SS-SSD strategy picks assets with higher skewness and kurtosis, as an attempt to increase the right tail and diminish the left one. 

\begingroup
\setlength{\tabcolsep}{6pt}
\renewcommand{\arraystretch}{0.5}
\begin{table}[H]
\caption{ S$\&$P 500 Industry weights}
\label{Inweights}
\begin{center}
\scalebox{0.8}{
\begin{tabular}{lccccc}
\toprule[1.5pt] 
 & {\small{}{}{}{}{}{}{}{}{}{}{}{}{}{}{}{}{}SS-SSD}  &  &  {\small{}{}{}{}{}{}{}{}{}{}{}{}{}{}{}{}{}MAXSER} &  &  \tabularnewline
\midrule 
{\small{}{}{}{}{}{}{}{}{}{}{}{}{}{}{}{}{} Weights}  & Average &  St. Dev&  Average &  St. Dev & \% overlap  \tabularnewline
\midrule 
{\small{}{}{}{}{}{}{}{}{}{}{}{}{}{}{}{}{}Capital Goods}  &  {\small{}{}{}{}{}{}{}{}{}{}{}{}{}{}{}{}{}3.43\%}  & {\small{}3.17\%}  &  {\small{}{}{}{}{}{}{}{}{}{}{}{}{}{}{}{}{}4.50\%}  & {\small{}3.26\%} & {\small{}{}{}{}{}{}{}{}{}{}{}{}{}{}{}{}{}29.24\%}  \tabularnewline
{\small{}{}{}{}{}{}{}{}{}{}{}{}{}{}{}{}{}Consumer Services}  &  {\small{}{}{}{}{}{}{}{}{}{}{}{}6.57\%}  & {\small{}4.11\%}  & {\small{}{}{}{}{}{}{}{}{}{}{}{}{}{}{}{}{}8.39\%}  & {\small{}5.97\%}  & {\small{}{}{}{}{}{}{}{}{}{}{}{}{}{}{}{}{}32.77\%} \tabularnewline
{\small{}{}{}{}{}{}{}{}{}{}{}{}{}{}{}{}{}Financial}  &  {\small{}{}{}{}{}{}{}{}{}{}{}{}{}{}{}{}{}3.60\%}  & {\small{}2.97\%}  & {\small{}{}{}{}{}{}{}{}{}{}{}{}{}{}{}{}{}4.73\%}  & {\small{}3.88\%}  &  {\small{}{}{}{}{}{}{}{}{}{}{}{}{}{}{}{}{}27.39\%}  \tabularnewline
{\small{}{}{}{}{}{}{}{}{}{}{}{}{}{}{}{}{}Consumer Staples}  &  {\small{}{}{}{}{}{}{}{}{}{}{}{}{}{}{}{}{}3.24\%} & {\small{}1.89\%} & {\small{}{}{}{}{}{}{}{}{}{}{}{}{}{}{}{}{}0\%}  & {\small{}-}  &{\small{}{}{}{}{}{}{}{}{}{}{}{}{}{}{}{}{}0\%} \tabularnewline
{\small{}{}{}{}{}{}{}{}{}{}{}{}{}{}{}{}{}Food}  &  {\small{}{}{}{}{}{}{}{}{}{}{}{}{}{}{}{}{}2.70\%}   & {\small{}1.25\%}  & {\small{}{}{}{}{}{}{}{}{}{}{}{}{}{}{}{}{}3.21\%}  & {\small{}2.67\%}  &  {\small{}{}{}{}{}{}{}{}{}{}{}{}{}{}{}{}{}45.34\%}  \tabularnewline
{\small{}{}{}{}{}{}{}{}{}{}{}{}{}{}{}{}{}Health care}  & {\small{}{}{}{}{}{}{}{}{}{}{}{}{}{}{}{}{}8.31\%}  & {\small{}5.41\%}  & {\small{}{}{}{}{}{}{}{}{}{}{}{}{}{}{}{}{}7.43\%}  & {\small{}4.74\%}  &  {\small{}{}{}{}{}{}{}{}{}{}{}{}{}{}{}{}{}51.25\%}   \tabularnewline
{\small{}{}{}{}{}{}{}{}{}{}{}{}{}{}{}{}{}Household}  &  {\small{}{}{}{}{}{}{}{}{}{}{}{}{}{}{}{}{}4.37\%}  & {\small{}3.26\%}  & {\small{}{}{}{}{}{}{}{}{}{}{}{}{}{}{}{}{}5.58\%}  & {\small{}4.12\%}  & {\small{}{}{}{}{}{}{}{}{}{}{}{}{}{}{}{}{}44.12\%} \tabularnewline
{\small{}{}{}{}{}{}{}{}{}{}{}{}{}{}{}{}{}IMedia}  &  {\small{}{}{}{}{}{}{}{}{}{}{}{}{}{}{}{}{}4.34\% }  & {\small{}3.29\%}  & {\small{}{}{}{}{}{}{}{}{}{}{}{}{}{}{}{}{}4.58\%}  & {\small{}3.66\%}  &  {\small{}{}{}{}{}{}{}{}{}{}{}{}{}{}{}{}{}51.25\%} \tabularnewline
{\small{}{}{}{}{}{}{}{}{}{}{}{}{}{}{}{}{}Pharm}  &  {\small{}{}{}{}{}{}{}{}{}{}{}{}{}{}{}{}{}5.69\%}  & {\small{}4.23\%} & {\small{}{}{}{}{}{}{}{}{}{}{}{}{}{}{}{}{}6.89\%}  & {\small{}5.47\%}  &  {\small{}{}{}{}{}{}{}{}{}{}{}{}{}{}{}{}{}45.96\%} \tabularnewline
{\small{}{}{}{}{}{}{}{}{}{}{}{}{}{}{}{}{}Retailing}  &  {\small{}{}{}{}{}{}{}{}{}{}{}{}{}{}{}{}{}19.43\%}  & {\small{}9.34\%} &  {\small{}{}{}{}{}{}{}{}{}{}{}{}{}{}{}{}{}17.21\%}  & {\small{}9.39\%} &  {\small{}{}{}{}{}{}{}{}{}{}{}{}{}{}{}{}{}61.56\%} \tabularnewline
{\small{}{}{}{}{}{}{}{}{}{}{}{}{}{}{}{}{}Software}  &  {\small{}{}{}{}{}{}{}{}{}{}{}{}{}{}{}{}{}16.21\%}   & {\small{}8.95\%} &  {\small{}{}{}{}{}{}{}{}{}{}{}{}{}{}{}{}{}14.51\%}  & {\small{}8.34\%} & {\small{}{}{}{}{}{}{}{}{}{}{}{}{}{}{}{}{}39.43\%} \tabularnewline
{\small{}{}{}{}{}{}{}{}{}{}{}{}{}{}{}{}{}Technology}  &  {\small{}{}{}{}{}{}{}{}{}{}{}{}{}{}{}{}{}12.79\%}  & {\small{}7.94\%}  &  {\small{}{}{}{}{}{}{}{}{}{}{}{}{}{}{}{}{}11.45\%}  & {\small{}7.38\%}  &  {\small{}{}{}{}{}{}{}{}{}{}{}{}{}{}{}{}{}37.94\%}  \tabularnewline
{\small{}{}{}{}{}{}{}{}{}{}{}{}{}{}{}{}{}Transportation}  &  {\small{}{}{}{}{}{}{}{}{}{}{}{}{}{}{}{}{}4.81\%}   & {\small{}3.76\%}  &  {\small{}{}{}{}{}{}{}{}{}{}{}{}{}{}{}{}{}5.62\%}  & {\small{}3.87\%} & {\small{}{}{}{}{}{}{}{}{}{}{}{}{}{}{}{}{}41.12\%} \tabularnewline
\bottomrule[1.5pt] 
\end{tabular}}
\end{center}
\noindent {\small{}{}{}{}{}{}{}{}{}{}{}{}{}{}{}{}{}}%
\noindent\parbox[c][0.7\totalheight][s]{1\textwidth}{%
\noindent {\footnotesize\begin{singlespace}Entries report the average and standard deviation of Industry weights  of the SS-SSD and the MAXSER portfolios in the major Industries of the S$\&$P 500 Index, as well as the \% of overlap, between assets selected by SS-SSD and MAXSER.\end{singlespace} } %
}
\end{table}
\endgroup

\begingroup
\setlength{\tabcolsep}{6pt}
\renewcommand{\arraystretch}{0.5}
\begin{table}[H]
\caption{Average Skewness and Kurtosis of the selected S$\&$P 500 Industry assets}
\label{SkewKurt}
\begin{center}
{\small{}{}{}{}{}{}{}{}{}{}{}{}{}{}{}{}{}}

\scalebox{0.8}{
\begin{tabular}{lcccc}
\toprule[1.5pt]
 & {\small{}{}{}{}{}{}{}{}{}{}{}{}{}{}{}{}{}SS-SSD} &  &  {\small{}{}{}{}{}{}{}{}{}{}{}{}{}{}{}{}{}MAXSER}  &    \tabularnewline
\midrule
{\small{}{}{}{}{}{}{}{}{}{}{}{}{}{}{}{}{} Weights}  & Skewness & Kurtosis &  Skewness & Kurtosis \tabularnewline
\midrule
{\small{}{}{}{}{}{}{}{}{}{}{}{}{}{}{}{}{}Capital Goods}    & {\small{}{}{}{}{}{}{}{}{}{}{}{}{}{}{}{}{}0.239}  & {\small{}2.339}  & {\small{}{}{}{}{}{}{}{}{}{}{}{}{}{}{}{}{}0.166}   & {\small{}1.799}  \tabularnewline
{\small{}{}{}{}{}{}{}{}{}{}{}{}{}{}{}{}{}Consumer Services}   & {\small{}{}{}{}{}{}{}{}{}{}{}{}{}{}{}{}{}{}{}-0.163}  & {\small{}2.098} & {\small{}{}{}{}{}{}{}{}{}{}{}{}-0.303}      & {\small{}1.614}   \tabularnewline
{\small{}{}{}{}{}{}{}{}{}{}{}{}{}{}{}{}{}Financial}   & {\small{}{}{}{}{}{}{}{}{}{}{}{}{}{}{}{}{}0.104}  & {\small{}2.360}  & {\small{}{}{}{}{}{}{}{}{}{}{}{}{}{}{}{}{}-0.027}           & {\small{}1.815}   \tabularnewline
{\small{}{}{}{}{}{}{}{}{}{}{}{}{}{}{}{}{}Consumer Staples}  & {\small{}{}{}{}{}{}{}{}{}{}{}{}{}{}{}{}{}0.122} & {\small{}1.460}   & {\small{}{}{}{}{}{}{}{}{}{}{}{}{}{}{}-}          & {\small{}-}  \tabularnewline
{\small{}{}{}{}{}{}{}{}{}{}{}{}{}{}{}{}{}Food}   & {\small{}{}{}{}{}{}{}{}{}{}{}{}{}{}{}{}{}0.188}   & {\small{}2.644}   & {\small{}{}{}{}{}{}{}{}{}{}{}{}{}{}{}{}{}0.171}                   & {\small{}2.034}  \tabularnewline
{\small{}{}{}{}{}{}{}{}{}{}{}{}{}{}{}{}{}Health care}    & {\small{}{}{}{}{}{}{}{}{}{}{}{}{}{}{}{}{}-0.203}  & {\small{}1.552}  &{\small{}{}{}{}{}{}{}{}{}{}{}{}{}{}{}{}{}-0.348}        & {\small{}0.348}   \tabularnewline
{\small{}{}{}{}{}{}{}{}{}{}{}{}{}{}{}{}{}Household}    & {\small{}{}{}{}{}{}{}{}{}{}{}{}{}{}{}{}{}-0.128}  & {\small{}3.377} & {\small{}{}{}{}{}{}{}{}{}{}{}{}{}{}{}{}{}-0.475}        & {\small{}2.367}    \tabularnewline
{\small{}{}{}{}{}{}{}{}{}{}{}{}{}{}{}{}{}IMedia}   & {\small{}{}{}{}{}{}{}{}{}{}{}{}{}{}{}{}{}0.148}  & {\small{}1.628}  & {\small{}{}{}{}{}{}{}{}{}{}{}{}{}{}{}{}{}0.053}                & {\small{}1.175}  \tabularnewline
{\small{}{}{}{}{}{}{}{}{}{}{}{}{}{}{}{}{}Pharm}   & {\small{}{}{}{}{}{}{}{}{}{}{}{}{}{}{}{}{}0.110}  & {\small{}1.325} & {\small{}{}{}{}{}{}{}{}{}{}{}{}{}{}{}{}{}-0.233}               & {\small{}0.327} \tabularnewline
{\small{}{}{}{}{}{}{}{}{}{}{}{}{}{}{}{}{}Retailing}  & {\small{}{}{}{}{}{}{}{}{}{}{}{}{}{}{}{}{}0.023}  & {\small{}1.194}   & {\small{}{}{}{}{}{}{}{}{}{}{}{}{}{}{}{}{}0.021}            & {\small{}0.725} \tabularnewline
{\small{}{}{}{}{}{}{}{}{}{}{}{}{}{}{}{}{}Software}   & {\small{}{}{}{}{}{}{}{}{}{}{}{}{}{}{}{}{}0.107}   & {\small{}1.652}  & {\small{}{}{}{}{}{}{}{}{}{}{}{}{}{}{}{}{}0.119}           & {\small{}1.194}  \tabularnewline
{\small{}{}{}{}{}{}{}{}{}{}{}{}{}{}{}{}{}Technology}     & {\small{}{}{}{}{}{}{}{}{}{}{}{}{}{}{}{}{}-0.256}  & {\small{}1.679}  & {\small{}{}{}{}{}{}{}{}{}{}{}{}{}{}{}{}{}-0.395}       & {\small{}1.138}   \tabularnewline
{\small{}{}{}{}{}{}{}{}{}{}{}{}{}{}{}{}{}Transportation}   & {\small{}{}{}{}{}{}{}{}{}{}{}{}{}{}{}{}{}0.086}   & {\small{}1.386}  & {\small{}{}{}{}{}{}{}{}{}{}{}{}{}{}{}{}{}0.096}   & {\small{}0.587} \tabularnewline
\bottomrule[1.5pt] 

\end{tabular}}
\end{center}
\noindent {\small{}{}{}{}{}{}{}{}{}{}{}{}{}{}{}{}{}}%
\noindent\parbox[c][0.7\totalheight][s]{1\textwidth}{%
\noindent {\footnotesize\begin{singlespace}Entries report the average skewness and kurtosis of assets selected by the SS-SSD and the MAXSER portfolios in the major Industries of the S$\&$P 500 Index.\end{singlespace}} %
}
\end{table}
\endgroup

\subsection{Rolling-Window Analysis}

\subsubsection{Diversification Loss}

We conduct out-of-sample backtesting experiments and we evaluate the optimal SS-SSD portfolios achieving a zero diversification loss in a rolling-window scheme. 
We use a window width of 240 monthly return observations. 
A stock is excluded from the asset pool if it has missing data in the 240-month training period; the number of stocks varies over time and can be smaller than the total number of constituents of the S$\&$P 500. 
Each month the portfolios are constructed using the monthly returns during the prior 240 months. The clock is advanced and the realized returns of the optimal portfolios are determined from the actual returns of the various assets. The same procedure is then repeated for the next time period and the ex post realized returns over the period from 01/2001 to 12/2020 (240 months) are computed.
The out-of-sample test is a real-time exercise avoiding a potential look-ahead bias and 
mimicking the way that a real-time investor acts in practice. 
\begin{singlespace}
\begin{figure}[H]
\caption{\textcolor{black}{\footnotesize{}{}{}{}The upper panel plots the number of stocks of the optimal SS-SSD portfolios through time that eliminate the diversification loss, as well as the number of stocks of the efficient MV portfolios. The lower panel plots the estimated expected loss of the optimal MAXSER portfolios corresponding to the inefficient SS-SSD portfolios with the same number of stocks as MAXSER. The grey areas are the NBER recession periods}}
\begin{minipage}[c]{1\linewidth}
\centering{}\includegraphics[scale=1,width=1\textwidth]{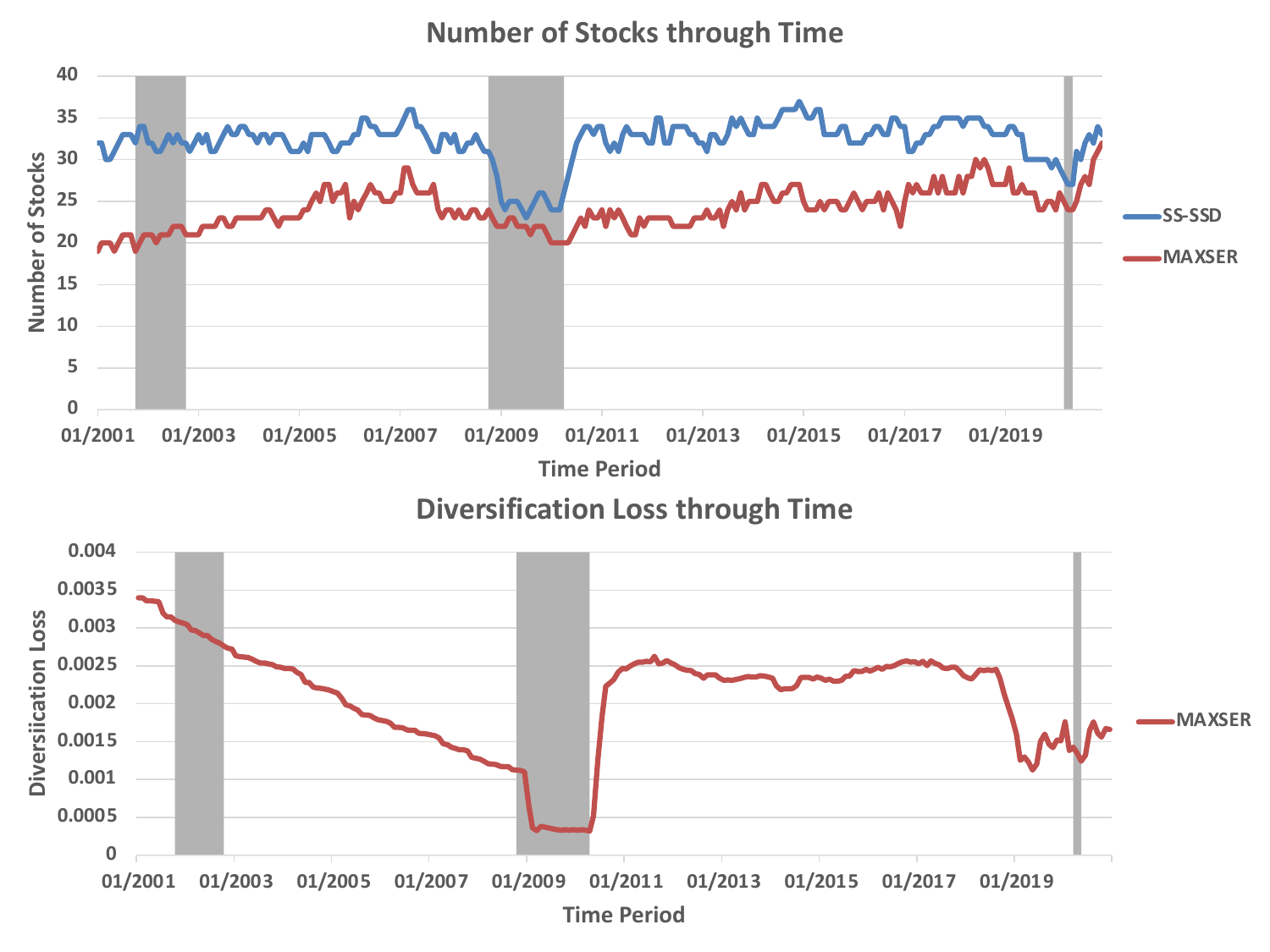}
\end{minipage}

\label{Fig2} 
\end{figure}
\end{singlespace}

We again compare the performance of the optimal SS-SSD portfolios with that of the MAXSER portfolios of Ao, Li, and Zheng (2019). The upper panel of Figure \ref{Fig2} plots the number of stocks of the optimal SS-SSD portfolios through time that eliminate the diversification loss, as well as the number of stocks of the efficient MAXSER portfolio. The lower panel plots the estimated expected loss of the optimal MAXSER corresponding to the inefficient SSD portfolios with the same number of stocks as MAXSER. The  diversification loss is zero for the efficient SS-SSD portfolios corresponding to the upper panel by construction. On a rolling-window basis, the number of assets in the SS-SSD portfolios is always higher than in the MAXSER portfolios. It shrinks to around 25 assets in the crisis periods of 2008-2009 and at the beginning of the Covid-19 period. Otherwise the number of assets in the SS-SSD portfolios is stable between 30 and 35.
The number of assets in the MAXSER portfolios is more volatile.

\begin{figure}[H]
 \caption{\textcolor{black}{\footnotesize{}{}{}{} Cumulative performance of the MAXSER, the SS-SSD, the $1/N$ and the S$\&$P 500 portfolios  for the out-of-sample period from January 2001 to December 2020. The grey areas are the NBER recession periods.}}
\centering{}
\includegraphics[scale=0.8,width=0.8\textwidth]{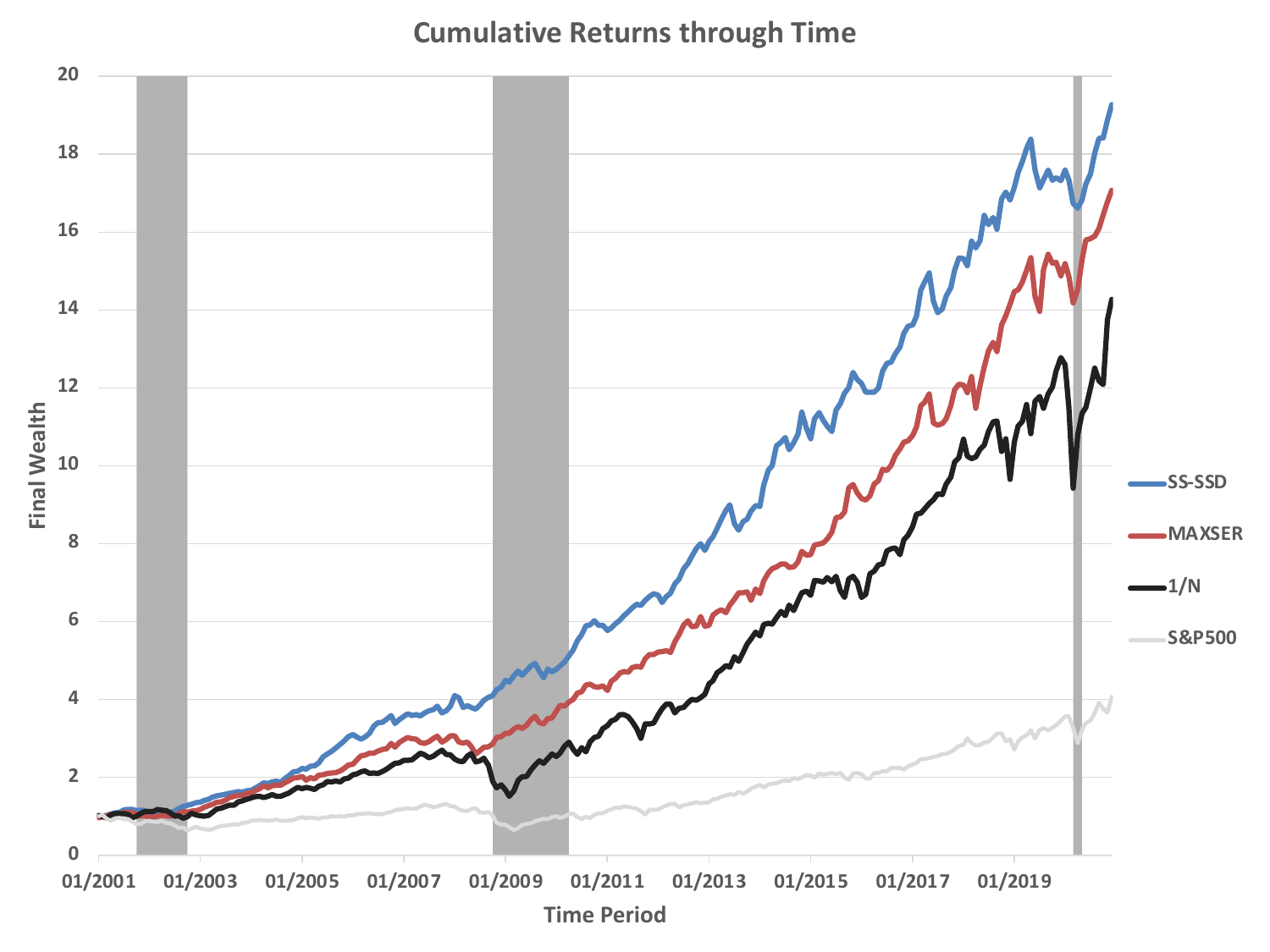}
\label{Fig3} 
\end{figure}

Figure \ref{Fig3} illustrates the out-of-sample cumulative returns of the SS-SSD, the MAXSER, the $1/N$ and the S$\&$P 500 portfolios during the period (January 2001 to December 2020). The grey areas are the NBER recession periods. We observe that the SS-SSD optimal portfolio has a 19.3 times higher value at the end of the holding period compared to the beginning, while the MAXSER portfolio has a 17.1 times higher value.  The $1/N$ portfolio follows with 14.3 higher value than at the beginning of the period. Finally,  the S$\&$P 500 portfolio exhibits the worst performance, with 4.2 higher value than the initial.

\begin{singlespace}
\end{singlespace}

Next, we compare the performance of the SS-SSD optimal portfolio with the performance of the MAXSER optimal portfolio using both nonparametric tests as well as parametric performance measures. 

\subsubsection{Nonparametric Stochastic Dominance Performance Test}

We use the pairwise (non-)dominance test of Anyfantaki et al.\ (2022), for a risk-adjusted comparison of the out-of-sample performance of the SS-SSD and MAXSER portfolios. 

The definition for second order stochastic non-dominance is the following:
\begin{defn}
\textbf{(Stochastic non-dominance)}: \emph{The SS-SSD portfolio $\lambda$} \emph{does not strictly second order stochastically dominate the MAXSER
portfolio} $\kappa$, say $\lambda\nsucc_{F}\kappa$, iff 
$
\exists z\in\mathbb{\mathcal{Z}}:\,D\left(z,\lambda,\kappa,F\right)>0, \text{ or } \forall z\in\mathbb{\mathcal{Z}}:\,D\left(z,\lambda,\kappa,F\right)=0.
$
\end{defn}
Strict second order stochastic non-dominance holds iff $\kappa$
achieves a higher expected utility for some non-decreasing and concave
utility function or achieves the same expected utility for every non-decreasing
and concave utility function. Equivalently,
strict stochastic non-dominance holds iff $\kappa$ is strictly preferred
to $\lambda$ by some risk averter, or every risk averter is indifferent
between them. 

We test the null hypothesis ${H'_0}$ vis-\`{a}-vis the alternative hypothesis :

\begin{description}
\item[$\mathrm{\textbf{H}}'_{0}$:] SS-SSD portfolio  $\lambda$  does not strictly second order stochastically dominate 
MAXSER portfolio $\kappa$.
\item[$\mathrm{\textbf{H}}'_{1}$:] SS-SSD Portfolio
$\lambda$ stochastically dominates MAXSER  portfolio $\kappa$.

\end{description}

For the pairwise test of the two portfolios, the test statistic is  $
\xi_T = \sup_{z\in\mathcal{Z}}D\left(z,\kappa,\lambda,F\right)
$.
To calculate the $p$-value, we use  block-boostrapping. The $p$-value
is approximated by $\tilde{p}_{j}=\frac{1}{R}\sum_{r=1}^{R}\{\xi_{T,r}^{\star}>\xi_{T}\},$
where $\xi_{T}$ is the test statistic, $\xi_{T,r}^{\star}$ is the bootstrap test statistic,
averaging over $R=1000$ replications. We reject the null hypothesis of non-dominance if the $p$-value is lower than $5\%$.
The test statistic $\xi_{T}$ is -0.0012, and the $p$-value is estimated at $4.4\%$. We thus reject the null hypothesis of non-dominance of portfolio SS-SSD over MAXSER.

\subsubsection{Performance Summary of the Optimal Portfolios}

We also compute  a number of parametric performance measures to compare the out-of-sample performance of the optimal portfolios. 
Apart from the performance measures we used in the in-sample analysis, we additionally compute the
Portfolio Turnover (PT), which measures the degree of rebalancing
required to implement each one of the two strategies. For any portfolio
strategy $P$, the portfolio turnover is defined as the average of
the absolute change of weights over the $T$ rebalancing points in
time and across the $N$ available assets:
$
\mathrm{PT}=\frac{1}{T}\sum_{t=1}^{T}\sum_{i=1}^{N}(|w_{P,i,t+1}-w_{P,i,t}|),
$
where $w_{P,i,t+1}$ and $w_{P,i,t}$ are the optimal weights of asset
$i$ under strategy $P$ (SS-SSD or MAXSER) at time $t$ and
$t+1$, respectively.

The performance of the portfolios is also assessed under the
risk-adjusted (net of transaction costs) returns measure of DeMiguel
et al.\ (2009) which is an indicator of how the proportional transaction
cost generated by the portfolio turnover affects the portfolio returns.
We use a transaction cost of 35 bps,  which is typical
in the literature. For this, the change in the net of transaction cost wealth $\mathrm{NW}_{P}$
of portfolio $P$ through time is defined as $
\mathrm{NW}_{P,t+1}=\mathrm{NW}_{P,t}(1+R_{P,t+1})[1-\mathrm{trc}\times\sum_{i=1}^{N}(|w_{P,i,t+1}-w_{P,i,t}|),
$ where $\mathrm{trc}$ is the proportional transaction cost and $R_{P,t+1}$
is the realized return of portfolio $P$ at time $t+1$.
Then, the portfolio return net of transaction costs is defined as
$
\mathrm{RTC}_{P,t+1}=\frac{\mathrm{NW}_{P,t+1}}{\mathrm{NW}_{P,t}}-1.
$

The return-loss measures the additional return needed so
that the MAXSER optimal portfolio performs equally well with the SS-SSD
portfolio is defined as $
\mathrm{R}_{Loss}=\frac{\mu_{\mathrm{SSD}}}{\sigma_{\mathrm{SSD}}}\times\sigma_{\mathrm{MAXSER}}-\mu_{\mathrm{MAXSER}},
$
where $\mu_{\mathrm{MAXSER}}$ and $\mu_{\mathrm{SSD}}$ are the out-of-sample mean of monthly
$\mathrm{RTC}$ for the MAXSER and the SS-SSD opportunity set,
and $\sigma_{\mathrm{MAXSER}}$ and $\sigma_{\mathrm{SSD}}$ are the corresponding standard
deviations.

\begingroup
\setlength{\tabcolsep}{6pt}
\renewcommand{\arraystretch}{0.5}
\begin{table}[H]
\caption{Out-of-sample performance: risk and performance measures }
\label{t-1}{\small{}{}{}{}{}{}{}{}{}{}{}{}{}{}{}{}{}
}%

\begin{center}
\scalebox{0.8}{
\begin{tabular}{lccc}
\toprule[1.5pt]
 & {\small{}{}{}{}{}{}{}{}{}{}{}{}{}{}{}{}{}SS-SSD}  & {\small{}{}{}{}{}{}{}{}{}{}{}{}{}{}{}{}{}MAXSER}& {\small{}{}{}{}{}{}{}{}{}{}{}{}{}{}{}{}{}$1/N$} \tabularnewline
\midrule 
{\small{}{}{}{}{}{}{}{}{}{}{}{}{}{}{}{}{}
Measures}  &  &  & \tabularnewline
\midrule
{\small{}{}{}{}{}{}{}{}{}{}{}{}{}{}{}{}{}Average}  & {\small{}{}{}{}{}{}{}{}{}{}{}{}{}{}{}{}{}0.0127 } & {\small{}{}{}{}{}{}{}{}{}{}{}{}{}{}{}{}{}0.0122}   & {\small{}{}{}{}{}{}{}{}{}{}{}{}{}{}{}{}{}0.0121 }\tabularnewline
{\small{}{}{}{}{}{}{}{}{}{}{}{}{}{}{}{}{}Standard Deviation} & {\small{}{}{}{}{}{}{}{}{}{}{}{}{}{}{}{}{}0.0239}  & {\small{}{}{}{}{}{}{}{}{}{}{}{}{}{}{}{}{}0.0258}   & {\small{}{}{}{}{}{}{}{}{}{}{}{}{}{}{}{}{}0.0450 }  \tabularnewline
{\small{}{}{}{}{}{}{}{}{}{}{}{}{}{}{}{}{}Sharpe Ratio}  & {\small{}{}{}{}{}{}{}{}{}{}{}{}{}{}{}{}{}0.4571} & {\small{}{}{}{}{}{}{}{}{}{}{}{}{}{}{}{}{}0.4056}    & {\small{}{}{}{}{}{}{}{}{}{}{}{}{}{}{}{}{}0.2313 } \tabularnewline
{\small{}{}{}{}{}{}{}{}{}{}{}{}{}{}{}{}{}Downside Sharpe Ratio} & {\small{}{}{}{}{}{}{}{}{}{}{}{}{}{}{}{}{}1.1188}  & {\small{}{}{}{}{}{}{}{}{}{}{}{}{}{}{}{}{}0.8614}   & {\small{}{}{}{}{}{}{}{}{}{}{}{}{}{}{}{}{}0.9311 }  \tabularnewline
{\small{}{}{}{}{}{}{}{}{}{}{}{}{}{}{}{}{}Value-at-Risk}  & {\small{}{}{}{}{}{}{}{}{}{}{}{}{}{}{}{}{}0.0295} & {\small{}{}{}{}{}{}{}{}{}{}{}{}{}{}{}{}{}0.0403}    & {\small{}{}{}{}{}{}{}{}{}{}{}{}{}{}{}{}{}0.0744 } \tabularnewline
{\small{}{}{}{}{}{}{}{}{}{}{}{}{}{}{}{}{}Expected Shortfall}  & {\small{}{}{}{}{}{}{}{}{}{}{}{}{}{}{}{}{}0.0476} & {\small{}{}{}{}{}{}{}{}{}{}{}{}{}{}{}{}{}0.0532}  & {\small{}{}{}{}{}{}{}{}{}{}{}{}{}{}{}{}{}0.1004 } \tabularnewline
{\small{}{}{}{}{}{}{}{}{}{}{}{}{}{}{}{}{}UP ratio}  & {\small{}{}{}{}{}{}{}{}{}{}{}{}{}{}{}{}{}1.2014} & {\small{}{}{}{}{}{}{}{}{}{}{}{}{}{}{}{}{}1.0864}    & {\small{}{}{}{}{}{}{}{}{}{}{}{}{}{}{}{}{}0.7704 } \tabularnewline
{\small{}{}{}{}{}{}{}{}{}{}{}{}{}{}{}{}{}Portfolio Turnover}  & {\small{}{}{}{}{}{}{}{}{}{}{}{}{}{}{}{}{}8.835\%}  & {\small{}{}{}{}{}{}{}{}{}{}{}{}{}{}{}{}{}8.477\%}  & {\small{}{}{}{}{}{}{}{}{}{}{}{}{}{}{}{}{}0.0}  \tabularnewline
{\small{}{}{}{}{}{}{}{}{}{}{}{}{}{}{}{}{}Return Loss}  & & {\small{}{}{}{}{}{}{}{}{}{}{}{}{}{}{}{}{}0.087\%}  &    {\small{}{}{}{}{}{}{}{}{}{}{}{}{}{}{}{}{}0.156\% } \tabularnewline
{\small{}{}{}{}{}{}{}{}{}{}{}{}{}{}{}{}{}Certainty Equivalent}  &  &  & \tabularnewline
\emph{\small{}{}{}{}{}{}{}{}{}{}{}{}{}{}{}{}{}Exponential Utility}{\small{}{}{}{}{}{}{}{}{}{}{}{}{}{}{}{}{}}  &  &    & \tabularnewline
{\small{}{}{}{}{}{}{}{}{}{}{}{}{}{}{}{}{}ARA=2}  & {\small{}{}{}{}{}{}{}{}{}{}{}{}{}{}{}{}{}1.211\%}  & {\small{}{}{}{}{}{}{}{}{}{}{}{}{}{}{}{}{}1.155\%}  &    {\small{}{}{}{}{}{}{}{}{}{}{}{}{}{}{}{}{}1.010\%}   \tabularnewline
{\small{}{}{}{}{}{}{}{}{}{}{}{}{}{}{}{}{}ARA=4}  & {\small{}{}{}{}{}{}{}{}{}{}{}{}{}{}{}{}{}1.152\%}  & {\small{}{}{}{}{}{}{}{}{}{}{}{}{}{}{}{}{}1.086\%}  &   {\small{}{}{}{}{}{}{}{}{}{}{}{}{}{}{}{}{}0.794\%}   \tabularnewline
{\small{}{}{}{}{}{}{}{}{}{}{}{}{}{}{}{}{}ARA=6}  & {\small{}{}{}{}{}{}{}{}{}{}{}{}{}{}{}{}{}1.091\%}  & {\small{}{}{}{}{}{}{}{}{}{}{}{}{}{}{}{}{}1.016\%}  &   {\small{}{}{}{}{}{}{}{}{}{}{}{}{}{}{}{}{}0.567\%}   \tabularnewline
\emph{\small{}{}{}{}{}{}{}{}{}{}{}{}{}{}{}{}{}Power Utility}{\small{}{}{}{}{}{}{}{}{}{}{}{}{}{}{}{}{}}  &  &   &  \tabularnewline
{\small{}{}{}{}{}{}{}{}{}{}{}{}{}{}{}{}{}RRA=2}  & {\small{}{}{}{}{}{}{}{}{}{}{}{}{}{}{}{}{}1.211\%}  &  {\small{}{}{}{}{}{}{}{}{}{}{}{}{}{}{}{}{}1.156\%}  &    {\small{}{}{}{}{}{}{}{}{}{}{}{}{}{}{}{}{}1.009\%}   \tabularnewline
{\small{}{}{}{}{}{}{}{}{}{}{}{}{}{}{}{}{}RRA=4}  & {\small{}{}{}{}{}{}{}{}{}{}{}{}{}{}{}{}{}3.725\%}  & {\small{}{}{}{}{}{}{}{}{}{}{}{}{}{}{}{}{}3.549\%}  &    {\small{}{}{}{}{}{}{}{}{}{}{}{}{}{}{}{}{}3.089\%}   \tabularnewline
{\small{}{}{}{}{}{}{}{}{}{}{}{}{}{}{}{}{}RRA=6}  & {\small{}{}{}{}{}{}{}{}{}{}{}{}{}{}{}{}{}6.366\%}  & {\small{}{}{}{}{}{}{}{}{}{}{}{}{}{}{}{}{}6.058\%} &     {\small{}{}{}{}{}{}{}{}{}{}{}{}{}{}{}{}{}5.257\%}   \tabularnewline
{\small{}{}{}{}{}{}{}{}{}{}{}{}{}{}{}{}{}Opportunity Cost}  &  &  & \tabularnewline
\emph{\small{}{}{}{}{}{}{}{}{}{}{}{}{}{}{}{}{}Exponential
Utility}{\small{}{}{}{}{}{}{}{}{}{}{}{}{}{}{}{}{}}  &  &    & \tabularnewline
{\small{}{}{}{}{}{}{}{}{}{}{}{}{}{}{}{}{}ARA=2}  & & {\small{}{}{}{}{}{}{}{}{}{}{}{}{}{}{}{}{}0.073\%}  &    {\small{}{}{}{}{}{}{}{}{}{}{}{}{}{}{}{}{}0.126\%}   \tabularnewline
{\small{}{}{}{}{}{}{}{}{}{}{}{}{}{}{}{}{}ARA=4}  & & {\small{}{}{}{}{}{}{}{}{}{}{}{}{}{}{}{}{}0.081\%}  &   {\small{}{}{}{}{}{}{}{}{}{}{}{}{}{}{}{}{}0.139\%}   \tabularnewline
{\small{}{}{}{}{}{}{}{}{}{}{}{}{}{}{}{}{}ARA=6}  & & {\small{}{}{}{}{}{}{}{}{}{}{}{}{}{}{}{}{}0.092\%}  &   {\small{}{}{}{}{}{}{}{}{}{}{}{}{}{}{}{}{}0.152\%}   \tabularnewline
\emph{\small{}{}{}{}{}{}{}{}{}{}{}{}{}{}{}{}{}Power
Utility}{\small{}{}{}{}{}{}{}{}{}{}{}{}{}{}{}{}{}}  &  &   &  \tabularnewline
{\small{}{}{}{}{}{}{}{}{}{}{}{}{}{}{}{}{}RRA=2}  & &  {\small{}{}{}{}{}{}{}{}{}{}{}{}{}{}{}{}{}0.070\%}  &    {\small{}{}{}{}{}{}{}{}{}{}{}{}{}{}{}{}{}0.132\%}   \tabularnewline
{\small{}{}{}{}{}{}{}{}{}{}{}{}{}{}{}{}{}RRA=4}  & & {\small{}{}{}{}{}{}{}{}{}{}{}{}{}{}{}{}{}0.079\%}  &    {\small{}{}{}{}{}{}{}{}{}{}{}{}{}{}{}{}{}0.144\%}   \tabularnewline
{\small{}{}{}{}{}{}{}{}{}{}{}{}{}{}{}{}{}RRA=6}  & & {\small{}{}{}{}{}{}{}{}{}{}{}{}{}{}{}{}{}0.091\%} &     {\small{}{}{}{}{}{}{}{}{}{}{}{}{}{}{}{}{}0.159\%}   \tabularnewline
\bottomrule[1.5pt]
\end{tabular}}

\noindent {\small{}{}{}{}{}{}{}{}{}{}{}{}{}{}{}{}{}}%
\end{center}
\noindent {\small{}{}{}{}{}{}{}{}{}{}{}{}{}{}{}{}{}}%
\noindent\parbox[c][0.7\totalheight][s]{1\textwidth}{%
\noindent {\footnotesize\begin{singlespace}Entries report the risk and performance
measures (Sharpe Ratio, Downside Sharpe Ratio, VaR, ES, UP Ratio, Portfolio
Turnover, Returns Loss, Opportunity Cost and Certainty Equivalent) for the SS-SSD, the MAXSER and for the $1/N$ portfolios.
The realized monhtly returns cover the period from January, 2001
to December, 2020.\end{singlespace}} %
}
\end{table}
\endgroup
Table \ref{t-1} reports the parametric performance measures (monthly) for
the MAXSER, the SS-SSD optimal portfolios and the $1/N$ portfolio for the sample period. 
The higher the value of each
one of these measures, the greater the investment opportunities for
the relative portfolio.

We observe that the Average, the Sharpe Ratios and the Downside Sharpe Ratios of the SS-SSD optimal portfolios are higher than those of the MAXSER optimal portfolios. It reflects an
increase in the risk-adjusted performance (i.e., an increase in the
expected return per unit of risk) and hence expands the investment
opportunities for risk-averse investors. The same is true for
the UP Ratio. The Value-at-Risk and the Expected Shortfall (with a positive sign for a loss) of the SS-SSD portfolios are lower, indicating lower downside losses. Furthermore, the MAXSER portfolios
induce slightly less portfolio turnover than the SS-SSD portfolios. 
The SS-SSD strategy may have more frequent rebalancing and incur higher transaction costs, but the additional performance justifies the additional cost; see Carroll et al.\ (2017).
The return-loss measure that takes into account transaction costs, is positive. 
The CEQ of the SS-SSD optimal portfolios is the highest in all cases.
Finally, the Opportunity Cost is always positive, indicating that a positive return should be added in the MAXSER or in the $1/N$ portfolio to achieve the same expected return with the SS-SSD portfolio.
The $1/N$ portfolio exhibits again the worst performance, dominated by both the SS-SSD as well as the MAXSER portfolios.

\begin{figure}[H]
\caption{\textcolor{black}{\footnotesize{}{}{}{} Average Industry weights through time. The upper panel plots the average Industry weights of the optimal SS-SSD portfolios, while the lower panel plots the average Industry weights of the optimal MAXSER portfolios, for the out-of-sample period from January 2001 to December 2020. The grey areas are the NBER recession periods.}}
\begin{minipage}[c]{1\linewidth}
\centering{}\includegraphics[scale=1,width=1\textwidth]{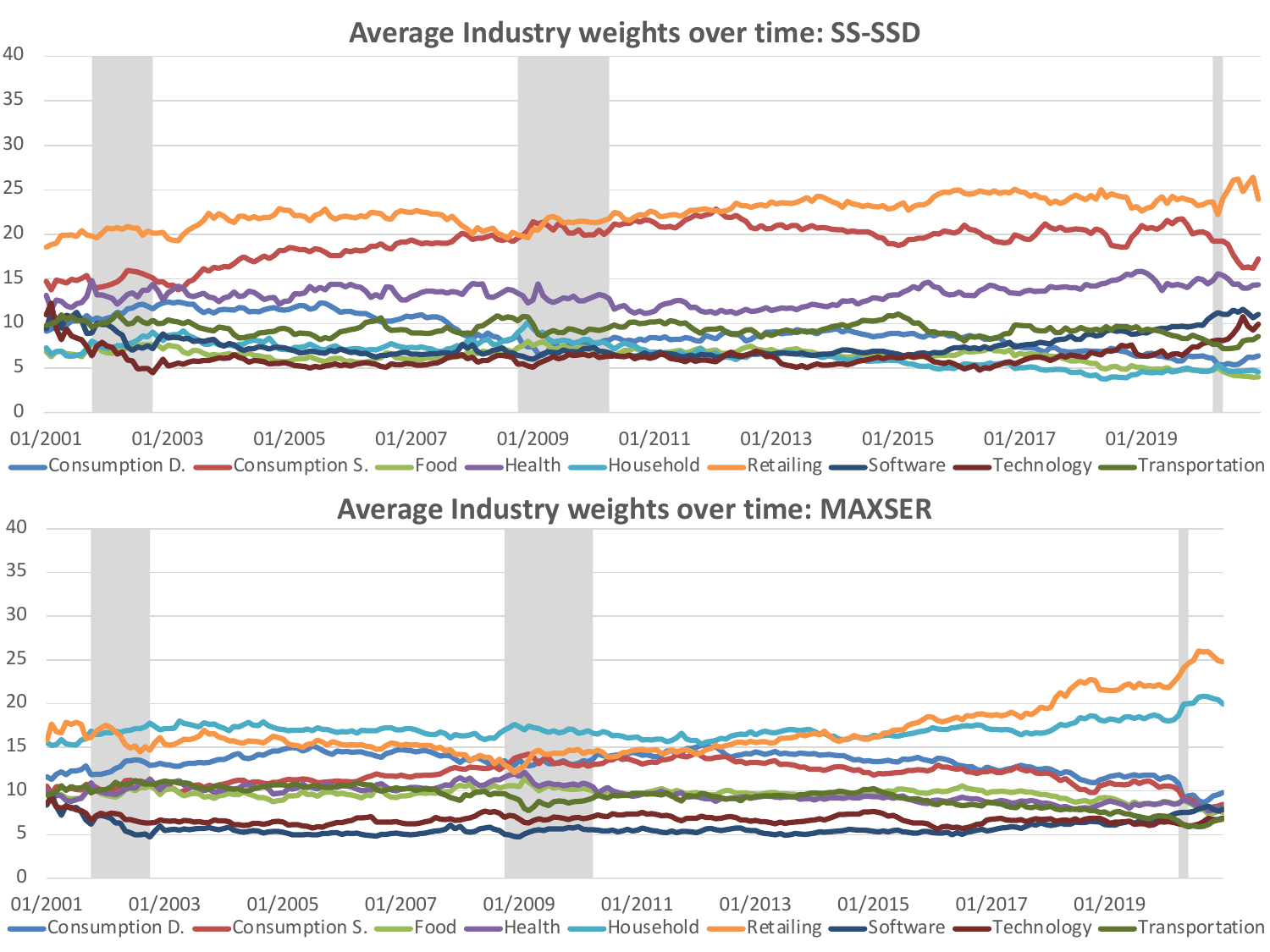}
\end{minipage}
\label{FigWeights} 
\end{figure}

Let us now analyze the composition of the SS-SSD and the MAXSER portfolios through time. 
Figure \ref{FigWeights}  reports the optimal average weights of the major Industries selected by each one of the two portfolios during the out-of-sample period. We observe that both portfolios are well diversified and invest in almost the same Industries,  with different overall weights. 
%The optimal SS-SSD portfolio invests mainly in 9 industry sectors with a larger weighting on small size,
%high book-to-market, and momentum stocks from the S\&P 500 index.

\begin{figure}[H]
\caption{\textcolor{black}{\footnotesize{}{}{}{}The upper panel plots the range (min, max, and quartiles) of the Jensen Alpha coefficients of the individual stocks of the optimal SS-SSD portfolios through time. The lower panel  plots the range (min, max, and quartiles)  of the Beta coefficients of the individual stocks of the optimal SS-SSD portfolios through time. The grey areas are the NBER recession periods. The CAPM is used for the estimation of the Jensen Alpha and Beta coefficients, using the previous 5 years of individual monthly returns (60 monthly return observations). }}
\begin{minipage}[c]{1\linewidth}
\centering{}\includegraphics[scale=1,width=1\textwidth]{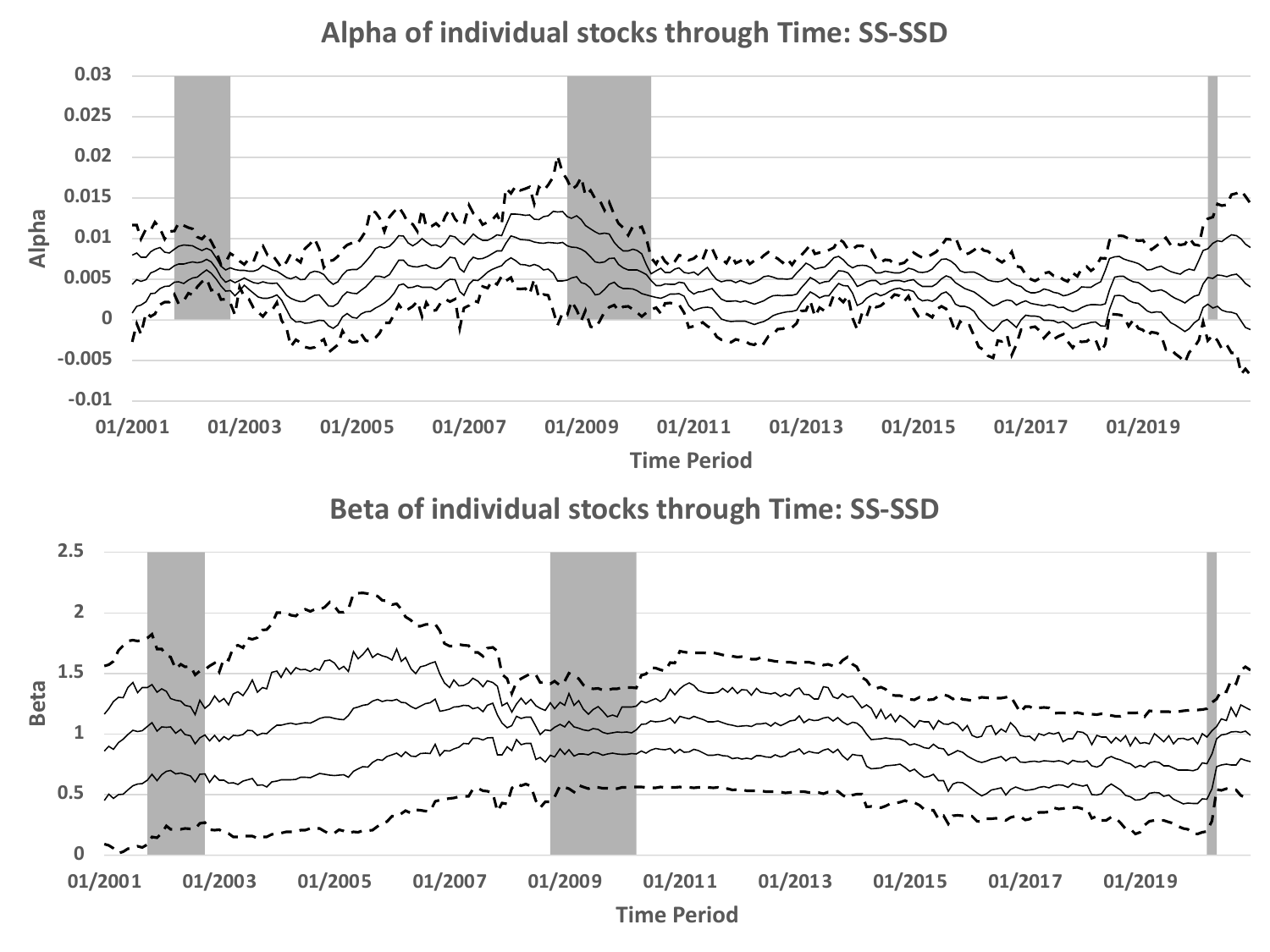}
\end{minipage}
\label{ABSSD} 
\end{figure}

\begin{figure}[H]
\caption{\textcolor{black}{\footnotesize{}{}{}{}The upper panel plots the range (min, max, and quartiles) of the Jensen Alpha coefficients of the individual stocks of the optimal MAXSER portfolios, and the lower panel  plots the range (min, max, and quartiles) of the Beta coefficients of the individual stocks of the optimal MAXSER portfolios, for the out-of-sample period from January 2001 to December 2020. The grey areas are the NBER recession periods. The CAPM is used for the estimation of the Jensen Alpha and Beta coefficients, using the previous 5 years of individual monthly returns (60 monthly return observations).}}
\begin{minipage}[c]{1\linewidth}
\centering{}
\centering{}\includegraphics[scale=1,width=1\textwidth]{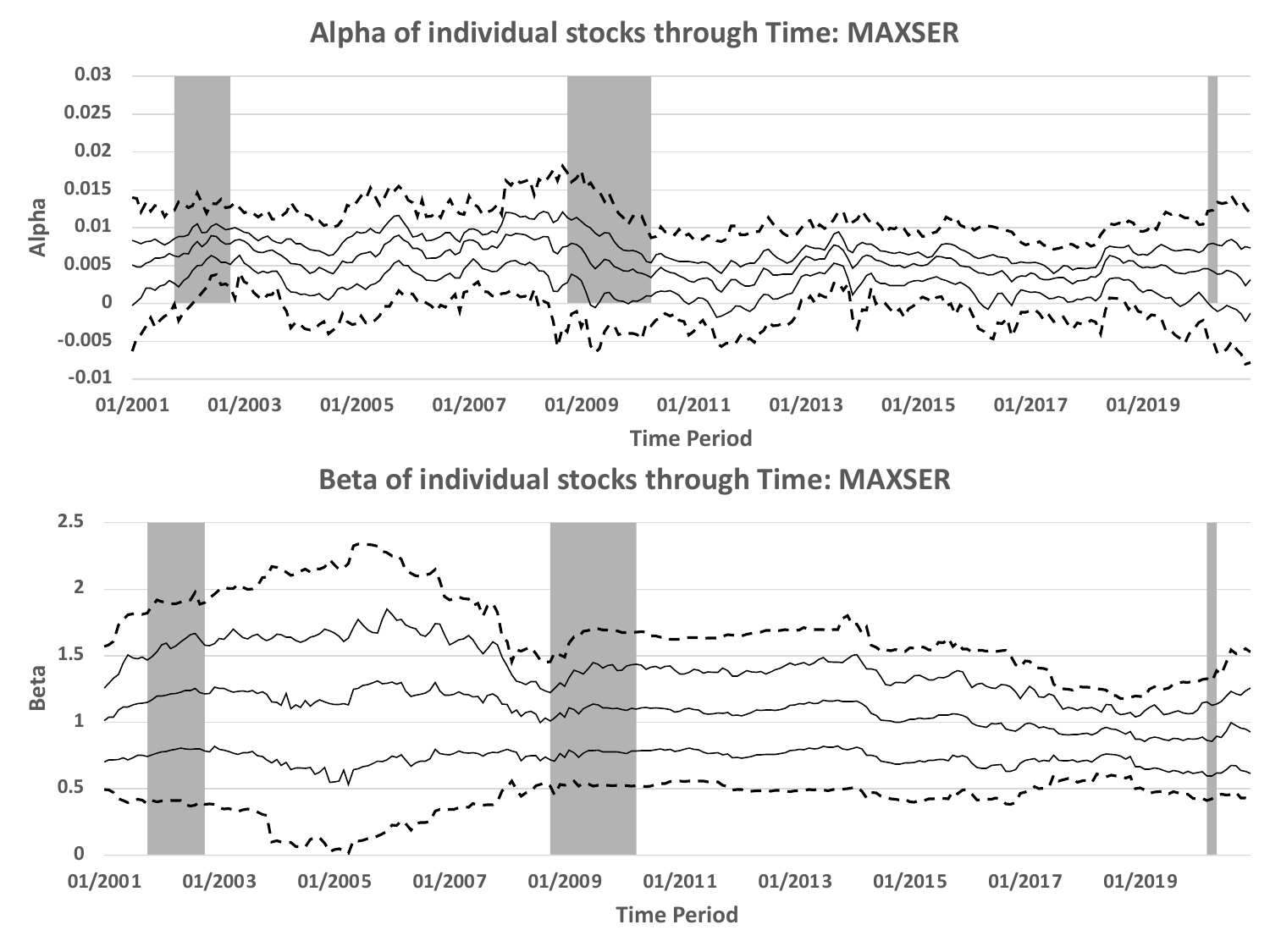}
\end{minipage}
\label{ABMAXSER} 
\end{figure}

We further analyze the characteristics of the SS-SSD and the MAXSER portfolios through time, by estimating the Jensen Alpha and Beta coefficients of the individual stocks of these portfolios, each month. Figures \ref{ABSSD} and \ref{ABMAXSER} exhibit the range (max, min, and quartiles) of the Jensen Alpha and Beta coefficients, estimated from the CAPM, of the individual stocks of these portfolios during the out-of-sample period. For the estimation of the Alpha and Beta coefficients, the previous 5 years of individual monthly returns have been used (60 monthly returns). We can observe that the Beta coefficients in the SS-SSD portfolios have a more defensive profile. The heterogeneity of Alpha and Beta coefficients is lower in the SS-SSD portfolios.

\pagebreak
\vspace*{2cm}

\begin{singlespace}
\end{singlespace}

\vspace*{2cm}

\begin{singlespace}
    
\end{singlespace}

\begin{singlespace}

\end{singlespace}

Finally, we investigate which factors explain the returns of the active
investors with SSD preferences. To do so, we start with the classical single factor model (CAPM), and we additionally use  five asset pricing models 
that are popular in the literature. 
First, we use, the Fama-French 6-factor model (2016), which is the Fama and French 3-factor model augmented by profitability (RMW - robust minus weak), investment (CMA - conservative minus aggressive), and momentum (UMD - up minus down).
Second, the q-6-factor model of Hou, Xue and Zhang (2015), including the original market and size factors of Fama-French model, augmented by a profitability (ROE - return on equity) and investment factor (I/A - investment to assets). Third,  the M4 4-factor model of Stambaugh and Yuan (2017) including the standard market and size factors along with two composite factors for profitability (PERF - performance) and investment (MGMT - management). 
Fourth, the Barillas and Shanken 6-factor model (2018), who use a  Bayesian approach,  suggesting the model of six factors including market, I/A, ROE, SMB, the value factor HMLm from Asness and Frazzini (2013), and UMD.
 Finally,  the 3-factor model  of Daniel, Hirshleifer, and Sun (2020) introducing behavioral-related factors such as the market factor augmented by long- and short-term mispricing factors (FIN and PEAD, respectively). The last is included to give an economic insight on behavioral influence. A brief description of the factors is given in the Appendix.

We consider linear regression models of the following form:
$R_{P,t}-R_{f,t}=a +\sum_{i}b_{i}R_{i,t}+e_{t}, $
where $R_{P,t} -R_{f,t}$ is the excess return of either the  SS-SSD or MAXSER optimal portfolio at period
$t$, $R_{i,t}$ is the return on the
$i$th factor and $e_t$ is the error term. If the exposures $b_{i}$
to the various factors capture all variation in expected
returns, the intercept $a$ is zero since the factors are tradable.
\vspace{9mm}
\begin{singlespace}

\begingroup
\setlength{\tabcolsep}{6pt}
\renewcommand{\arraystretch}{0.5}
\begin{table}[H]
\caption{Single factor model (CAPM)}
\begin{center}
\scalebox{0.8}{
\begin{tabular}{lrr}
\toprule[1.5pt] 
 & $a$  & $R_{M}-R_{F}$    \tabularnewline
\midrule 
SS-SSD  &   &  \tabularnewline
\midrule[1.5pt] 
Coef.  & 0.0119  & -0.0077     \tabularnewline
$t$-stat  & 7.186  & -0.201      \tabularnewline
$p$-value & 0.0  & 0.8411     \tabularnewline
\midrule
MAXSER  &   &  \tabularnewline
\midrule[1.5pt] 
Coef.  & 0.0105  & -0.0148     \tabularnewline
$t$-stat  & 5.772  & -0.348      \tabularnewline
$p$-value & 0.0  & 0.7275      \tabularnewline
\midrule 
$1/N$ &   &  \tabularnewline
\midrule[1.5pt]
Coef.  & 0.0067  & 0.9499     \tabularnewline
$t$-stat  & 9.543  & 58.086      \tabularnewline
$p$-value & 0.0  & 0.0    \tabularnewline
\bottomrule[1.5pt] 
\end{tabular}}
\end{center}
\noindent {\small{}{}{}{}{}{}{}{}{}{}{}{}{}{}{}{}{}}%
\noindent\parbox[c][0.7\totalheight][s]{1\textwidth}{%
\noindent {\footnotesize\begin{singlespace}Entries report the coefficients, their respective $t$-statistics and
 $p$-values for the SS-SSD portfolio (upper panel), the MAXSER portfolio (second panel), and for the $1/N$ portfolio (lower panel). The dataset spans 01/2001-12/2020 for optimal portfolios computed with 240-month  windows rolled over one month.\end{singlespace}}}
\label{t-2} 
\end{table}
\endgroup
\end{singlespace}

\begin{singlespace}
\begingroup
\setlength{\tabcolsep}{6pt}
\renewcommand{\arraystretch}{0.5}
\begin{table}[H]
\caption{Daniel, Hirshleifer, and Sun (2020), 3-factor model}
\begin{center}
\scalebox{0.8}{
\begin{tabular}{lrrrr}
\toprule[1.5pt] 
 & $a$  & $R_{M}-R_{F}$  & $\mathrm{PEAD}$  & $\mathrm{FIN}$  \tabularnewline
\midrule 
SS-SSD  &   & && \tabularnewline
\midrule[1.3pt] 
Coef.  & 0.0120  & 0.0293  & -0.0643  & 0.0776    \tabularnewline
$t$-stat  & 6.7823  & 0.6026  & -0.7328  & 1.4454    \tabularnewline
$p$-value & 0.0  & 0.5474  & 0.4654  & 0.1498     \tabularnewline
\midrule  
MAXSER  &   & && \tabularnewline
\midrule[1.3pt] 
Coef.  & 0.0097  & 0.0740  & 0.0359 & 0.1362    \tabularnewline
$t$-stat  & 5.0996  & 1.4214  & 0.3817  & 2.3676    \tabularnewline
$p$-value & 0.0  & 0.1567  & 0.7030  & 0.0188    \tabularnewline
\midrule  
$1/N$ &   & && \tabularnewline
\midrule[1.3pt]  
Coef.  & 0.0066  & 0.9722  & -0.0720  & 0.0745    \tabularnewline
$t$-stat  & 8.9460  & 48.1833  & -1.9779  & 3.3402    \tabularnewline
$p$-value & 0.0  & 0.0  & 0.0492  & 0.0010     \tabularnewline
\bottomrule[1.5pt] 
\end{tabular}}
\end{center}
\noindent {\small{}{}{}{}{}{}{}{}{}{}{}{}{}{}{}{}{}}%
\noindent\parbox[c][0.7\totalheight][s]{1\textwidth}{%
\noindent {\footnotesize\begin{singlespace}Entries report the coefficients, their respective $t$-statistics and
 $p$-values for the SS-SSD portfolio (upper panel), the MAXSER portfolio (second panel), and for the $1/N$ portfolio (lower panel). The dataset spans 01/2001-12/2018 for optimal portfolios computed with 240-month  windows rolled over one month.\end{singlespace}}}
\label{t-2} 
\end{table}
\end{singlespace}
\begin{singlespace}
\begin{table}[H]
\caption{Barillas and Shanken (2018),  6-factor model}
\begin{center}
\scalebox{0.8}{
\begin{tabular}{lrrrrrrr}
\toprule[1.5pt]
 & $a$  & $R_{M}-R_{F}$  & SMB  & R-IA  & R-ROE  & HMLm  & UMD \tabularnewline
\midrule 
SS-SSD  &   & &&&&& \tabularnewline
\midrule[1.3pt]
Coef.  & 0.0116 & 0.0065  & -0.0808  & -0.0634  & 0.0666  & 0.1202 & 0.0261 \tabularnewline
$t$-stat  & 6.7464  & 0.1350 & -1.0955  & -0.5364  & 0.6409  & 1.4021 & 0.4103 \tabularnewline
$p$-value & 0.0  & 0.8927  & 0.2745  & 0.5922  & 0.5222  & 0.1623 & 0.6820 \tabularnewline
\midrule
MAXSER  &   & &&&&& \tabularnewline
\midrule[1.3pt] 
Coef.  & 0.0104  & 0.0004  & -0.0695  & 0.0492 & 0.0123  & 0.0466 & 0.0148 \tabularnewline
$t$-stat  & 5.4458  & 0.0079 & -0.8498  & 0.3756  & 0.1071  & 0.4898 & 0.2104 \tabularnewline
$p$-value & 0.0  & 0.9937  & 0.3964  & 0.7075  & 0.9148  & 0.6248 & 0.8336 \tabularnewline
\midrule
$1/N$  &   & &&&&& \tabularnewline
\midrule[1.3pt]
Coef.  & 0.0057 & 0.9156  & 0.1945  & 0.0336  & 0.0943  & 0.1824& 0.0314 \tabularnewline
$t$-stat  & 10.353  & 59.506 & 8.228  & 0.887  & 2.831 & 6.642 & 1.542 \tabularnewline
$p$-value & 0.0  & 0.0  & 0.0  & 0.3759  & 0.0  & 0.0 & 0.1246 \tabularnewline
\bottomrule[1.5pt] \end{tabular}}
\end{center}
\noindent {\small{}{}{}{}{}{}{}{}{}{}{}{}{}{}{}{}{}}%
\noindent\parbox[c][0.7\totalheight][s]{1\textwidth}{%
\noindent {\footnotesize\begin{singlespace}Entries report the coefficients and their respective $t$-statistics and
 $p$-values for the SS-SSD portfolio (upper panel), the MAXSER portfolio (second panel), and for the $1/N$ portfolio (lower panel). The dataset spans 01/2001-12/2020 for optimal portfolios computed with 240-month  windows rolled over one month.\end{singlespace}}}
\label{t-3} 
\end{table}
\end{singlespace}

\begin{singlespace}
\begin{table}[H]
\caption{Fama-French (2016), 6-factor model}
\begin{center}
\scalebox{0.8}{
\begin{tabular}{lrrrrrrr}
\toprule[1.5pt] 
 & $a$  & $R_{M}-R_{F}$  & SMB  & HML  & RMW  & CMA  & Mom\tabularnewline
\midrule 
SS-SSD  &   & &&&&& \tabularnewline
\midrule[1.3pt]
Coef.  & 0.0121  & -0.0040  & -0.0708  & 0.0770  & 0.0153  & -0.0503 & -0.0139 \tabularnewline
$t$-stat  & 6.8617  & -0.0788  & -0.9930  & 0.9747  & 0.1512  & -0.4443 & -0.3544 \tabularnewline
$p$-value & 0.0  & 0.9372  & 0.3218  & 0.3308  & 0.8799  & 0.6572 & 0.7234 \tabularnewline
\midrule 
MAXSER  &   & &&&&& \tabularnewline
\midrule[1.3pt]
Coef.  & 0.0104  & 0.0044  & -0.0586  & -0.0472  & 0.0180  &0.1355 & -0.0154 \tabularnewline
$t$-stat  & 5.3465  & 0.0795  & -0.7455  & -0.5413  & 0.1605  & 1.0840 & -0.3541 \tabularnewline
$p$-value & 0.0  & 0.9367  & 0.4567  & 0.5888 & 0.8726  & 0.2796 & 0.7236 \tabularnewline
\midrule 
$1/N$  &   & &&&&& \tabularnewline
\midrule[1.3pt]
Coef.  & 0.0056  & 0.9274  & 0.2262  & 0.0427  & 0.1507  & 0.0877 & -0.0511 \tabularnewline
$t$-stat  & 10.009  & 58.145 & 9.976  & 1.697  & 4.671  & 2.433 & -4.083 \tabularnewline
$p$-value & 0.0  & 0.0  & 0.0 & 0.0910  & 0.0  & 0.0158 & 0.0 \tabularnewline
\bottomrule[1.5pt] 
\end{tabular}}
\end{center}
\noindent\parbox[c][0.7\totalheight][s]{1\textwidth}{%
\noindent{\footnotesize\begin{singlespace}Entries report the coefficients and their respective $t$-statistics and
 $p$-values for the SS-SSD portfolio (upper panel), the MAXSER portfolio (second panel), and for the $1/N$ portfolio (lower panel). The dataset spans 01/2001-12/2020 for optimal portfolios computed with 240-month  windows rolled over one month.\end{singlespace}}}

\label{t-4} 
\end{table}
\end{singlespace}

%\begin{center}
\begin{singlespace}
\begin{table}[H]
\caption{\footnotesize Stambaugh and Yuan(2017), M4  4-factor model}
\begin{center}
\scalebox{0.8}{
\begin{tabular}{lrrrrr}
\toprule[1.5pt]
 & $a$  & $R_{M}-R_{F}$  & SMB  & MGMT  & PERF  \tabularnewline
\midrule
SS-SSD  &   & &&& \tabularnewline
\midrule[1.3pt]
Coef.  & 0.0129  & 0.0026  & -0.0498  & 0.1094  & -0.0520  \tabularnewline
$t$-stat  & 6.7428  & 0.0464  & -0.6478  & 1.5526  & -1.1453  \tabularnewline
$p$-value & 0.0  & 0.9630  & 0.5179  & 0.1222  & 0.2535   \tabularnewline
\midrule 
MAXSER  &   & &&& \tabularnewline
\midrule[1.3pt]
Coef.  & 0.0105  & 0.0045  & -0.0131  & 0.1789 & -0.0710   \tabularnewline
$t$-stat  & 5.2433  & 0.0777  & -0.1626  & 2.4328  & -1.4994  \tabularnewline
$p$-value & 0.0  & 0.9381  & 0.8710  & 0.0159 & 0.1355   \tabularnewline
\midrule
$1/N$  &   & &&& \tabularnewline
\midrule[1.3pt]
Coef.  & 0.0063  & 0.9090  & 0.2510  & 0.0789  & -0.0200  \tabularnewline
$t$-stat  & 9.259  & 46.482  & 9.202  & 3.157  & -1.243  \tabularnewline
$p$-value & 0.0  & 0.0  & 0.0  & 0.0  & 0.2153   \tabularnewline
\bottomrule[1.5pt] 
\end{tabular}}
\end{center}
\noindent\parbox[c][0.7\totalheight][s]{1\textwidth}{%
\noindent{\footnotesize\begin{singlespace}Entries report the coefficients and their respective $t$-statistics and
 $p$-values for the SS-SSD portfolio (upper panel), the MAXSER portfolio  (second panel), and for the $1/N$ portfolio (lower panel). The dataset spans 01/2001-12/2016 for optimal portfolios computed with 240-month  windows rolled over one month.\end{singlespace}}}
\label{t-5} 
\end{table}
\end{singlespace}

%\begin{center}
\begin{singlespace}
\begin{table}[H]
\caption{Hou, Xue and Zhang (2015),  q-4-factor model}
\begin{center}
\scalebox{0.8}{
\begin{tabular}{lrrrrr}
\toprule[1.5pt]
 & $a$  & $R_{M}-R_{F}$  & ME  & IA  & ROE  \tabularnewline
\midrule
SS-SSD  &   & &&& \tabularnewline
\midrule[1.3pt]
Coef.  & 0.0122 & 0.0263  & -0.0452  & 0.0503 & 0.0331   \tabularnewline
$t$-stat  & 6.9438 & 0.5218  & -0.6424  & 0.5270  & 0.4361  \tabularnewline
$p$-value & 0.0  & 0.6024  & 0.5213  & 0.5987  & 0.6632   \tabularnewline
\midrule
MAXSER  &   & &&& \tabularnewline
\midrule[1.3pt]
Coef.  & 0.0106  & 0.0301  & -0.0521  & 0.1102 & 0.0181  \tabularnewline
$t$-stat  & 5.5931  & 0.5541  & -0.6879  & 1.0715  & 0.2213  \tabularnewline
$p$-value & 0.0  & 0.5801  & 0.4923  & 0.2852  & 0.8251   \tabularnewline
\midrule
$1/N$  &   & &&& \tabularnewline
\midrule[1.3pt]
Coef.  & 0.0063& 0.9016  & 0.2028 & 0.1513 & -0.0386   \tabularnewline
$t$-stat  & 10.398 & 51.890  & 8.382  & 4.602  & -1.475  \tabularnewline
$p$-value & 0.0  & 0.0  & 0.0  & 0.0  & 0.1416   \tabularnewline
\bottomrule[1.5pt]
\end{tabular}}
\end{center}
 \noindent\parbox[c][0.7\totalheight][s]{1\textwidth}{%
\noindent{\footnotesize\begin{singlespace}Entries report the coefficients and their respective $t$-statistics and
 $p$-values for the SS-SSD portfolio (upper panel), the MAXSER portfolio (second panel), and for the $1/N$ portfolio (lower panel). The dataset spans 01/2001-12/2020 for optimal portfolios computed with 240-month  windows rolled over one month.\end{singlespace}} }
\label{t-6} 
\end{table}
\endgroup
\end{singlespace}

Tables~\ref{t-2} -~\ref{t-6} report the coefficient estimates of
the factor models, as well as their respective $t$-statistics and $p$-values.
The results indicate that none of the factor models could explain the performance of the two strategies. In particular, a close to zero market loading indicates a market neutral exposure.
The  intercept $a$ is statistically different from zero in all cases.

For all factor models, we observe that the beta market is smaller than one (defensive) for both portfolios as expected.
When the Fama and French 6-factor model is used, the negative sign for the SMB factor loading and positive sign for the HML factor loading correspond to an additional defensive tilt of the SS-SSD portfolio returns. 
Defensive strategies overweight large value stocks and underweight small growth stocks (Novy-Marx (2016)).

We also observe that the only factors that are significant for the MAXSER returns are the FIN factor of the 3-factor model 
of Daniel, Hirshleifer, and Sun (2020), and the MGMT factor of the Stambaugh and Yuan(2017), four-factor model. The FIN factor (long-horizon financing factor) exploits the information in manager decisions to issue or repurchase equity in response to persistent mispricing, while the MGMT, or Management factor, is the excess returns of stocks with high ranking on management-related anomalies over the return of those with low ranking.
On the other hand, there is no statistically significant factor that explains the returns of the SS-SSD portfolios.

Finally, in order to understand whether the results are explained by the long-short nature of the factors, we construct long-only factors for the Fama and French 5-factor model.  Table~\ref{t-7} confirms that the performance of the SS-SSD and MAXSER optimal portfolios is not explained by traditional factors even if we consider their long-only legs.

\begin{singlespace}
\begin{table}[H]
\caption{Fama-French (2015), 5-factor model (long-only)}
\begin{center}
\scalebox{0.8}{
\begin{tabular}{lrrrrrr}
\toprule[1.5pt] 
 & $a$  & $R_{M}-R_{F}$  & SMB  & HML  & RMW  & CMA  \tabularnewline
\midrule 
SS-SSD  &   & &&&& \tabularnewline
\midrule[1.3pt]
Coef.  & 0.0116  & -0.0119  & -0.0104  & 0.0745  & -0.0532  & -0.0121  \tabularnewline
$t$-stat  & 6.6020  & -0.3053  & -0.1836  & 1.1470 & -0.7859  & -0.1246  \tabularnewline
$p$-value & 0.0  & 0.7604  & 0.8545  & 0.2526  & 0.4328  & 0.9010  \tabularnewline
\midrule 
MAXSER  &   & &&&& \tabularnewline
\midrule[1.3pt]
Coef.  & 0.0106  & -0.0171  & -0.0419  & -0.0014  & 0.0141  &0.0454  \tabularnewline
$t$-stat  & 5.4283  & -0.3990  & -0.6711  & -0.0198  & 0.1885  & 0.4224  \tabularnewline
$p$-value & 0.0  & 0.6903 & 0.5029  & 0.9842 & 0.8507  & 0.6731 \tabularnewline
\midrule 
$1/N$  &   & &&&& \tabularnewline
\midrule[1.3pt]
Coef.  & 0.0065  & 0.9488  & 0.0136  & 0.0551  & -0.0071  & -0.0400  \tabularnewline
$t$-stat  & 8.8016  & 58.4311 & 0.5779  & 2.0299  & -0.2495  & -0.9850  \tabularnewline
$p$-value & 0.0  & 0.0  & 0.5639 & 0.0436  & 0.8032  & 0.3257  \tabularnewline
\bottomrule[1.5pt] 
\end{tabular}}
\end{center}
\noindent\parbox[c][0.7\totalheight][s]{1\textwidth}{%
\noindent{\footnotesize\begin{singlespace}Entries report the coefficients and their respective $t$-statistics and
 $p$-values for the SS-SSD portfolio (upper panel), the MAXSER portfolio (second panel), and for the $1/N$ portfolio (lower panel). The dataset spans 01/2001-12/2020 for optimal portfolios computed with 240-month  windows rolled over one month.\end{singlespace}}}

\label{t-7} 
\end{table}
\end{singlespace}

The results seem to indicate that other factors drive the performance of these portfolios. We also observe that most of the factors in all factor models are significant in the case of the $1/N$ portfolio (apart from the long-only factors), with a positive significant  market loading.
This observation exemplifies that the two dynamic strategies are markedly different from the naive $1/N$ one.

\section{Concluding Remarks}
Our new methodology designed to target sparse spanning portfolios shows that we can often limit ourselves to a subset of a large investment opportunity set without sacrificing expected utility because of under-diversification. It also reveals that a sparse mean-variance portfolio selection (MAXSER) yields under-diversification w.r.t.\ an optimal sparse spanning portfolio. This paper focuses on second-order stochastic dominance but could be modified to accommodate higher-order stochastic dominance. We could then check whether the empirical findings extend in such settings as well.

The methodology avoids the use of LASSO-type regularizations on the stochastic dominance inequalities. It does not require fine tuning regularization parameters. Its asymptotic behavior is known whether sparse spanning holds of not. Importantly, it  enables the investigation of the relation between under-diversification loss and the sparsity (cardinality) constraint.

The FSS greedy algorithm technology can be felt as time consuming especially if it is employed in resampling frameworks to get suitable statistical inference. Even though our fast subsampling methodology avoids this, it could however be of interest to alleviate its associated numerical cost,  and provide paths for further research. 
One example is the possibility of exploiting the geometric realization of $\mathcal{L}_{p,q}$ as a sub-simplex of $\Lambda$, when the latter is a simplex for large enough $p$ - see Edelsbrunner (2014). Another example concerns the investigation of the existence of suitable smooth approximations of the Russell-Seo utilities, for the subsequent use of greedy algorithms that exploit smoothness; e.g.\ the Orthogonal Matching Pursuit in Elenberg et al.\ (2018).

In relation to the cost of resampling, it could be of interest to approximate the upper tail behavior of the limiting distribution of the under-diversification loss. The latter could be related to extensions of approximations of the analogous probabilities for the supremum of Gaussian random fields via topological features of the underlying parameter space like its Euler characteristic-see for example Takemura and Kuriki (2003). Another approach, especially when testing for sparse spanning, is via the combined use of Empirical Likelihood Ratio statistics with conservative chi-squared based rejection regions formed by moment selection; see for example Arvanitis and Post (2024).

The sparse spanning methodology could be used as an alternative selection framework to identify the factors out of a large set of factors and anomalies (for example, the set of 153 factors of Jensen, Kelly and Pedersen (2023), or the set of 193 factors of Hou, Chen, and Zhang (2020, 2021) that explain the returns of funds. Chen et al.\ (2023) impose a sparsity assumption via a regularized regression approach, the adaptive LASSO estimator, from the machine learning literature, to select a model of 9 factors.

\section*{Appendix}

The appendix contains the proofs of our results and the list of factors used in the empirical application.

\section*{Proofs}
\begin{proof}[Proof of Lemma \ref{lem:Span_Char}]
The result is obtained by exploiting the continuity
of $D$ w.r.t.\ its first triplet of arguments, and the compactness
of the parameter space $K\times\Lambda$. It evolves by iteratively
establishing that $\inf_{K}\sup_{z\in Z}D\left(z,\kappa,\lambda,\mathbb{P}\right)$
is continuous in $\lambda$, which then implies that it has a maximizer.
Specifically, $D\left(z,\kappa,\lambda,\mathbb{P}\right)$ is continuous
in $\left(z,\kappa,\lambda\right)$ (w.r.t.\ the product of the Euclidean
topology on $\mathbb{R}$, $l_{1}$ on $K,\:\Lambda$, respectively),
due to the continuity of\linebreak{}
 $\left(z-\sum_{i=0}^{\infty}\kappa_{i}X_{t}^{(i)}\right)_{+}-\left(z-\sum_{i=0}^{\infty}\lambda_{i}X_{t}^{(i)}\right)_{+}$,
Assumption \ref{assu:UMom} and Dominated Convergence. The CMT implies
that $\sup_{z\in Z}D\left(z,\kappa,\lambda,\mathbb{P}\right)$ is
continuous in $\left(\kappa,\lambda\right)$. We have that $K\underset{\text{SSD}}{\nsucceq}\Lambda$
iff $\exists\lambda^{\star}\in\Lambda-K$ such that $\forall\kappa\in K$,
$\sup_{z\in Z}D\left(z,\kappa,\lambda^{\star},\mathbb{P}\right)>0$.
The compactness of $K$ and the continuity of $\sup_{z\in Z}D\left(z,\kappa,\lambda^{\star},\mathbb{P}\right)$
on the second argument imply that the latter holds iff\linebreak{}
 $\inf_{K}\sup_{z\in Z}D\left(z,\kappa,\lambda^{\star},\mathbb{P}\right)>0$.
The compactness of $K$ also implies via Theorem\textcolor{black}{{}
3.4 of Molchanov (2006) that $\inf_{K}\sup_{z\in Z}D\left(z,\kappa,\lambda,\mathbb{P}\right)$
is continuous w.r.t.\ its third argument. Hence},\textcolor{black}{{}
$\inf_{K}\sup_{z\in Z}D\left(z,\kappa,\lambda^{\star},\mathbb{P}\right)>0$
is equivalent to } $\sup_{\Lambda}\inf_{K}\sup_{z\in Z}D\left(z,\kappa,\lambda,\mathbb{P}\right)>0$. 
\end{proof}
\begin{proof}[Proof of Lemma \ref{lem:Characterizations}]
It follows by Lemma \ref{lem:Span_Char} and the monotonicity
of $\Lambda$\textcolor{black}{{} as a function of $p$.} 
\end{proof}
\begin{proof}[Proof of Lemma \ref{lem:Helpful_Compactness}]
The proof evolves in the following steps: (i) we
majorize $\sup_{z\in Z}D\left(z,\kappa,\lambda,\mathbb{P}\right)$
by the supremum of $\int_{Z}D\left(z,\kappa,\lambda,\mathbb{P}\right)d\cdot$
w.r.t.\ a set of linear operators, (ii) we validate a $\max$-$\min$
result to interchange the order of optimization operators for $\inf_{K}\sup_{\cdot}\int_{Z}D\left(z,\kappa,\lambda,\mathbb{P}\right)d\cdot$,
(iii) we use an appropriate topology for $\mathcal{L}_{p,q}$ and establish
appropriate continuity and generalized convexity properties for $\inf_{K}\sup_{F\in\mathcal{P}\left(Z\right)}\int_{Z}D\left(z,\kappa,\lambda,\mathbb{P}\right)dF\left(z\right)$
as a function on $\Lambda\times\mathcal{L}_{p,q}$, so that we validate a
$\max$-$\min$ result to interchange the order of the outer pair
of optimization operators in $\inf_{\mathcal{L}_{p,q}}\sup_{\Lambda}\sup_{\cdot}\inf_{K}\int_{Z}D\left(z,\kappa,\lambda,\mathbb{P}\right)d\cdot$,
(iv) Analogously to (iii), we validate a $\max$-$\min$ result to
interchange the order of the middle pair of optimization operators
in $\sup_{\Lambda}\inf_{\mathcal{L}_{p,q}}\sup_{\cdot}\inf_{K}\int_{Z}D\left(z,\kappa,\lambda,\mathbb{P}\right)d\cdot$,
(v) we finally use the extreme point properties of the set of linear
operators in (i) and the $\max$-$\min$ inequality to obtain the
result. Specifically, for (i), consider the space $\mathcal{P}\left(Z\right)$
comprised by the probability distributions that are supported on $Z$,
and equipped with the weak topology. The space is convex and contains
the degenerate distributions on the elements of $Z$ as its extreme
points. Then, by Theorem 15.9 of Aliprantis and Border (2006), we deduce
that
 $\sup_{z\in Z}D\left(z,\kappa,\lambda,\mathbb{P}\right)\leq\sup_{F\in\mathcal{P}\left(Z\right)}\int_{Z}D\left(z,\kappa,\lambda,\mathbb{P}\right)dF\left(z\right)$.
For (ii), we have that due to Assumption \ref{assu:UMom},
and the Lipschitz continuity property of $\left(\cdot\right)_{+}$,
we have that $\sup_{Z,\Lambda^{2}}\left|D\left(z,\kappa,\lambda,\mathbb{P}\right)\right|\leq2\max_{i}\mathbb{E}\left[\left|X^{(i)}\right|\right]<+\infty$,
hence the linear functional $F\rightarrow\int_{Z}D\left(z,\kappa,\lambda,\mathbb{P}\right)dF\left(z\right)$
is also continuous w.r.t.\ $F$ for all $\kappa,\lambda$, due to the
Portmanteau Lemma. Furthermore, $\mathbb{E}\left[\left(z-\sum_{i=0}^{\infty}\kappa{}_{i}X_{t}^{(i)}\right)_{+}\right]$
is convex in $\kappa$, due to the convexity and monotonicity of $\left(\cdot\right)_{+}$
and the linearity of $z-\sum_{i=0}^{\infty}\kappa{}_{i}X_{t}^{(i)}$ w.r.t.\
$\kappa$. Hence, since $K\in\mathcal{L}_{p,q}$ is closed and $\Lambda$
is compact, the dual version of the Kneser-Fan Theorem (see Theorem
4.2' of Sion (1958)) implies that 
$
\inf_{K}\sup_{F\in\mathcal{P}\left(Z\right)}\int_{Z}D\left(z,\kappa,\lambda,\mathbb{P}\right)dF\left(z\right)=\sup_{F\in\mathcal{P}\left(Z\right)}\inf_{K}\int_{Z}D\left(z,\kappa,\lambda,\mathbb{P}\right)dF\left(z\right).
$
For (iii), equip $\mathcal{L}_{p,q}$ with the PK-topology
(see Definition 3.1.4 of Klein and Thompson (1984)). Due to Theorem
4.3.4-5 of Klein and Thompson (1984), $\mathcal{L}_{p,q}$ is compact.
Due to Theorem 3.4 of Klein and Thompson (1984) and the boundedness
and continuity of $D\left(\cdot,\cdot,\cdot,\mathbb{P}\right)$, the
mapping
 $\inf_{K}\int_{Z}D\left(z,\kappa,\lambda,\mathbb{P}\right)dF\left(z\right):\mathcal{L}_{p,q}\times\Lambda\rightarrow\mathbb{R}$
is jointly continuous for all $F$. Then, the boundedness of $D\left(\cdot,\cdot,\cdot,\mathbb{P}\right)$
and the CMT imply that $\sup_{F\in\mathcal{P}\left(Z\right)}\inf_{K}\int_{Z}D\left(z,\kappa,\lambda,\mathbb{P}\right)dF\left(z\right):\mathcal{L}_{p,q}\times\Lambda\rightarrow\mathbb{R}$
is also jointly continuous.

For any $t\in\left(0,1\right)$ and any $K_{1},K_{2}\in\mathcal{L}_{p,q}$,
we have that 
\[
\begin{array}{c}
t\inf_{K_{1}}\int_{Z}D\left(z,\kappa,\lambda,\mathbb{P}\right)dF\left(z\right)+\left(1-t\right)\inf_{K_{2}}\int_{Z}D\left(z,\kappa,\lambda,\mathbb{P}\right)dF\left(z\right)\\
\geq\min\left[\inf_{K_{i}^{\star}}\int_{Z}D\left(z,\kappa,\lambda,\mathbb{P}\right)dF\left(z\right),i=1,2\right]
\end{array},
\]
where $K_{i}^{\star}$ is any element of $\mathcal{L}_{p,q}$ of support
$q$ that contains $K_{i},\:i=1,2$. Analogously, we obtain from the
previous and the monotonicity of $\sup$ 
\[
\begin{array}{c}
t\sup_{F\in\mathcal{P}\left(Z\right)}\inf_{K_{1}}\int_{Z}D\left(z,\kappa,\lambda,\mathbb{P}\right)dF\left(z\right)+\left(1-t\right)\sup_{F\in\mathcal{P}\left(Z\right)}\inf_{K_{2}}\int_{Z}D\left(z,\kappa,\lambda,\mathbb{P}\right)dF\left(z\right)\\
\geq\min_{i=1,2}\inf_{K_{i}^{\star}}\int_{Z}D\left(z,\kappa,\lambda,\mathbb{P}\right)dF\left(z\right)\\
\sup_{F\in\mathcal{P}\left(Z\right)}\left[t\inf_{K_{1}}\int_{Z}D\left(z,\kappa,\lambda,\mathbb{P}\right)dF\left(z\right)+\left(1-t\right)\inf_{K_{2}}\int_{Z}D\left(z,\kappa,\lambda,\mathbb{P}\right)dF\left(z\right)\right]\geq\\
\sup_{F\in\mathcal{P}\left(Z\right)}\min_{i=1,2}\inf_{K_{i}^{\star}}\int_{Z}D\left(z,\kappa,\lambda,\mathbb{P}\right)dF\left(z\right)
\end{array},
\]
and the previous pair of displays implies that the mapping $\inf_{K}\int_{Z}D\left(z,\kappa,\lambda,\mathbb{P}\right)dF\left(z\right):\mathcal{L}_{p,q}\rightarrow\mathbb{R}$
is convex-like for all $\left(F,\lambda\right)$, and the mapping
$\sup_{F\in\mathcal{P}\left(Z\right)}\inf_{K}\int_{Z}D\left(z,\kappa,\lambda,\mathbb{P}\right)dF\left(z\right):\mathcal{L}_{p,q}\rightarrow\mathbb{R}$
is convex-like for all $\lambda$ (see Section 2 of Sion (1958)).
For any $t\in\left(0,1\right)$ and any $\lambda_{1},\lambda_{2}\in\Lambda$
we have that due to Theorem 15.9 of Aliprantis and Border (2006) 
\[
\begin{array}{c}
t\sup_{F\in\mathcal{P}\left(Z\right)}\inf_{K}\int_{Z}D\left(z,\kappa,\lambda_{1},\mathbb{P}\right)dF\left(z\right)+\left(1-t\right)\sup_{F\in\mathcal{P}\left(Z\right)}\inf_{K}\int_{Z}D\left(z,\kappa,\lambda_{2},\mathbb{P}\right)dF\left(z\right)\\
=t\sup_{z\in Z}\inf_{K}D\left(z,\kappa,\lambda_{1},\mathbb{P}\right)+\left(1-t\right)\sup_{z\in Z}\inf_{K}D\left(z,\kappa,\lambda_{2},\mathbb{P}\right)
\end{array},
\]
and the rhs of the previous display is less than or equal to $\max_{\lambda}\sup_{z\in Z}\inf_{K}D\left(z,\kappa,\lambda,\mathbb{P}\right)$
and the maximum exists due to the joint continuity and boundedness
of $D\left(\cdot,\cdot,\cdot,\mathbb{P}\right)$, the CMT and the
compactness of $\Lambda$. Hence, the mapping $\sup_{F\in\mathcal{P}\left(Z\right)}\inf_{K}\int_{Z}D\left(z,\kappa,\lambda,\mathbb{P}\right)dF\left(z\right):\Lambda\rightarrow\mathbb{R}$
again see Section 2 of Sion (1958)).

For (iv), for any $t\in\left(0,1\right)$
and any $F_{1},F_{2}\in\mathcal{P}\left(Z\right)$, we have that 
\[
\begin{array}{c}
t\inf_{K}\int_{Z}D\left(z,\kappa,\lambda,\mathbb{P}\right)dF_{1}\left(z\right)+\left(1-t\right)\inf_{K}\int_{Z}D\left(z,\kappa,\lambda,\mathbb{P}\right)dF_{2}\left(z\right)\\
\geq\inf_{K}\int_{Z}D\left(z,\kappa,\lambda,\mathbb{P}\right)d\left[tF_{1}\left(z\right)+\left(1-t\right)F_{2}\left(z\right)\right]
\end{array},
\]
and thereby the mapping $\inf_{K}\int_{Z}D\left(z,\kappa,\lambda,\mathbb{P}\right)dF\left(z\right):\mathcal{P}\left(Z\right)\rightarrow\mathbb{R}$
is concave and hence concave-like for all $K\in\mathcal{L}_{p.q}$
and $\lambda$. Using the previous and applying twice the dual version
of the Kneser-Fan Theorem, we jointly obtain the required
results in steps (iii)-(iv) as,
\[
\begin{array}{c}
\inf_{\mathcal{L}_{p,q}}\sup_{\Lambda}\sup_{F\in\mathcal{P}\left(Z\right)}\inf_{K}\int_{Z}D\left(z,\kappa,\lambda,\mathbb{P}\right)dF\left(z\right)\\
=\sup_{\Lambda}\inf_{\mathcal{L}_{p,q}}\sup_{F\in\mathcal{P}\left(Z\right)}\inf_{K}\int_{Z}D\left(z,\kappa,\lambda,\mathbb{P}\right)dF\left(z\right)\\
=\sup_{\Lambda}\sup_{F\in\mathcal{P}\left(Z\right)}\inf_{\mathcal{L}_{p,q}}\inf_{K}\int_{Z}D\left(z,\kappa,\lambda,\mathbb{P}\right)dF\left(z\right).
\end{array}
\]
Finally, for (v), again due to Theorem 15.9 of Aliprantis
and Border (2006), we get \break
$
\sup_{\Lambda}\sup_{F\in\mathcal{P}\left(Z\right)}\inf_{\mathcal{L}_{p,q}}\inf_{K}\int_{Z}D\left(z,\kappa,\lambda,\mathbb{P}\right)dF\left(z\right)=\sup_{\Lambda}\sup_{z\in Z}\inf_{\mathcal{L}_{p,q}}\inf_{K}D\left(z,\kappa,\lambda,\mathbb{P}\right).
$
The result follows by the $\max$-$\min$ inequality. 
\end{proof}
\begin{proof}[Proof of Theorem \ref{thm:ELT}]
For (a), we use the Ergodic Theorem uniformly
in $\lambda$ and continuously in $z$. Specifically,
we derive the limiting behavior of $\frac{1}{T}\sum_{t=0}^{T}\left(z-\sum_{i=0}^{\infty}\lambda_{i}X_{t}^{(i)}\right)_{+}$
from the locally uniform in $z$ and uniform in $\lambda$, version
of the Ergodic Theorem applied on the function $\frac{1}{T}\sum_{t=0}^{T}\left(z-\sum_{i=0}^{\infty}\lambda_{i}X_{t}^{(i)}\right)_{+}$,
noting that it is applicable due to Assumption \ref{assu:UMom} and the $l_{1}$ boundedness of $\Lambda_{\infty}$. Continuously uniform convergence then implies continuous hypo-convergence by Molchanov (2006).

For (b)-(c), (i) we establish
that the associated set of functions has an integrable envelope, (ii) we use the fact that the
associated sets of functions-which admit generalized derivatives w.r.t.\
the sample arguments-are bounded subsets of a weighted Sobolev space, and thus have controllable bracketing
entropy numbers, and (iii) we use the above and the time series properties
of $X$ to verify the validity of an appropriate FCLT or maximal inequality. For (i) we have that due to Jensen's inequality,
\begin{equation}
\begin{array}{c}
\mathbb{E}\left[\sup_{z,\kappa}\left(\mu^{T}\left(\begin{array}{c}
\left(z-\sum_{i=0}^{\infty}\kappa_{i}X^{(i)}\right)_{+}-\left(z-\sum_{i=0}^{\infty}\lambda_{i}X^{(i)}\right)_{+}\\
X^{T}\mathbb{I}\left\{ z\geq\sum_{i=0}^{\infty}\lambda_{i}X^{(i)}\right\} \left(\kappa-\lambda\right)
\end{array}\right)\right)^{2+\varepsilon}\right]\\
\leq C\mathbb{E}\left[\left(\sup_{z,\kappa}\left(\left(z-\sum_{i=0}^{\infty}\kappa_{i}X^{(i)}\right)_{+}-\left(z-\sum_{i=0}^{\infty}\lambda_{i}X^{(i)}\right)_{+}\right)\right)^{2+\varepsilon}\right]\\
+C\mathbb{E}\left[\left(\sup_{z,\kappa}\left(X^{T}\mathbb{I}\left\{ z\geq\sum_{i=0}^{\infty}\lambda_{i}X^{(i)}\right\} \left(\kappa-\lambda\right)\right)\right)^{2+\varepsilon}\right]\\
\leq C\mathbb{E}\left[\left(\sup_{z,\kappa}\left(\left(\sum_{i=0}^{\infty}\left(\lambda_{i}-\kappa_{i}\right)X_{t}^{(i)}\right)\right)\right)^{2+\varepsilon}\right]%\\
+C\mathbb{E}\left[\max_{i}\left(X^{(i)}\mathbb{I}\left\{ z\geq\sum_{i=0}^{\infty}\lambda_{i}X^{(i)}\right\} \right)^{2+\varepsilon}\right]\\
\leq 2^{1+\varepsilon}C\left(\mathbb{E}\left[\left(\sup_{z,\kappa}\left(\left(\sum_{i=0}^{\infty}\lambda_{i}X^{(i)}\right)\right)\right)^{2+\varepsilon}\right]+\mathbb{E}\left[\left(\sup_{z,\kappa}\left(\left(\sum_{i=0}^{\infty}\kappa_{i}X^{(i)}\right)\right)\right)^{2+\varepsilon}\right]\right)\\
+C\mathbb{E}\left[\max_{i}\left(\left|X^{(i)}\right|\right)^{2+\varepsilon}\right]\leq2^{1+\varepsilon}C\mathbb{E}\left[\max_{i}\left(X^{(i)}\right)^{2+\varepsilon}\right]<+\infty.
\end{array}\label{eq:help0}
\end{equation}

For (ii), we have that for any $l\geq1$, $\delta>0$,
the function class\linebreak 
$\mathcal{M}_{1}:=\left\{ \mathbb{R}^{\left\lfloor q\left(\ln T+1\right)\right\rfloor }\ni x\rightarrow 
\left(z-x^{T}\lambda\right)_{+}-\left(z-x^{T}\kappa\right)_{+}\right\}$, 
as well as the function class\linebreak$\mathcal{M}_{2}:=\left\{ \mathbb{R}^{\left\lfloor q\left(\ln T+1\right)\right\rfloor }\ni x\rightarrow x^{T}\mathbb{I}\left\{ z\geq\sum_{i=0}^{\left\lfloor q\left(\ln T+1\right)\right\rfloor }\kappa_{i}^{*}X^{(i)}\right\} \left(\kappa-\kappa^{*}\right),z,\kappa,\kappa^{*}\right\}$
are bounded subsets of the weighted Sobolev space $H_{l}^{1}\left(\mathbb{R}^{\left\lfloor q\left(\ln T+1\right)\right\rfloor },\left\langle x\right\rangle ^{2+\delta}\right)$, i.e. the semi-normed space $\left\{ f:\mathbb{R}^{\left\lfloor q\left(\ln T+1\right)\right\rfloor }\rightarrow\mathbb{R},\:\begin{array}{c}
\left\Vert f\right\Vert _{l,2+\delta,\mu}:=\\
\left(\int_{\mathbb{R}^{\left\lfloor q\left(\ln T+1\right)\right\rfloor }}\left[\left|\frac{f\left(x\right)}{\left(1+\left\Vert x\right\Vert \right)^{2+\delta}}\right|^{l}+\left|D\frac{f\left(x\right)}{\left(1+\left\Vert x\right\Vert \right)^{2+\delta}}\right|^{l}\right]d\mu\right)^{\nicefrac{1}{l}}<+\infty
\end{array}\right\} $, where $D$ denotes partial derivation in the sense of distributions,
and $\mu$ denotes the Lebesgue measure on $\mathbb{R}^{\left\lfloor q\left(\ln T+1\right)\right\rfloor }$ - see
 3.3.2 of Nickl and Potcher (2007), due to the $l_{1}$-boundedness
of $K$. In the notation of the aforementioned paper, choosing $l$
such that $\frac{\left\lfloor q\left(\ln T+1\right)\right\rfloor }{l}\rightarrow0$,
$r=2+\varepsilon$ and $\gamma=3+\delta$, $\beta=2+\delta$, $\mathfrak{M}$
the set of finite dimensional distributions of $X^{\infty}$, we have that, due to Corollary 4.2 of Nickl
and Potcher (2007), and for large enough $T$, the bracketing entropy
of $\mathcal{M}_{i},\: i=1,2$, as a function of $\epsilon>0$, is universally bounded from above by $c\epsilon^{-\lfloor q(\ln T+1)\rfloor}$
for some universal constant $c>0$. 

Then, from (i) above, (ii) the fact that $\beta_{k}\sim b^{k}$,
and (iii)  the fact that the class has an $L^{2+\varepsilon}\left(\mathbb{P}\right)$-integrable
envelope due to (\ref{eq:help0}), we get that Theorems 1 and 2 of Doukhan, Massart,
and Rio (1995) are applicable and the results in (b) and (c) follow since $\frac{\ln {p \choose \left\lfloor q\left(\ln T+1\right)\right\rfloor }}{\sqrt{T}}+\frac{\left\lfloor q\left(\ln T+1\right)\right\rfloor}{\sqrt{T}}\rightarrow 0$. The latter holds since $\frac{\ln p}{\sqrt{T}}\rightarrow 0$ via Stirling's approximation on factorials and first order Taylor expansions on the logarithms.
\end{proof}
\begin{proof}[Proof of Theorem \ref{thm:Guarantees}]
The proof works by (i) establishing that the empirical
LPM,\linebreak{}
$\frac{1}{T}\sum_{t=0}^{T}\left(z-\sum_{i=0}^{\infty}\kappa_{i}X_{t}^{(i)}\right)_{+}$,
satisfies uniformly over $z$, w.h.p.\ the weak sub-modularity property
of Elenberg et al.\ (2018), so that (ii) the guarantees results on
the Forward Selection Algorithm of the aforementioned paper hold w.h.p..
We do so by (iii) establishing that the first order Taylor expansion
restricted on appropriate parts of the empirical LPM, is approximated
by the analogous expansion of its population counterpart, uniformly
over $z$, w.h.p.. Given (i), the statistical guarantees for the overall
optimization problem follow (iv) by standard results on approximation
of optimization problems and the CMT. 

First, the restricted smoothness condition that implies the middle lower bound for the submodularity ratio in Theorem 1 of Elenberg et al.\ (2018) holds trivially in our case , namely $M_{\Omega}$ can be fixed to one in their notation, since $\Lambda$ and thereby $K$ have simplicial structures. Indeed, the denominator in the aforementioned upper bound can be chosen as any real number that satisfies the r.h.s.\ of the inequality that appears in Definition 3 of Elenberg et al.\ (2018), for any pair of  elements of $K$ that satisfies the following: the $\ell_{0}$ norm of the difference equals at most one. %Since $K$ has a simplicial structure, 
%that restriction to be less than one 
The $\ell_{0}$ norm being less than or equal to 1 means that the vectors should disagree at most at one coordinate, and thus they must agree at the remaining coordinates by construction.
Since the weights sum to one, 
it is only possible if the aforementioned pair contains the same vector, which implies that the r.h.s.\ of the inequality holds trivially for any possible $M_{\Omega}$.
%Since $K$ has a simplicial structure, the $\ell_{0}$ norm of the difference of two vectors with entries (weights) restricted to  $[0,1]$ is always less or equal to one, which implies that the r.h.s.\ of the inequality holds trivially for any possible $M_{\Omega}$.

Then, we recall some notation mainly from convex
analysis. Specifically, in what follows\textcolor{black}{{} $\partial$
denotes the sub-gradient of an arbitrary real valued convex function
defined on a locally convex space (see Ch.\ D of Hiriart-Urruty
and Lemar\'{e}chal (2004)-HUL). }Besides, for $\mathbb{Q:=P},\mathbb{P}_{T}$
and $\mathbb{E}_{\mathbb{Q}}$ denoting integration w.r.t.\ $\mathbb{Q}$,
$\mathbb{E}_{\mathbb{Q}}\left[\left(z-\sum_{i=0}^{\infty}\kappa{}_{i}X_{0,i}\right)_{+}\right]$
is convex in the second argument due to the convexity and monotonicity
of $\left(\cdot\right)_{+}$ and the linearity of $z-\sum_{i=0}^{\infty}\kappa{}_{i}X_{t}^{(i)}$
w.r.t.\ $\kappa$.

Then, for any $\kappa\in\Lambda$ we
obtain the inclusion {$g_{z,T}\left(\kappa\right):=\frac{1}{T}\sum_{t=0}^{T}X_{t}\mathbb{I}_{z\geq\sum_{i=0}^{\infty}\kappa_{i}X_{0,i}}{}_{+}\in\partial\mathbb{E}_{\mathbb{P}_{T}}\left(z-\sum_{i=0}^{\infty}\kappa_{i}X_{0,i}\right)_{+}$
due to Theorems 4.1.1. and 4.2.1 of HUL. Furthermore, due to }Theorem
1 of Savare (1996)\textcolor{black}{{} and the fact that $X_{t}$
has a continuous density, we have that $g_{z}\left(\kappa\right):=\frac{\partial\mathbb{E}\left(z-\sum_{i=0}^{\infty}\kappa{}_{i}X_{0,i}\right)_{+}}{\partial\kappa}=\mathbb{E}\left[X_{t}\mathbb{I}_{z\geq\sum_{i=0}^{\infty}\kappa_{i}X_{t}^{(i)}}\right]$.}
Then, for any $\left(\kappa^{\star},\kappa\right)\in\Lambda_{\left(\left\lfloor q\left(\ln T+1\right)\right\rfloor \right)}$,
define the Taylor expansion $\mathcal{E}_{z,T}\left(\kappa^{\star},\kappa\right):=\frac{1}{T}\sum_{t=0}^{T}\left(z-\sum_{i=0}^{\infty}\kappa_{i}^{\star}X_{t}^{(i)}\right)_{+}-\frac{1}{T}\sum_{t=0}^{T}\left(z-\sum_{i=0}^{\infty}\kappa_{i}X_{t}^{(i)}\right)_{+}-\left(\kappa^{\star}-\kappa\right)'g_{z,T}\left(\kappa\right)$,
and similarly the Taylor expansion $\mathcal{E}_{z}\left(\kappa^{\star},\kappa\right):=\mathbb{E}\left[\left(z-\sum_{i=0}^{\infty}\kappa_{i}^{\star}X_{t}^{(i)}\right)_{+}\right]-\mathbb{E}\left[\left(z-\sum_{i=0}^{\infty}\kappa_{i}X_{t}^{(i)}\right)_{+}\right]-\left(\kappa^{\star}-\kappa\right)'g_{z}\left(\kappa\right)$. 

Working towards (iii), and using the previous theorem,
we have that for any $\delta>0$ and any $C>0,0<\epsilon<\frac{1}{4}$:
\begin{equation}
\begin{array}{c}
\mathbb{P}\left(\sup_{z}\sup_{\Lambda_{\left(\left\lfloor q\left(\ln\left(T+1\right)\right)\right\rfloor \right)},\left\Vert \kappa-\kappa^{\star}\right\Vert >\frac{C}{T^{\epsilon}}}\left(\frac{1}{\left\Vert \kappa-\kappa^{\star}\right\Vert ^{2}}\left|\mathcal{E}_{z,T}\left(\kappa,\kappa^{\star}\right)-\mathcal{E}_{z}\left(\kappa,\kappa^{\star}\right)\right|\right)\geq\delta\right)\\
\leq\mathbb{P}\left(\sup_{z}\sup_{\Lambda_{\left(\left\lfloor q\left(\ln\left(T+1\right)\right)\right\rfloor \right)},\left\Vert \kappa-\kappa^{\star}\right\Vert >\frac{C}{T^{\epsilon}}}\left(T^{2\epsilon}\left|\mathcal{E}_{z,T}\left(\kappa,\kappa^{\star}\right)-\mathcal{E}_{z}\left(\kappa,\kappa^{\star}\right)\right|\right)\geq\frac{\delta}{C}\right)\\
\leq\mathbb{P}\left(\sup_{z}\sup_{\Lambda_{\left(\left\lfloor q\left(\ln\left(T+1\right)\right)\right\rfloor \right)}}\left|\sqrt{T}D\left(z,\kappa,\kappa^{\star},\mathbb{P}_{T}-\mathbb{P}\right)\right|\geq\frac{2\delta T^{\frac{1}{2}-2\epsilon}}{3C}\right)\\
+\mathbb{P}\left(\sup_{z}\sup_{\Lambda_{\left(\left\lfloor q\left(\ln\left(T+1\right)\right)\right\rfloor \right)}}\left|G_{T}\left(z,\kappa,\kappa^{\star}\right)\right|\geq\frac{\delta T^{\frac{1}{2}-2\epsilon}}{3C}\right)=o\left(1\right),
\end{array}\label{eq:help-1}
\end{equation}
where the final equality in (\ref{eq:help-1}) follows from the
first two parts of Theorem \ref{thm:ELT}, the Lipschitz property of $D$ w.r.t. the parameters, the fact that $T^{\frac{1}{2}-2\epsilon}\rightarrow+\infty$,
and the Portmanteau Theorem.

Due to the bounds on the eigenvalues of $\mathbb{E}\left(z-\sum_{i=0}^{\infty}\kappa_{i}X_{0,i}\right)_{+}$
of Assumption \ref{assu:SCS}, Theorem 6.1.2 of HUL, Paragraph 1.3.(d)
in Ch.\ 4 of Hiriart-Urruty and Lemar\'{e}chal (2013), and the dual
form of Remark 1 of Elenberg et al.\ (2018),  we have that uniformly
w.r.t.\ $z$ and for any $\left(\kappa^{\star},\kappa\right)\in\Lambda_{\left(\left\lfloor q\left(\ln T+1\right)\right\rfloor \right)}$,
$\frac{m_{\left\lfloor q\left(\ln\left(T+1\right)\right)\right\rfloor }}{2}\left\Vert \kappa-\kappa^{\star}\right\Vert ^{2}\leq\mathcal{E}_{z,T}\left(\kappa^{\star},\kappa\right)$.
Due to this, and (\ref{eq:help-1}), uniformly w.r.t.\ $z$ and for
any $\left(\kappa^{\star},\kappa\right)\in\Lambda_{\left(\left\lfloor q\left(\ln T+1\right)\right\rfloor \right)}\cap\left\{ \left\Vert \kappa-\kappa^{\star}\right\Vert >\frac{C}{T^{\epsilon}}\right\} $,
$
\frac{m_{\left\lfloor q\left(\ln\left(T+1\right)\right)\right\rfloor }+o_{p}\left(1\right)}{2}\left\Vert \kappa-\kappa^{\star}\right\Vert ^{2}\leq\mathcal{E}_{z,T}\left(\kappa^{\star},\kappa\right),\:\text{w.h.p}.,
$
where the $o_{p}\left(1\right)$ terms are independent of $z,\lambda$.
Thus (i) is established.

Then, for (ii), by noting that Theorem 1 of Elenberg
et al.\ (2018) is also valid if the gradient in its proof is substituted
by any fixed element of the sub-gradient, and using the previous display,
the inclusion $\Lambda_{\left(\left\lfloor q\left(\ln T+1\right)\right\rfloor \right)}\cap\left\{ \left\Vert \kappa-\kappa^{\star}\right\Vert >\frac{C}{T^{\epsilon}}\right\} \subseteq\Lambda_{\left(\left\lfloor q\left(\ln T+1\right)\right\rfloor \right)}$
and the discussion immediately after Remark 1 of Elenberg et al.\ (2018),
we get that w.h.p. $\mathcal{K}^{\mathrm{FS}}\left(\Lambda,\mathcal{L}_{p,q},z,\mathbb{P}_{T},q\ln T\right)\leq\left(1-\frac{1}{T^{\gamma_{T}}}\right)\inf_{\mathcal{L}_{p,q}}\inf_{K}\frac{1}{T}\sum_{t=0}^{T}\left(z-\sum_{i=0}^{\infty}\kappa_{i}X_{t}^{(i)}\right)_{+}$, where $\gamma_{T}:=m_{\left\lfloor q\left(\ln\left(T+1\right)\right)\right\rfloor }+o_{p}\left(1\right)$,
establishing (ii). Finally, working towards (iv),
note that Assumption \ref{assu:SCS}, Theorem \ref{thm:ELT}, the
PK-convergence of $\Lambda_{\left(\left\lfloor q\left(\ln T+1\right)\right\rfloor \right)}\cap\left\{ \left\Vert \kappa-\kappa^{\star}\right\Vert >\frac{C}{T^{\epsilon}}\right\} $
to $\Lambda_{\left(\left\lfloor q\left(\ln T+1\right)\right\rfloor \right)}$,
and the CMT imply then (\ref{eq:cons}). The final result follows
from the dual version of Theorem 3.4 (Ch.\ 5, p.\ 338) of Molchanov
(2006) and the CMT. 
\end{proof}
\begin{proof}[Proof of Theorem (\ref{thm:delta})]
The strategy of the proof evolves as: (i) we establish the $o_{p}(\frac{1}{\sqrt{T}})$ approximation of the empirical optimum by the FS solution for the particular choice of $r$. Then, (ii) we  apply the generalized Delta
method (see C\'arcamo et al.\ (2020)) on the relevant empirical process.

For (i), first, due to Theorem 1 of Elenberg
et al.\ (2018), the results of Theorem \ref{thm:Guarantees} are valid
since $r=q\left(\ln T\right)^{\epsilon}$. Using the final result of Theorem \ref{thm:ELT} and Theorem 3.4 (Ch.\
5, p.\ 338) of Molchanov (2006), then we have the approximation \linebreak{}
$\sqrt{T}\left|\inf\frac{1}{T}\sum_{t=0}^{T}\left(z-\sum_{i=0}^{\infty}\kappa_{i}X_{i,t}\right)_{+}-\inf_{\mathrm{csupp}\left(\kappa\right)\leq q}\frac{1}{T}\sum_{t=0}^{T}\left(z-\sum_{i=0}^{\infty}\kappa_{i}X_{i,t}\right)_{+}\right|=o_{p}\left(1\right)$,
where the remainder is independent of $z$ and the
first empirical infimum is derived via forward selection, since this first empirical infimum is tight. For (ii), we can then use part (b)  of Theorem \ref{thm:ELT} which establishes  the limiting behavior of the empirical process $\sqrt{T}D\left(z,\kappa,\lambda,\mathbb{P}_{T}-\mathbb{P}\right)$,  Theorem 2.1 and Proposition 2.1 of C\'arcamo et al.\ (2020) and the chain rule for Hadamard directional differentiability (see Shapiro (1990)) to
the optimizations appearing in the empirical process. The limits in Theorem 2.1 C\'arcamo et al.\ (2020) are in our case the convex sets of optimizers of the functions involved, due to the convexity properties of those functions and of the parameter spaces involved. 
\end{proof}
\begin{proof}[Proof of Proposition \ref{prop:subs_1}]
The proof proceeds as follows: (i) we establish the
weak convergence of the scaled-by-$b_{T}$ discrepancy between the
subsampling empirical process, evaluated at any convergent subsequence
of the FSS optimizers, and the population optimum, to
the $\sup\inf$ of the Gaussian process appearing in the previous
result over $\Gamma^{\star}$, (ii) we establish conservativeness by showing that the cdf of the weak limit is continuous
at its $1-\alpha$ quantile.

For (i), we have that from the weak convergence to
the empirical process in the proof of Theorem \ref{thm:delta}, and
applying Proposition 7.3.1 of Politis, Romano and Wolf (1999), we obtain
that 
$
\sqrt{b_{T}}\left(\mathbb{E}^{\star}\left[D\left(z,\kappa,\lambda,\mathbb{P}_{t,b_{T}}\right)\right]-D\left(z,\kappa,\lambda,\mathbb{P}_{T}\right)\right)\rightsquigarrow\mathcal{G}\left(z,\lambda,\kappa\right),
$
in $\ell^{\infty}\left(Z\times\Lambda_{\infty}\times\Lambda_{\infty}\right)$,
where $\mathbb{E}^{\star}\left[\cdot\right]$ denotes expectation
w.r.t.\ the empirical distribution of $D\left(z,\kappa,\lambda,\mathbb{P}_{t,b_{T}}\right)$
across $t=1,\dots,T-b_{T}+1$. 

In what follows, we also denote with $\left(T\right)$
the index set of the subsequence of $\kappa_{z,T}$ associated with
the examined accumulation point, for notational simplicity. Due to
that (see the proof of Theorem \ref{thm:delta}), $\sqrt{T}\left|\inf\frac{1}{T}\sum_{t=0}^{T}\left(z-\sum_{i=0}^{\infty}\kappa_{i}X_{i,t}\right)_{+}-\inf_{\mathrm{csupp}\left(\kappa\right)\leq q}\frac{1}{T}\sum_{t=0}^{T}\left(z-\sum_{i=0}^{\infty}\kappa_{i}X_{i,t}\right)_{+}\right|=o_{p}\left(1\right)$
uniformly in $z$, the definition of $\kappa_{z,T}$, and that $\frac{b_{T}}{T}\rightarrow0$,
we have that \linebreak{}
$\sqrt{b_{T}}\left(\inf_{\mathrm{csupp}\left(\kappa\right)\leq q}D\left(z,\kappa,\lambda,\mathbb{P}_{T}\right)-D\left(z,\kappa_{z,T},\lambda,\mathbb{P}_{T}\right)\right)=o_{p}\left(1\right)$
uniformly on $Z\times\Lambda_{\infty}$. It implies that $\sqrt{b_{T}}\left(\sup_{Z\times\Lambda}\inf_{\mathrm{csupp}\left(\kappa\right)\leq q}D\left(z,\kappa,\lambda,\mathbb{P}_{T}\right)-\sup_{Z\times\Lambda}D\left(z,\kappa_{z,T},\lambda,\mathbb{P}_{T}\right)\right)=o_{p}\left(1\right)$.
Employing a) the use of Skorokhod representations, applicable due
to Theorem 3.7.25 of Gin\'{e} and Nickl, (2016), b) the convergence
above, c. Theorem 3.4 of Molchanov (2006), d)
Theorem 2.1 and Proposition 2.1 of C\'arcamo et al. (2020)-working similarly to the proof of Theorem
\ref{thm:delta}, e)
the fact that $\left(\kappa_{z,T}\right)_{z}$ are
optimizers of $\mathcal{K}^{\mathrm{FS}}\left(\Lambda,\mathcal{\mathcal{L}}_{p,q},z,\mathbb{P}_{T},r_{T}\left(q\right)\right)-\mathcal{J}\left(\Lambda,z,\mathbb{P}_{T}\right)$,
which due to Theorem \ref{thm:Guarantees} converges to the deterministic
$\mathcal{K}\left(\Lambda^{\infty},\mathcal{\mathcal{L}}_{\infty,q},z,\mathbb{P}\right)-\mathcal{L}\left(\Lambda_{\infty},z,\mathbb{P}\right)$,
and thereby $\left(\kappa_{z,T}\right)_{z}$ are asymptotically independent
to the process $\sqrt{b_{T}}(\sup_{Z\times\Lambda}\mathbb{E}^{\star}\left[D\left(z,\kappa_{z,T},\lambda,\mathbb{P}_{t,b_{T}}\right)\right]$\break $-M^{\mathrm{FS}}\left(\Lambda,\mathcal{\mathcal{L}}_{p,q},\mathbb{P}_{T},q\left(\ln T\right)^{\epsilon}\right))$,
and f) the fact that $\frac{b_{T}}{T}\rightarrow0$,
we then obtain the convergence 
$\sqrt{b_{T}}\left(\sup_{Z\times\Lambda}\mathbb{E}^{\star}\left[D\left(z,\kappa_{z,T},\lambda,\mathbb{P}_{t,b_{T}}\right)\right]-M^{\mathrm{FS}}\left(\Lambda,\mathcal{\mathcal{L}}_{p,q},\mathbb{P}_{T},q\left(\ln T\right)^{\epsilon}\right)\right)\rightsquigarrow\sup\inf_{\Gamma^{\star}}\mathcal{G}\left(z,\lambda,\kappa\right).$
For (ii), first, the definition of $\Gamma^{\star}$ implies
that $\sup\inf_{\Gamma^{\star}}\mathcal{G}\left(z,\lambda,\kappa\right)\geq\sup\inf_{\Gamma}\mathcal{G}\left(z,\lambda,\kappa\right)$.
Then conservativeness follows from this inequality as long as the
cdf of $\sup\inf_{\Gamma^{\star}}\mathcal{G}\left(z,\lambda,\kappa\right)$
is continuous at its $1-\alpha$ quantile. From Lemma
18.15 of van der Vaart (2000), we have that for $\mu,v\in\Gamma^{\star}$
and $\mathcal{G}_{\mu},\mathcal{G}_{v}$ the Gaussian process $\mathcal{G}$
evaluated there,
$
0\leq\sigma^{2}:=\sup_{\mathrm{\Gamma^{\star}}}\mathbb{E}\left[\mathcal{G}_{\mu}^{2}\right]\leq\sup_{\mu,v\in\Gamma^{\star}}\mathbb{E}\left[\left(\mathcal{G}_{\mu}-\mathcal{G}_{v}\right]^{2}\right)<+\infty.
$
Hence due to the zero mean function of $\mathcal{G}_{\mu}$, and Furnique's
inequality (see Relation (1,1) in Samorodnitsky (1991)), we have that
for $0<\varepsilon<1$, there exists a $\kappa\left(\varepsilon\right)$,
such that 
$
\mathbb{E}\left[\sup_{\mathrm{\Gamma^{\star}}}\mathcal{G}_{\mu}^{2}\right]=\int_{0}^{+\infty}\mathbb{P}\left(\sup_{\Gamma^{\star}}\left|\mathcal{G}_{\mu}\right|>\sqrt{y}\right)dy
\leq2\kappa\left(\varepsilon\right)\int_{0}^{+\infty}\exp\left(\frac{-\left(1-\varepsilon\right)}{2\sigma^{2}}y\right)dy<+\infty.
$
Then, Ch.\ 2 of Nualart (2006), (see the remark after the proof of Proposition
2.1.11 (p.\ 109)), implies the existence of the square integrable Malliavin
derivative for $\mathcal{G}_{\mu}$. Nualart (2006) implies then that
the Malliavin derivative of $\mathcal{G}_{\mu}$ equals zero only
at trivial triplets. The previous imply the validity of Assumption
1 of Arvanitis, Scaillet and Topaloglou (2019) for $\mathcal{T}=\left\{ 0\right\} $
in their notation, when trivial triplets exist, and $\mathcal{T}=\emptyset$
when trivial triplets do not exist. In the latter case, Theorem 1 of
Arvanitis, Scaillet and Topaloglou (2019) implies (\ref{eq:cons-1}),
for any $\alpha\in\left(0,1\right)$. In the former case, ND assumes
the existence of the non trivial $\left(\lambda^{\star},\kappa^{\star},z^{\star}\right)\in\Gamma^{\star}$ for
which we have that
$
\mathbb{P}\left(\sup\inf_{\Gamma^{\star}}\mathcal{G}\left(z,\lambda,\kappa\right)>0\right)\geq\mathbb{P}\left(\sup\inf_{\Gamma^{\star}}\mathcal{G}\left(z,\lambda,\kappa^{\star}\right)>0\right)
\geq\mathbb{P}\left(\mathcal{G}\left(z^{\star},\lambda^{\star},\kappa^{\star}\right)>0\right)=\frac{1}{2},
$
due to non-degeneracy and zero mean Gaussianity. The result then follows
again from Theorem 1 of Arvanitis, Scaillet and Topaloglou (2019), and
(\ref{eq:exact}) follows from the previous by noting that in this special
case, $\Gamma=\Gamma^{\star}$ due to Theorem 3.4
of Molchanov (2006).
\end{proof}

\section*{List of Factors}

We consider 6 different factor models:

1. The CAPM:
\begin{itemize}
\item Market (RM): Market excess return over the risk-free rate.
\end{itemize}

2. The Daniel, Hirshleifer, and Sun (DHS-2020) consists of the following 3 factors
\begin{itemize}
\item Market (RM): Market excess return over the risk-free rate.
\item The long-horizon financing factor (FIN) exploits the information in managers decisions to issue or repurchase equity in response to persistent mispricing.
\item The short-horizon earnings surprise factor (PEAD) is motivated by investor inattention and evidence of short-horizon underreaction, and captures short-horizon mispricing.
\end{itemize}

3. The Barillas and Shanken (2018), 6-factor model
\begin{itemize}
\item Market (RM): Market excess return over the risk-free rate.
\item Profitability (ROE): difference between the return on a portfolio of high return on equity (ROE) stocks and the return on a portfolio of low return on equity stocks
\item Investment (I/A):  difference between the return on a portfolio of low-investment stocks and the return on a portfolio of high-investment stocks
\item Size (SMB): Excess return of small firms over that of the large ones.
\item Value (HMLm): Based on book-to-market rankings that use the most recent monthly stock price in the denominator.
\item Momentum (UMD): Equal-weight average of firms with the highest 30 percent eleven-month returns lagged one month minus the equal-weight average of firms with the lowest 30 percent eleven-month returns lagged one month.
\end{itemize}

4. The Fama-French model (FF6 - 2016) consists of the following 6 factors:
\begin{itemize}
\item Market (RM): Market excess return over the risk-free rate.

\item Size (SMB): Excess return of small firms over that of the large ones.

\item Value (HML): Excess return of high book-to-market stocks over those
with low book-to-market.

\item Operating Profitability (RMW): Excess returns
of firms with high profitability over those with low.

\item Investment (CMA): Excess returns
of firms with low investment over those with high.

\item Momentum (Mom): winners minus losers.
\end{itemize}

5. The Stambaugh-Yuan (M4-2016) construct their factors from the same universe
with that used in FF5, although they adopt an approach that takes
into account the commonality that is present in 11 well-documented
anomalies. Their model (M4) comprises 4 factors:

\begin{itemize}
\item Market (RM): Market excess return over the risk-free rate, calibrated
however to the set of the 11 stock anomalies.

\item Size (SMB): Excess return of small firms over that of the large
ones, calibrated again to the set of anomalies.

\item Management (MGMT): Excess returns of stocks with high ranking on
management-related anomalies (Net Stock Issues, Composite Equity Issues,
Accruals, Net Operating Assets, Asset Growth, Investment to Assets)
over the return of those with low ranking.

\item Performance (PERF): Excess returns of stocks with high ranking
on ``performance''-related anomalies (Distress, O-Score, Momentum,
Gross Profitability, Return on Assets) over the return of those with
low ranking.
\end{itemize}

6. The Hou, Xue and Zhang (2015) q-4-factor model.
\begin{itemize}
\item Market (RM): Market excess return over the risk-free rate.

\item Size (SMB): Excess return of small firms over that of the large ones.

\item Profitability (ROE): difference between the return on a portfolio of high return on equity (ROE) stocks and the return on a portfolio of low return on equity stocks.

\item Investment (I/A):  difference between the return on a portfolio of low-investment stocks and the return on a portfolio of high-investment stocks.
\end{itemize}

\newpage

\begin{center}
\Large \textbf{ONLINE APPENDIX}\\
Sparse spanning portfolios and under-diversification \\
 with second-order stochastic dominance\\
Stelios Arvanitis, Olivier Scaillet, Nikolas Topaloglou
\end{center}

In Appendix A, we present a result that justifies the characterization of $M\left(\Lambda,\mathcal{L}_{p,q},\mathbb{P}\right)$ as optimal diversification loss, and an interpretation of the sparse optimal portfolios as approximately efficient. We gather Monte Carlo experiment  to assess the finite sample properties of our procedure for sparse SSD
spanning in Appendix B.

\section*{A. Theory}
\subsection*{Approximate Sparse Spanning}

We consider the optimization problem $M\left(\Lambda,\mathcal{L}_{p,q},\mathbb{P}\right):=\sup_{z\in Z}\sup_{\Lambda}\inf_{\mathcal{L}_{p,q}}\inf_{K}D\left(z,\kappa,\lambda,\mathbb{P}\right)$.
Even if $M\left(\Lambda,\mathcal{L}_{p,q},\mathbb{P}\right)>0$, so
that there exists no $K$ with $\mathrm{csupp}\left(K\right)\leq q$
for which SS-SSD holds, any solution to this problem has an interpretation
as an approximate sparse spanning subset of $\Lambda$ in the sense
of an expected utility loss as stated in the next proposition. For
$\mathcal{P}\left(Z\right)$ the set of probability distributions
(or equivalently cdfs) supported on $Z$, and for any $\mathbb{Q}$ there,
consider the Russell-Seo increasing and concave utility (see Russell
and Seo (1989)) $u_{\mathbb{Q}}\left(x\right):=\int_{Z}\min\left(0,x-z\right)d\mathbb{Q}$. 
\begin{prop}
\label{prop:ASSSp}$K\in\mathcal{L}_{p,q}$ does not solve $\sup_{\Lambda}\sup_{z\in Z}\inf_{\mathcal{L}_{p,q}}\inf_{K}D\left(z,\kappa,\lambda,\mathbb{P}\right)$,
iff there exists some $\lambda\in\Lambda$ and some $u_{\mathbb{Q}}$ such
that $\mathbb{E}\left[u_{\mathbb{Q}}\left(\sum_{i=0}^{\infty}\lambda_{i}X^{(i)}\right)\right]-\mathbb{E}\left[u_{\mathbb{Q}}\left(\sum_{i=0}^{\infty}\kappa_{i}X^{(i)}\right)\right]>M\left(\Lambda,\mathcal{L}_{p,q},\mathbb{P}\right)$,
for any $\kappa\in K$. 
\end{prop}
\begin{proof}
%[Proof of Proposition \ref{prop:ASSSp}]
$K$ does not solve $\sup_{z\in Z}\sup_{\Lambda}\inf_{\mathcal{L}_{p,q}}\inf_{K}D\left(z,\kappa,\lambda,\mathbb{P}\right)$
iff\\
 $\sup_{z\in Z}\sup_{\Lambda}\inf_{K}D\left(z,\kappa,\lambda,\mathbb{P}\right)>M\left(\Lambda,\mathcal{K}_{p,q},\mathbb{P}\right)$.
Using the same argument as in the proof of Lemma \ref{lem:Helpful_Compactness}
the latter is equivalent to $\sup_{F\in\mathcal{P}\left(Z\right)}\sup_{\Lambda}\inf_{K}\int_{Z}D\left(z,\kappa,\lambda,\mathbb{P}\right)dF\left(z\right)>M\left(\Lambda,\mathcal{L}_{p,q},\mathbb{P}\right)$.
Now, due to Fubini's Theorem we have that 
\[
\begin{array}{c}
\sup_{F\in\mathcal{P}\left(Z\right)}\sup_{\Lambda}\inf_{K}\int_{Z}D\left(z,\kappa,\lambda,\mathbb{P}\right)dF\left(z\right)\\
=\sup_{F\in\mathcal{P}\left(Z\right)}\sup_{\Lambda}\inf_{K}\int_{Z}\mathbb{E}\left[\left(z-\sum_{i=0}^{\infty}\kappa_{i}X^{(i)}\right)_{+}\right]-\mathbb{E}\left[\left(z-\sum_{i=0}^{\infty}\lambda_{i}X^{(i)}\right)_{+}\right]dF\left(z\right)\\
=\sup_{F\in\mathcal{P}\left(Z\right)}\sup_{\Lambda}\inf_{K}\int_{Z}\mathbb{E}\left(\left[\min\left(0,\sum_{i=0}^{\infty}\lambda_{i}X^{(i)}-z\right)-\min\left(0,\sum_{i=0}^{\infty}\kappa_{i}X^{(i)}-z\right)\right]\right)dF\left(z\right)\\
=\sup_{F\in\mathcal{P}\left(Z\right)}\sup_{\Lambda}\inf_{K}\mathbb{E}\left[\int_{Z}\min\left(0,\sum_{i=0}^{\infty}\lambda_{i}X^{(i)}-z\right)-\min\left(0,\sum_{i=0}^{\infty}\kappa_{i}X^{(i)}-z\right)dF\left(z\right)\right]\\
=\sup_{F\in\mathcal{P}\left(Z\right)}\left[\sup_{\Lambda}\mathbb{E}\left(u_{\mathbb{Q}}\left(\sum_{i=0}^{\infty}\lambda_{i}X^{(i)}\right)\right)-\sup_{K}\mathbb{E}\left(u_{\mathbb{Q}}\left(\sum_{i=0}^{\infty}\kappa_{i}X^{(i)}\right)\right)\right],
\end{array}
\]
and the result follows. 
\end{proof}

Hence, $M\left(\Lambda,\mathcal{L}_{p,q},\mathbb{P}\right)$ is the
optimal expected utility difference that the elements of any sparse
subset of $\Lambda$ of support dimension equal to $q$ can achieve
w.r.t.\ the elements of $\Lambda$ uniformly over the Russell-Seo utilities. It is thus interpreted as the minimal expected utility diversification loss occurring from ignoring investment opportunities of support greater than $q$ uniformly over the set of increasing and concave utilities.
The solutions to $\sup_{\Lambda}\sup_{z\in Z}\inf_{\mathcal{L}_{p,q}}\inf_{K}D\left(z,\kappa,\lambda,\mathbb{P}\right)$
are those subsets that actually achieve this optimality bound. We are interested in the investigation of $M\left(\Lambda,\mathcal{L}_{p,q},\mathbb{P}\right)$ as a function of $q$. This is obviously in any case non increasing and bounded below by zero to which it converges as $q\rightarrow\infty$; it is though of further interest to examine whether zero is approximately achieved for small values of $q$. Lemma
\ref{lem:Helpful_Compactness} along with the monotonicity of $\left(\Lambda_{p}\right)$
implies also that as $p\rightarrow +\infty$, $M\left(\Lambda,\mathcal{L}_{p,q},\mathbb{P}\right)\rightarrow M\left(\Lambda_{\infty},\mathcal{L}_{\infty,q},\mathbb{P}\right)$. Besides, due to the transitivity property of the relation, it is impossible for sparse spanning to hold as $p\rightarrow +\infty$ without holding for every $p>q$.

\subsection*{Sparse Approximately Efficient Elements}

For any $z\in Z$, since every increasing and concave utility up to a translation constant is represented via a convex combination of the Russell-Seo utilities  (see Russell
and Seo (1989)), and due to the utility representations in Proposition \ref{prop:ASSSp},
any solution to $\inf_{\Lambda}\mathbb{E}\left[\left(z-\sum_{i=0}^{\infty}\lambda_{i}X_{t}^{(i)}\right)_{+}\right]$
is an efficient element of $\Lambda$; it must be non-dominated
as an optimizer of the utility that corresponds to $F$ that concentrates
its mass on $z$. Thus any portfolio that results from
the solution to $\inf_{\mathcal{L}_{p,q}}\inf_{K}\mathbb{E}\left[\left(z-\sum_{i=0}^{\infty}\kappa_{i}X_{t}^{(i)}\right)_{+}\right]$
must be a sparse element of $\Lambda$ of support at most $q$. That
sparse element optimally approximates the efficient element since \textcolor{black}{{}$\sup_{\Lambda}\mathbb{E}\left[u_{\mathbb{Q}}\left(\sum_{i=0}^{\infty}\lambda_{i}X^{(i)}\right)\right]-\sup_{\mathcal{L}_{p,q}}\sup_{K}\mathbb{E}\left[u_{\mathbb{Q}}\left(\sum_{i=0}^{\infty}\kappa_{i}X^{(i)}\right)\right]$ is less than or equal to\linebreak$\sup_{\Lambda}\left[u_{\mathbb{Q}}\left(\sum_{i=0}^{\infty}\lambda_{i}X^{(i)}\right)\right]-\mathbb{E}\left[u_{\mathbb{Q}}\left(\sum_{i=0}^{\infty}\kappa_{i}X^{(i)}\right)\right]$,
for any $\kappa$ of support at most $q$. It is also an efficient
element of maximizer over $\mathcal{L}_{p,q}$ of $\sup_{K}\mathbb{E}\left[u_{\mathbb{Q}}\left(\sum_{i=0}^{\infty}\kappa_{i}X^{(i)}\right)\right]$.
When $M\left(\Lambda,\mathcal{L}_{p,q},\mathbb{P}\right)\leq0$, the
solution to}
 \textcolor{black}{$\inf_{\mathcal{L}_{p,q}}\inf_{K}\mathbb{E}\left[\left(z-\sum_{i=0}^{\infty}\kappa_{i}X_{t}^{(i)}\right)_{+}\right]$
is also efficient in $\Lambda$. When $K\in\mathcal{L}_{p,q}$ maximizes}
 \textcolor{black}{$\sup_{K}\mathbb{E}\left[u_{\mathbb{Q}}\left(\sum_{i=0}^{\infty}\kappa_{i}X^{(i)}\right)\right]$
uniformly in $z$, but does not span $\Lambda$, then there necessarily
exist efficient elements of $\Lambda$ that are not in $K$. Then
the portfolio that solves $\sup_{K}\mathbb{E}\left[u_{\mathbb{Q}}\left(\sum_{i=0}^{\infty}\kappa_{i}X^{(i)}\right)\right]$
uniformly in $z$ is by construction an efficient element of $K$
that minimizes $\mathbb{E}\left[u_{\mathbb{Q}}\left(\sum_{i=0}^{\infty}\lambda_{i}X^{(i)}\right)\right]-\mathbb{E}\left[u_{\mathbb{Q}}\left(\sum_{i=0}^{\infty}\kappa_{i}X^{(i)}\right)\right]$
uniformly w.r.t.\ the efficient set of $\Lambda$ and the Russell-Seo
utilities. Interestingly, it is an efficient element of $K$ that
maximizes a utility that corresponds to a distribution $F$ that concentrates
its mass on some threshold $z$. }

As $p\rightarrow +\infty$, any accumulation point of the solution to
$\inf_{\mathcal{L}_{p,q}}\inf_{K}\mathbb{E}\left[\left(z-\sum_{i=0}^{\infty}\kappa_{i}X_{t}^{(i)}\right)_{+}\right]$
is a $q$-sparse approximate efficient element of $\Lambda_{\infty}$.
If it is unique and independent of $z$, then it is also a portfolio
bound for the set of $q$-sparse portfolios (for the concept of portfolio
bounds on finite dimensional portfolio spaces, see Arvanitis et al.\
(2021)). In this case, every efficient element of $\Lambda_{\infty}$
is approximated by the same $q$-sparse approximate efficient element
of $\Lambda_{\infty}$. If $\inf_{\Lambda_{\infty}}\mathbb{E}\left[\left(z-\sum_{i=0}^{\infty}\lambda_{i}X_{t}^{(i)}\right)_{+}\right]$
has also a unique solution independent of $z$, then this is also
a portfolio bound of potentially infinite support of $\Lambda_{\infty}$.
When SS-SSD holds and $q$ is large enough, then those two bounds
coincide, and thereby $\Lambda_{\infty}$ admits a $q$-sparse bound.

\section*{B. Monte Carlo Experiments}
\setcounter{equation}{0}\def\theequation{A.\arabic{equation}}
\setcounter{table}{0}\def\thetable{A.\arabic{table}}

We gauge the finite sample properties of our sparse SSD Spanning methodology via two Monte Carlo (MC) experiments.
We rely on data generated processes driven by multivariate Gaussian distributions in i.i.d.\ settings.

\subsection*{First Experiment}

The first MC experiment is based on a problem with $N = p$ = 49, 100, 500 mutually i.i.d.\ normally distributed assets with $T$ = 300, 500, 1000 observations. Mutual i.i.d.-ness is used to invoke the aforementioned argument by Samuelson (1967) and it can be empirically motivated by the analysis of hedged returns of well-diversified portfolios. The number of assets are selected to match the results of the empirical application, where we use 49, 100, and 500 assets. 

The covariance matrix is fitted to the historical monthly returns of three datasets: the 49 Industry portfolios from Kenneth French's web page, the 100 assets of the FTSE 100 index, and the 500 assets of the S$\&$P 500 index. The mean vector in each case is calculated from the historical monthly returns of these assets. 

Based on the empirical application, we set $q=13$, for $N=49$,  $q=25$, for $N=100$, and finally, $q=45$, for $N=500$. We set the weights of the $N-q$ assets to zero to get sparse SSD spanning. For each combination of $N$ and $T$, we repeat 500 times the sparse selection  procedure described in the main text, and check how many times we get a number of assets close to $q$ on average. We additionally compute the average estimated loss across the Monte Carlo samples. Table A.1 exhibits the results of the first MC experiment. They show that our sparse SD Spanning methodology is accurate in recovering the number of assets and the expected utility loss.
 
\begingroup
\setlength{\tabcolsep}{6pt}
\renewcommand{\arraystretch}{0.5}
\begin{table}[h]
\caption{First Experiment}
\begin{center}
\scalebox{0.8}{
\begin{tabular}{llll} 
%\caption{First Monte Carlo Experiment.}
\toprule[1.5pt]
  Sample size $T$ & 300 & 500 & 1000
  \\ \midrule
Case 1: $N = 49$, $q = 13$ &  &  &  
\\ \midrule
Assets selected:     & & & \\
Average number    & 11.45 & 12.04 & 12.54 \\
St Deviation           & 1.18   & 1.12 & 1.13   \\
Variability of the Loss:  & & & \\
Average  Loss        & 0.003 & 0.001 & 0.0007  \\
Standard Error       & $10^{-4}$ & $10^{-4}$  &  $10^{-4}$ \\
\\ \midrule
Case 2: $N  = 100$, $q = 25$ &  &  &  
\\ \midrule
Assets selected:     & & & \\
Average number    & 22.57 & 23.02  & 23.88 \\
St Deviation           & 1.33  & 1.30 & 1.29  \\
Variability of the Loss:  & & & \\
Average  Loss        & 0.0009 & 0.0005 & 0.0002   \\
Standard Error       & $10^{-4}$ & $10^{-5}$ & $10^{-5}$ \\
   \\ \midrule
Case 3: $N = 500$, $q = 45$ &  &  &  
\\ \midrule
Assets selected:     & & & \\
Average number    & 42.3 & 42.85 & 43.34 \\
St Deviation           & 1.68  & 1.57 & 1.54  \\
Variability of the Loss:  & & & \\
Average   Loss       & 0.0008 & 0.0004 & 0.0001   \\
Standard Error       & $10^{-4}$ & $10^{-5}$ & $10^{-5}$ \\
  \\ \bottomrule[1.5pt]
\end{tabular}
}
\end{center}
\noindent\parbox[c][0.7\totalheight][s]{1\textwidth}
{%

\noindent {\footnotesize{}{}{}{}{}{}{}\begin{singlespace}The experiment is based on a problem with $N = 49$, 100, 500 normally distributed assets and $T$ = 300, 500, 1000 time series observations. 
We compute the average number of assets selected and the standard deviations of these. We also measure the variability of the loss, by computing the average loss and the standard error of the loss.\end{singlespace}}}
\label{MC1}
\end{table}
\endgroup

$ $

\subsection*{Second Experiment}

The second MC experiment is based on a problem with $N = p = 50$  jointly normally distributed assets with $T$ = 300, 500, 1000 observations. In this experiment, we evaluate the expected utility loss if $q$ is lower than the minimal spanning support size, or $q$ equals exactly the minimal spanning support size. Specifically, we consider a set A of 5 asset returns with equal means $\mu_A=0.3$ and  equal standard deviations $\sigma_A=0.15$, and a set B of 5 asset returns with equal means  $\mu_B=0.15$ and  equal standard deviations $\sigma_B=0.1$. Since $(\mu_A - \mu_B)/ (\sigma_B - \sigma_A) < 0 $, there is no portfolio in set A that dominates any portfolio in set B by SSD, and vice versa. The other 40 generated asset returns have equal means $\mu=0.1$ and equal standard deviations $\sigma=0.5$. The correlation coefficient of all $N$ asset returns is set to $\rho_{i,j}=0.001$ for any pairs of $i,j=1,\dots,N,\:i\neq j$. Any convex combination of assets that belong to sets A and B dominate any portfolio constructed from the other 40 assets by SS-SSD. We set $q$ equal to either 5 (no spanning) or 10 (spanning). For each $T$, we repeat 500 times the sparse procedure described in main text, and we compute the average number of selected assets and average estimated loss.  Table A.2 exhibits the results. They show that our sparse SSD Spanning methodology is also accurate in recovering the number of assets and the expected utility loss when the cardinality constraint is binding, here $q$=5.
Under a Gaussian design (elliptical distribution), SS-SSD corresponds to sparse MV-spanning. So the good performance of our methodology shows that we can also use it to get sparse MV-spanning portfolios even if the true data generating process is not sparse (asymptotic statistical guarantee).

\begingroup
\setlength{\tabcolsep}{6pt}
\renewcommand{\arraystretch}{0.5}
\begin{table}[H]
\caption{Second Experiment}
\begin{center}
\scalebox{0.8}{
\begin{tabular}{llll} \toprule[1.5pt] 
  Sample size $T$ & 300 & 500 & 1000
  \\ \midrule
Case 1: $q = 5$ &  &  &   
\\ \midrule
Assets selected:     & & & \\
Average number    & 4.82 & 4.88 & 4.94 \\
St Deviation           & $0.0135$   & $0.0108$ &$0.0086$  \\
Variability of the Loss:  & & & \\
Average  Loss        & 0.022 & 0.015 & 0.009  \\
Standard Error       & 0.0003 & 0.0003 & 0.0002 \\
\\ \midrule
Case 2: $q = 10$ &  &  &  
\\ \midrule
Assets selected:     & & & \\
Average number    & 9.86 & 9.91 & 9.96 \\
St Deviation           & $0.0116$  & $0.0102$ & $0.0086$  \\
Variability of the Loss:  & & & \\
Average  Loss        & $0.007$ & $0.004$ & $0.001$  \\
Standard Error       & $10^{-4}$ & $10^{-4}$ & $10^{-4}$ \\
\\ \bottomrule[1.5pt]
\end{tabular}}
\end{center}
\noindent\parbox[c][0.7\totalheight][s]{1\textwidth}
{%

\noindent {\footnotesize{}{}{}{}{}{}{}\begin{singlespace}The experiment is based on a problem with $N = 50$ normally distributed assets and $T$ = 300, 500, 1000 time series observations. 
We compute the average number of assets selected and the standard deviations of these. We also measure the variability of the loss, by computing the average loss and the standard error of the loss.\end{singlespace}}}
\label{MC2}
\end{table}
\endgroup

$ $ 

\noindent \textbf{Additional References}
\\

\noindent  Arvanitis, S., Post, T. and Topaloglou,
N., 2021. Stochastic Bounds for Reference Sets in Portfolio Analysis.
Management Science 67(12), 7291-7950.

\noindent Samuelson P. A., 1967. General Proof that Diversification Pays. Journal of  Financial and  Quantitative Analysis 2(1),1-13.


\begin{thebibliography}{10}

\bibitem{Al}Aliprantis C. D., and K. C. Border, 2006. \textit{Infinite
Dimensional Analysis: A Hitchhiker's Guide}. Springer.

\bibitem{Anderson_2013} Anderson, A., 2013. Trading and Under-Diversification. Review of Finance 17, 1699-1741.

\bibitem{key-48}Andrews, D. W., 1994. Empirical Process Methods in
Econometrics, in \textit{Handbook of Econometrics}, Vol.\ 4, Edited
by R.F.\ Engle and D.L.\ McFadden, 2247-2294.

\bibitem{Anyf} Anyfantaki, S., Maasoumi, E., Ren, J. and N. Topaloglou, 2022. 
Evidence of Uniform Inefficiency in Market Portfolios based on Dominance Tests. Journal of Business and Economic Statistics 40(3), 937-949.

\bibitem{Ao_Li_Zheng_2019} Ao, M., Li, Y., and Zheng, X., 2019.
Approaching Mean-Variance Efficiency for Large Portfolios. Review of
Financial Studies 32(7), 2890-2919.

\bibitem{AHP} Ardia, D., Laurent, S., and Sessinou,
R., 2024. High-Dimensional Mean-Variance Spanning Tests. Working Paper HEC Montr\'eal

\bibitem{ART}Artzner, P., Delbaen, F., Eber, J-M., and Heath, D., 1998. Coherent Measures of Risk. Mathematical Finance 9(3), 203-228.

\bibitem{AHP} Arvanitis, S., Hallam, M. S., Post, T., and Topaloglou,
N., 2019. Stochastic Spanning. Journal of Business and Economic Statistics
37(4), 573-585.

\bibitem{ap23}Arvanitis, S. and Post, T., 2024. Generalized Stochastic Arbitrage Opportunities. Management Science 70(7), 4167-4952.



\bibitem{AST19} Arvanitis, S., Scaillet, O., and Topaloglou, N.,
2020a. Spanning Analysis of Stock Market Anomalies under Prospect Stochastic
Dominance. Forthcoming, Management Science.

\bibitem{AST19} Arvanitis, S., Scaillet, O., and Topaloglou, N.,
2020b. Spanning Tests for Markowitz Stochastic Dominance. Journal of
Econometrics 217(2), 291-311.

\bibitem{at}Arvanitis, S., and Topaloglou, N., 2017. Testing for
Prospect and Markowitz Stochastic Dominance Efficiency. Journal of
Econometrics 198(2), 253-270.

\bibitem{key-3} Athayde G., and Flores, R., 2004. Finding a Maximum
Skewness Portfolio: A General Solution to Three-Moments Portfolio
Choice. Journal of Economic Dynamics and Control 28, 1335-1352.

\bibitem{key-Bar}Barillas, F. and Shanken, J.,   2016. Which Alpha?
Review of Financial Studies 30(4), 1316-1338.

\bibitem{BAWA}Bawa, V., 1975. Optimal Rules for Ordering Uncertain Prospects. Journal of Financial Economics 2, 95-121.

%\bibitem{key-55}Beer, G., and Lucchetti, R., 1991. Convex Optimization
%and the Epi-Distance Topology. Transactions of the American Mathematical
%Society 327(2), 795-813.

\bibitem{key-14-1} Berk, J., 1997. Necessary Conditions for the CAPM.
Journal of Economic Theory 73, 245-257.

\bibitem{Bertsimas} Bertsimas, D., and Tsitsiklis, J., 1997. \textit{Introduction
to Linear Optimization}. 1st edition, Athena Scientific, Belmont.

\bibitem{Bian_etal_2017} Bian, A., Buhmann, J., Krause, A., and Tschiatschek,
S., 2017. Guarantees for Greedy Maximization of Non-submodular Functions
with Applications. In Proceedings of the 34th International Conference
on Machine Learning, 498-507.

\bibitem{Borw}Borwein, J., and Lewis, A.S., 2010. \textit{Convex Analysis
and Nonlinear Optimization: Theory and Examples}. Springer Science
and Business Media.

\bibitem{Brodie_etal_2009} Brodie, J., Daubechies, I., DeMol, C., Giannone, D., and Loris,
I., 2009. Sparse and Stable Markowitz Portfolios. Proceedings of the National Academy of Sciences
106(30), 12267-12272.

\bibitem{Buchbinder_Feldman_2018} Buchbinder, N., and Feldman, M.,
2018. Submodular Functions Maximization Problems - A Survey in \textit{Handbook
of Approximation Algorithms and Metaheuristics}, 2nd Ed., Ed.\ T.\ Gonzalez,
Vol.\ 1, Chapman and Hall, London, 753-788.

\bibitem{Calvet_etal_2007} Calvet, L., Campbell, J., and Sodini, P., 2007. Down or Out: Assessing the Welfare Costs of
Household Investment Mistakes. Journal of Political Economy 115(5), 707-747.

\bibitem{Campbell_2006} Campbell, J., 2006. Household Finance. Journal of Finance 61(4), 1553-1604.

\bibitem{Caner_etal_2023} Caner, M., Medeiros, M., and Vasconcelos, G., 2023. Sharpe Ratio Analysis in High Dimensions: Residual-Based Nodewise Regression in Factor Models. Journal of Econometrics 235(2), 393-417.

\bibitem{carc}C\'arcamo, J., Cuevas, A. and Rodríguez, L.A., 2020. Directional Differentiability for Supremum-Type Functionals: Statistical Applications. Bernoulli 26(3), 2143-2175.

\bibitem{key-18} Chamberlain, G., 1983. A Characterization of the
Distributions that Imply Mean-Variance Utility Functions. Journal
of Economic Theory 29, 185-201.

\bibitem{Chen} Chen, Y., Li, S.Z., Tang Y., and Zhou G, 2023. Anomalies as New Hedge Fund Factors: A Machine Learning Approach. Working paper.


%\bibitem{key-5}Cong, Y., Chen, B., and Zhou, M., 2017. Fast Simulation
%of Hyperplane-Truncated Multivariate Normal Distributions. Bayesian
%Analysis 12(4), 1017-1037.

\bibitem{DHS}Daniel, K., Hirshleifer, D., and L. Sun, 2020. Short- and Long-Horizon 
Behavioral Factors. Review of Financial Studies 33(4), 1673-1736.


\bibitem{key-20} Danthine, J.P., and Donaldson, J., 2014. \textit{Intermediate
Financial Theory}. 3rd edition, North-Holland, Amsterdam.

\bibitem{Das_Kempe_2011} Das, A. and Kempe, D., 2011. Submodular
Meets Spectral: Greedy Algorithms for Subset Selection, Sparse Approximation
and Dictionary Selection. In Proceedings of the 28th International
Conference on Machine Learning, 1057-1064.

\bibitem{Dav}Davidson, J., 1994. \textit{Stochastic Limit Theory:
An Introduction for Econometricians}. Oxford University Press.

\bibitem{DeMiguel_etal_2009}DeMiguel, V., Garlappi, L., Nogales,
F. J., and Uppal, R., 2009. A Generalized Approach to Portfolio Optimization:
Improving Performance by Constraining Portfolio Norms. Management
Science 55(5), 798-812.

\bibitem{key-1M}DeMiguel, V., Garlappi, L., and Uppal, R. (2009).
Optimal versus Naive Diversification: How Inefficient is the 1/n Portfolio
Strategy? Review of Financial Studies, 22, 1915-1953.


\bibitem{key-23} De Roon, F. , Nijman, T., and Werker, B., 2001.
Testing for Mean-Variance Spanning with Short Sales Constraints and
Transaction Costs: the Case of Emerging Markets. Journal of Finance
56, 721-742.

\bibitem{key-46}Dimmock, S. G., Kouwenberg, R., Mitchell, O. S.,
and Peijnenburg, K., 2021. Household Portfolio Underdiversification
and Probability Weighting: Evidence from the Field. Review of Financial
Studies 34(9), 4524-4563.

\bibitem{DMR}Doukhan, P., Massart, P., and Rio, E., 1995. Invariance
Principles for Absolutely Regular Empirical Processes. Annales de
l'IHP Probabilit\'{e}s et Statistiques 31(2), 393-427.

\bibitem{edd}Edelsbrunner, H., 2014. \textit{A Short Course in Computational
Geometry and Topology} (No. Mathematical Methods). Berlin, Germany:
Springer.

\bibitem{EKDN}Elenberg, E. R., Khanna, R., Dimakis, A. G., and Negahban,
S., 2018. Restricted Strong Convexity Implies Weak Submodularity.
Annals of Statistics 46(68), 3539-3568.

\bibitem{Elton_Gruber_1977} Elton, E., and Gruber, M., 1977. Risk
Reduction and Portfolio Size: An Analytical Solution. Journal of Business
50(4), 415-437.

\bibitem{EK}Estrada, R., and Kanwal, R. P., 2012. \textit{A Distributional
Approach to Asymptotics: Theory and Applications}. Springer Science
and Business Media.


\bibitem{Evans_Archer_1968} Evans, J., and Archer, S., 1968. Diversification
and the Reduction of Dispersion-an Empirical Analysis. Journal of
Finance 23(5), 761-767.

\bibitem{key-F3}Fama, E. F. and French, K. R. 2016. Dissecting Anomalies
with a Five-Factor Model. Review of Financial Studies 29(1), 69-103.


\bibitem{Fan_etal_2015} Fan, J., Liao, Y., and Shi, X., 2015. Risks
of Large Portfolios. Journal of Econometrics 186(2), 367-387.

\bibitem{Fan_etal_2012} Fan, J., Zhang, J., and Yu, K., 2012. Vast
Portfolio Selection with Gross-Exposure Constraints. Journal of the
American Statistical Association 107(498), 592-606.

%\bibitem{FS}Fang, Z., and Santos, A., 2014. Inference on Directionally
%Differentiable Functions. arXiv preprint arXiv:1404.3763.

\bibitem{fish}Fishburn, P., 1977. Mean-Risk Analysis with Risk Associated with Below-Target Returns. American Economic Review 67, 116-126.

\bibitem{Francq_Zakoian_2011} Francq, C., and Zakoian, J.M., 2011.
 \textit{GARCH Models: Structure, Statistical Inference and Financial Applications}.
John Wiley and Sons, New York.

\bibitem{GAO}Gao, R., Chen, X. and Kleywegt, A.J., 2017. Distributional
Robustness and Regularization in Statistical Learning. arXiv preprint
arXiv:1712.06050.

\bibitem{gine}Gin\'{e}, E., and Nickl, R., 2016. \textit{Mathematical
Foundations of Infinite-Dimensional Statistical Models (Vol.\ 40)}.
Cambridge University Press.

\bibitem{key-26}Hadar, J., and Russell, W., 1969. Rules for Ordering
Uncertain Prospects. American Economic Review 59, 2-34.

\bibitem{key-27} Hanoch, G., and Levy, H., 1969. The Efficiency Analysis
of Choices Involving Risk. Review of Economic Studies 36, 335-346.

\bibitem{HUL}Hiriart-Urruty, J. B., and Lemar\'{e}chal, C., 2004.
\textit{Fundamentals of Convex Analysis}, Springer.

\bibitem{HUL-1}Hiriart-Urruty, J. B., and Lemar\'{e}chal, C., 2013.
\textit{Convex Analysis and Minimization Algorithms I: Fundamentals
(Vol.\ 305)}. Springer Science and Business media.

%\bibitem{key-16}Horv\'{a}th, L., Kokoszka, P., and Zitikis, R., 2006.
%Testing for Stochastic Dominance using the Weighted McFadden-Type
%Statistic. \textit{\textcolor{black}{Journal of Econometrics}} 133(1),
%191-205.

\bibitem{Hong_Scaillet_2006} Hong, H., and Scaillet, O., 2006, A Fast Subsampling Method for Nonlinear Dynamic Models. Journal of Econometrics 133, 557-578. 

\bibitem{key-Hou1}Hou, K., Chen, X., and Zhang, L., 2015. Digesting
Anomalies, Review of Financial Studies 28, 650-705.

\bibitem{key-Hou2}Hou, K., Chen, X., and Zhang, L., 2020. Replicating Anomalies, Review of Financial Studies 33(5), 2019-2133.

\bibitem{key-Hou2}Hou, K., Chen, X., and Zhang, L., 2021. An Augmented q-Factor Model with Expected Growth. Review of Finance 25(1), 1-41.

\bibitem{key-29} Huberman, G., and Kandel, S., 1987. Mean-Variance
Spanning. Journal of Finance 42, 873-888.

\bibitem{hwang}Hwang, S.G., 2004. Cauchy's Interlace Theorem for Eigenvalues of Hermitian Matrices. The American Mathematical Monthly 111(2), 157-159.

\bibitem{Jagannathn_Ma_2003}Jagannathan, R., and Ma, T., 2003. Risk
Reduction in Large Portfolios: Why Imposing the Wrong Constraints
Helps. Journal of Finance 58(4), 1651-84.

\bibitem{jensen} Jensen T. I., Kelly B., and Pedersen L. H., 2023. Is there a replication crisis in Finance? Journal of Finance 78(5), 2465-2518.

\bibitem{key-31} Jobson, J., and Korkie, B., 1989. A Performance
Interpretation of Multivariate Tests of Asset Set Intersection, Spanning
and Mean-Variance Efficiency. Journal of Financial and Quantitative
Analysis 24, 185-204.

\bibitem{key-33} Kan, R., and Zhou, G., 2012. Tests of Mean-Variance
Spanning. Annals of Economics and Finance 13(1), 145-193.

\bibitem{KP}Kim, J. and Pollard, D., 1990. Cube Root Asymptotics. Annals of Statistics 18(1), 191-219.

\bibitem{KT}Klein, E., and A. Thompson, 1984. \textit{Theory of Correspondences}.
Wiley-Interscience, New York.

%\bibitem{Knight}Knight, K., 1999. Epi-convergence in Distribution
%and Stochastic Equi-semicontinuity, Unpublished manuscript, 37, 28-72.

\bibitem{Klemkosky_Martin_1975} Klemkosky, R., and Martin, J., 1975.
The Effect of Market Risk on Portfolio Diversification. Journal of
Finance 30(1), 147-154.

\bibitem{key-Koi} Koijen, R., and Yogo, M., 2019. A Demand System
Approach to Asset Pricing. Journal of Political Economy 127(4), 1475-1515.

\bibitem{key-36} Kuosmanen, T., 2004. Efficient Diversification According
to Stochastic Dominance Criteria. Management Science 50, 1390-1406.

\bibitem{Ledoit_Wolf_2017} Ledoit, O., and Wolf, M., 2017. Nonlinear
Shrinkage of the Covariance Matrix for Portfolio Selection: Markowitz
Meets Goldilocks. Review of Financial Studies 30, 4349-88.

\bibitem{Linton_2005} Linton, O., Maasoumi, E., and  Whang, Y.J., 2005. Consistent Testing for Stochastic Dominance under General Sampling Schemes. Review of Economic Studies 72, 735-765.

\bibitem{Linton_2014} Linton, O., Post, T., and Whang, Y.J., 2014. Testing for the Stochastic Dominance Efficiency of a Given Portfolio. Econometrics Journal 17(2), 59-74.

\bibitem{Liu_2014} Liu, H.,  2014. Solvency Constraint, Underdiversification,
and Idiosyncratic Risks. Journal of Financial
and Quantitative Analysis 49(2), 409-430.

\bibitem{key-39} Markowitz, H., 1952. Portfolio Selection. Journal
of Finance 7(1), 77-91.

\bibitem{Meinhausen_Buhlmann_2006f} Meinshausen, N., and Buehlmann,
P., 2006. Consistent Neighbourhood Selection for High-dimensional
Graphs with the Lasso. Annals of Statistics 34(3), 2006.

\bibitem{key-40} Mencia, J., and Sentana, E., 2009. Multivariate
Location-Scale Mixtures of Normals and Mean-Variance-Skewness Portfolio
Allocation. Journal of Econometrics 153, 105-121.

\bibitem{Molch}Molchanov, I., 2006. \textit{Theory of Random Sets}.
Springer.

\bibitem{Nemhauser_Wolsey_1978} Nemhauser, G., and Wolsey, L., 1978.
Best Algorithms for Approximating the Maximum of a Submodular Set
Function. Mathematics of Operations Research 3(3), 177-188.

\bibitem{Nemhauser_Wolsey_Fisher_1978} Nemhauser, G., Wolsey, L.,
and Fisher, M., 1978. An Analysis of Approximations for Maximizing
Submodular Set Functions - I. Mathematical Programming 14, 265-294.

\bibitem{Nest}Nesterov, Y., 1998, \textit{Introductory Lectures on
Convex Optimization: A Basic Course}. Springer.

\bibitem{key-1}Nickl, R., and P\"{o}tscher, B. M., 2007. Bracketing
Metric Entropy Rates and Empirical Central Limit Theorems for Function
Classes of Besov-and Sobolev-type. Journal of Theoretical Probability,
20(2), 177-199.

\bibitem{Novy}Novy-Marx, R., 2013, The Other Side of Value: The Gross Profitability Premium, Journal of Financial
Economics 108 (1), 1-28.

\bibitem{nual}Nualart, D., 2006. \textit{The Malliavin Calculus and Related Topics}. Springer Probability and its Applications.

\bibitem{key-42} Owen, J., and Rabinovitch, R., 1983. On the Class
of Elliptical Distributions and their Applications to the Theory of
Portfolio Choice. Journal of Finance 38, 745-752.

\bibitem{key-1}Politis, D. N., Romano, J. P., and Wolf, M., 1999.
\textit{Subsampling}. Springer Science and Business Media.

\bibitem{key-44} Post, T., 2003. Empirical Tests for Stochastic Dominance
Efficiency. Journal of Finance 58, 1905-1932.

\bibitem{key-43} Penaranda, F., and Sentana, E., 2012. Spanning Tests
in Portfolio and Stochastic Discount Factor Mean-Variance Frontiers:
A Unifying Approach. Journal of Econometrics 170, 303-324.

\bibitem{Rad}Radulovi\'{c}, D., 1998. On the Subsample Bootstrap
Variance Estimation. Test 7(2), 295-306.

\bibitem{rob} Robin, J.M. and Smith, R.J., 2000. Tests of Rank. Econometric Theory 16(2),151-175.

\bibitem{key-46} Rothschild, M., and Stiglitz, J., 1970. Increasing
Risk: I. A Definition. Journal of Economic Theory 2(3), 225-243.

\bibitem{key-6}Rio, E., 2017. \textit{Asymptotic Theory of Weakly
Dependent Random Processes (Vol. 80)}, Berlin: Springer.

\bibitem{RS}Russell, W.R. and Seo, T.K., 1989. \textit{Representative
Sets for Stochastic Dominance Rules}, in \textit{Studies in the Economics
of Uncertainty}: In honor of Josef Hadar, Springer-Verlag.

%\bibitem{SW} Salinetti, G., and Wets, R. J., 1977. On the Relations
%between Two Types of Convergence for Convex Functions. Journal of
%Mathematical Analysis and Applications 60(1), 211-226.

\bibitem{key-6-1}\textcolor{black}{Samorodnitsky, G., 1991. Probability
Tails of Gaussian Extrema}. Stochastic Processes
and their Applications 38(1), 55-84.

\bibitem{Sav}Savare, G., 1996. On the Regularity of the Positive
Part of Functions. Nonlinear Analysis, Theory, Methods and
Applications 27(9), 1055-1074.

%\bibitem{ST}Scaillet, O., and Topaloglou, N., 2010. Testing for Stochastic
%Dominance Efficiency. Journal of Business and Economic Statistics
%28, 169-180.

\bibitem{shap}Shapiro, A., 1990. On Concepts of Directional Differentiability. Journal of Optimization Theory and Applications 66, 477-487.

\bibitem{key-48} Simaan, Y., 1993. Portfolio Selection and Asset
Pricing-Three-Parameter Framework. Management Science 39, 568-577.


\bibitem{Sion}Sion, M., 1958. On General Minimax Theorems. Pacific
Journal of Mathematics 8(1), 171-176.

\bibitem{key-2S}Sortino, F. and van der Meer, R. 1991. Downside Risk-Capturing
What's at Stake in Investment Situations. Journal of Portfolio Management,
17, 27-31.

\bibitem{key-3Sta}Stambaugh, R.F., and Yuan, Y., 2016, Mispricing Factors. Review of Financial Studies 30, 1270-1315.

\bibitem{TK}Takemura, A. and Kuriki, S., 2003. Tail Probability via Tube Formula when the Critical Radius is Zero. Bernoulli 9(3), 535-558.

\bibitem{Trench}Trench, W.F., 1999. Asymptotic Distribution of the Spectra of a Class of Generalized Kac-Murdock-Szego Matrices. Linear Algebra and its Applications 294, 181-192.

\bibitem{Tibshirani_1996} Tibshirani, R., 1996. Regression Shrinkage
and Selection via the Lasso. Journal of the Royal Statistical Society:
Series B 58(1), 267-288.

\bibitem{Vaart_AS}  van der Vaart, A.W., 2000. \textit{Asymptotic Statistics}.
Cambridge University Press.

\bibitem{vvw} van der Vaart, A. W., and Wellner, J. A., 1996. \textit{Weak
Convergence and Empirical Processes}. Springer, New York, NY.

\bibitem{VanNieuwerburgh_2010} Van Nieuwerburgh, S., and Weldkamp, L., 2010. Information Acquisition and Under-diversification.
Review of  Economic Studies 77, 779-805.

\bibitem{Wain_HDS}Wainwright, M. J., 2019. \textit{High-Dimensional
Statistics: A Non-Asymptotic Viewpoint (Vol.\ 48)}. Cambridge University
Press.

\bibitem{Zhang_2009} Zhang, T., 2009. On the Consistency of Feature
Selection using Greedy Least Squares Regression. Journal of Machine
Learning Research 10, 555-568.

\bibitem{Zhao_Yiu_2006} Zhao, P., and Yiu, B., 2006. On Model Selection
Consistency of Lasso. Journal of Machine Learning Research 7, 2541-2563. 

\bibitem{key-3Z} Ziemba, W., 2005. The Symmetric Downside Risk Sharpe
Ratio. Journal of Portfolio Management 32(1), 108-122.
\end{thebibliography}
\end{document}